\documentclass[10pt]{article}

\usepackage{amsmath,amsmath,amssymb}
\usepackage{mathrsfs}
\usepackage{multirow}
\usepackage{graphicx}
\usepackage{authblk}
\usepackage{indentfirst}
\usepackage{multicol}  
\usepackage{tabu}
\usepackage{tabularx}
\usepackage{url}
\usepackage{fancyhdr}
\usepackage[numbers]{natbib}
\usepackage{lineno}
\usepackage{makecell}
\usepackage{fancyhdr}
\usepackage{colortbl}
\usepackage{caption}

\usepackage{hyperref}
\hypersetup{colorlinks=true,linkcolor=blue,anchorcolor=blue,citecolor=blue}

\title{Manifold Topological Deep Learning for Biomedical Data}

\author[1]{Xiang Liu} 
\author[1]{Zhe Su} 
\author[2]{Yongyi Shi} 
\author[3]{Yiying Tong} 
\author[2]{Ge Wang} 
\author[1,4,5]{Guo-Wei Wei \thanks{Corresponding author: weig@msu.edu}}
\affil[1]{Department of Mathematics, Michigan State University, MI, 48824, USA}
\affil[2]{Biomedical Imaging Center, Rensselaer Polytechnic Institute, NY, 12180, USA}
\affil[3]{Computer Science and Engineering, Michigan State University, MI 48824, USA}
\affil[4]{Department of Electrical and Computer Engineering, Michigan State University, MI 48824, USA}
\affil[5]{Department of Biochemistry and Molecular Biology, Michigan State University, MI 48824, USA}

\date{}

\begin{document}
\maketitle

\paragraph{Abstract} 
Recently, topological deep learning (TDL), which integrates algebraic topology with deep neural networks, has achieved tremendous success in processing point-cloud data, emerging as a promising paradigm in data science. However, TDL has not been developed for data on differentiable manifolds, including images, due to the challenges posed by differential topology. We address this challenge by introducing manifold topological deep learning (MTDL)
for the first time. To highlight the power of Hodge theory rooted in differential topology, we consider a simple convolutional neural network (CNN) in MTDL. In this novel framework, original images are represented as smooth manifolds with vector fields that are decomposed into three orthogonal components based on Hodge theory. These components are then concatenated to form an input image for the CNN architecture. The performance of MTDL is evaluated using the MedMNIST v2 benchmark database, which comprises 717,287 biomedical images from eleven 2D and six 3D datasets. MTDL significantly outperforms other competing methods, extending TDL to a wide range of data on smooth manifolds.

  
\newpage
	
\section{Introduction}
Topological deep learning (TDL) is an emerging field that integrates topological methods with deep learning techniques to perform learning tasks such as regression, classification, and representation learning \cite{hajij2022topological}. Unlike traditional black-box deep learning models, TDL models offer greater interpretability by leveraging topological features and representations to explicitly capture the underlying geometric and structural properties of data. Since its introduction in 2017 \cite{cang2017topologynet}, TDL has rapidly evolved, leading to a diverse set of methods and models. Besides its methodological advances, TDL has been successfully applied in various domains, including biology, chemistry, materials science, neuroscience, and social networks \cite{nguyen2019mathematical,papamarkou2024position}. Current TDL models mainly focus on combinatorial data structures, such as point clouds and graphs. Compared to combinatorial data, differentiable manifold data contains richer geometric information and is more suitable for analysis using methods from differential topology, such as differential forms and differential operators. These methods enable the study of continuous, smooth phenomena that cannot be adequately captured through purely combinatorial approaches. Despite this advantage, there are currently no TDL models designed for differentiable manifold data. 

Differentiable manifolds, such as curves and surfaces, are ubiquitous in real-world data. For example, DNA chains, object surfaces, and images are all natural examples of differentiable manifolds. Thus, extending TDL to differentiable manifold data is both meaningful and necessary. However, two primary challenges hinder this extension: firstly, although the images are inherently manifold data, it is nontrivial to rigorously model them as differentiable manifolds while preserving essential differentiable and topological properties. Secondly, designing an efficient model that combines the mathematical methods from differential topology with deep learning poses theoretical and computational challenges.

With the advancements in Topological Data Analysis \cite{carlsson2009topology,edelsbrunner2010computational}, particularly the remarkable successes achieved by persistent homology \cite{nguyen2020mathdl}, several studies have employed topological methods for manifold data analysis. For instance, using simplicial complexes or cubical complexes to model images \cite{ziou2002generating} and using curves or knots to represent amino acid chains \cite{shen2024knot}, then computing persistent homology for data analysis. While these methods are powerful in capturing the topological structures of manifold data across various scales, they exhibit limitations in capturing the smooth differentiable information within the manifold data, which can be addressed by incorporating methods from differential topology, such as vector fields, differential forms, and differential operators. Moreover, although differentiable manifolds have been utilized in manifold topological learning \cite{singh2023topological}, such as in modeling protein-ligand complexes \cite{su2024persistent}, these methods have not yet been extended to deep learning architectures.

Recently, a discrete topology-preserving Hodge theory for differentiable manifolds embedded in Cartesian grids has been introduced \cite{su2024hodge} and successfully applied to single-cell RNA velocity analysis \cite{su2024hodge2}. This theory provides an efficient way for modeling images as differentiable manifolds since images are naturally embedded in Cartesian grids. On the other hand, the MedMNIST v2 dataset offers a standard and reliable benchmark for evaluating model performance in medical image classification. The MedMNIST v2 dataset contains twelve 2D datasets and six 3D datasets, covering major medical data modalities, the data scale ranges from 100 to 100000, and the task type includes binary, multi-class classification, ordinal regression, and multi-label classification, making it highly suitable for assessing the efficiency, robustness, and generalizability of models \cite{yang2023medmnist}. 

Here, we introduce, for the first time, a Manifold Topological Deep Learning (MTDL) Model use the de Rham-Hodge theory, a landmark of the 20th Century's mathematics. MTDL integrates the discrete Hodge theory from differential topology, the Transformer encoder architecture, and convolutional operations, providing a novel framework for extending TDL to differentiable manifold data. In the MTDL model, the input image is represented as a discrete differentiable manifold and a vector field defined on this manifold. The Hodge Laplacian theory is then employed to decompose the vector field into three orthogonal components: curl-free, divergence-free, and harmonic parts. These components are concatenated to form a new image representation, which is passed to the CNN architecture for the prediction task. We evaluate MTDL on the MedMNIST v2 dataset, including 717,287  images from eleven 2D datasets and six 3D datasets. MTDL significantly outperforms other models, establishing MTDL as an efficient framework for TDL on differentiable manifold data.

\section{Results}
\subsection{Overview of MTDL}
\begin{figure}[ht]
    \centering
    \includegraphics[width=0.9\textwidth]{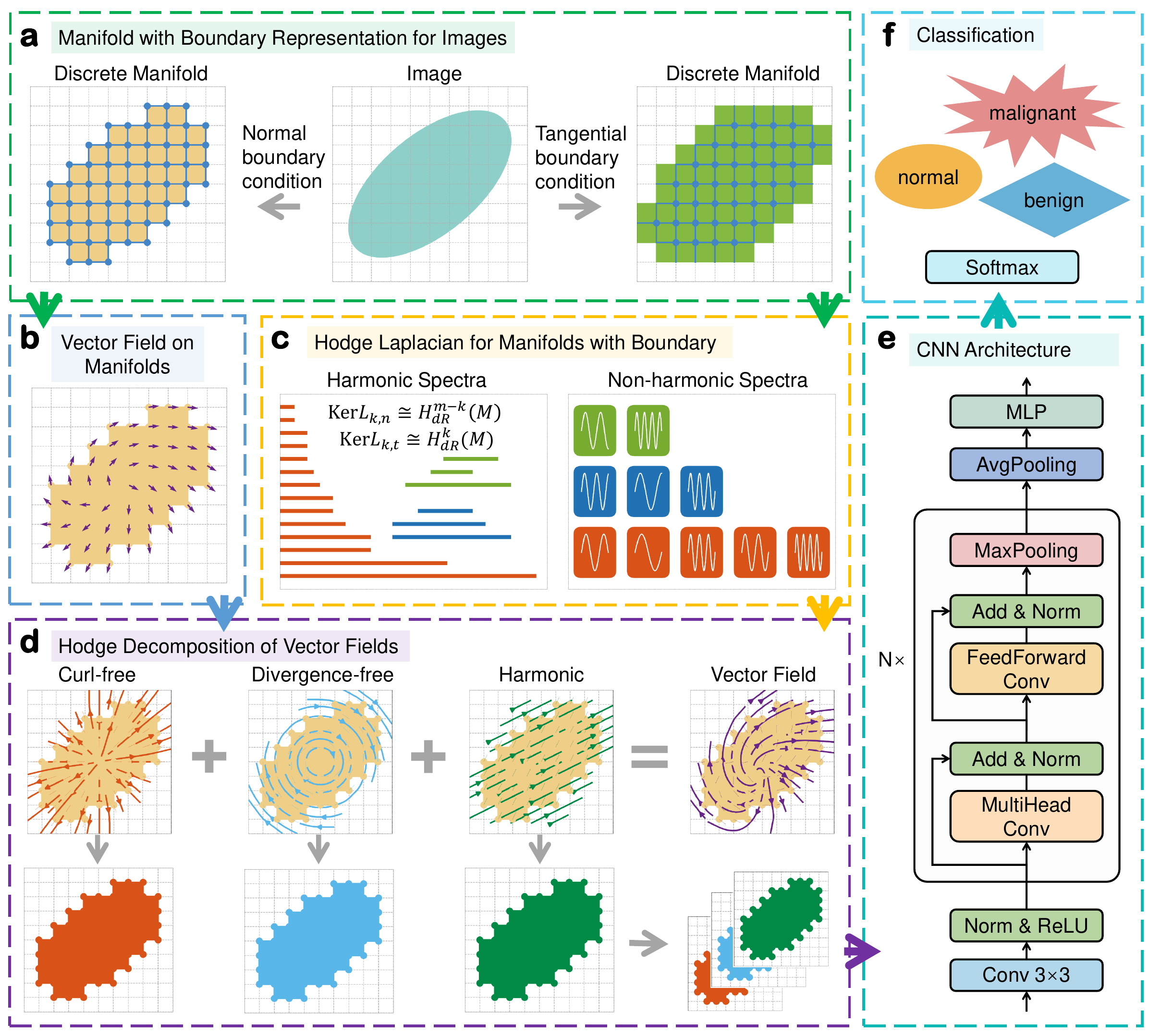}
    \caption{Model architecture of MTDL. The original image is first modeled as a discrete manifold on Cartesian grids under specific boundary conditions (\textbf{a}). A vector field is then constructed on the manifold (\textbf{b}). Using the discrete Hodge Laplacian for manifolds with boundary (\textbf{c}), this vector field is decomposed into three orthogonal components: curl-free, divergence-free, and harmonic parts (\textbf{d}). These components are subsequently concatenated to form a multi-channel image, which serves as the input of CNN for the classification task (\textbf{e}). }
    \label{fig:model-architecture}
\end{figure}
The discrete Hodge theory on Cartesian grids provides an approach for decomposing an image into three distinct components, each capturing different geometric and topological features. To leverage this theory in TDL, we propose a MTDL model that integrates discrete Hodge theory, Transformer encoder architecture, and convolutional operations for image classification. The architecture of MTDL is illustrated in Fig. \ref{fig:model-architecture}. As shown in the figure, the original image is first represented as a discrete manifold on Cartesian grids under normal or tangential boundary conditions (Fig. \ref{fig:model-architecture}\textbf{a}). This manifold representation establishes a mathematical formalization of the images, serving as the groundwork for further analysis using Hodge theory, such as using the harmonic spectra of Hodge Laplacian to detect the loop structures of the manifold (Fig. \ref{fig:model-architecture}\textbf{c}). Subsequently, a vector field that encodes the image information is constructed on the discrete manifolds (Fig. \ref{fig:model-architecture}\textbf{b}). There are several methods for constructing vector fields from images (Supplementary Information), each method provides a specific perspective on the image's content and structure. The generated vector field is then decomposed into three orthogonal components through the Hodge decomposition. These components, including curl-free, divergence-free, and harmonic parts, are concatenated to form a multi-channel representation of the decomposed images (Fig. \ref{fig:model-architecture}\textbf{d}). Finally, the resulting representation is fed into the CNN for image classification (Fig. \ref{fig:model-architecture}\textbf{e}). This CNN is based on the Transformer encoder architecture by adding a maxpooling operation and replacing the multihead attention and feedforward layers with convolution operations. 

\subsection{Evaluation of MTDL}
\subsubsection{Dataset}
The MedMNIST v2 dataset \cite{yang2023medmnist} is an updated version of the original MedMNIST dataset \cite{yang2021medmnist}. It is an MNIST-like collection of standardized biomedical images comprising twelve 2D datasets and six 3D datasets that cover primary medical imaging modalities, such as X-ray, Optical Coherence Tomography (OCT), Ultrasound, Computed Tomography (CT), Electron Microscope, and Magnetic Resonance Angiography (MRA). These datasets support a wide range of classification tasks, including binary classification, multi-class classification, ordinal regression, and multi-label classification. The data sizes range from 100 to 100,000 samples. In total, MedMNIST v2 includes 708,069 2D images and 9,998 3D images, with standard train-validation-test
splits provided for all datasets.

Among these datasets, BreastMNIST2D is derived from a dataset of 780 breast ultrasound images \cite{al2020dataset}. The original dataset has been reported to contain certain inconsistencies that could significantly impact model performance \cite{pawlowska2023re}. To ensure the validity and reliability of our evaluation, we exclude this dataset and utilize the remaining eleven 2D datasets along with all six 3D datasets for assessing our model's performance. The image resolutions we used are $224\!\times\! 224$ for 2D images and $64\!\times\! 64\!\times\! 64$ for 3D images. Further details about the datasets can be found in the Supplementary Information.

\subsubsection{Evaluation Protocols}
We use the MedMNIST v2 split training and validation sets to train and select hyperparameters and report the results of the test set. Accuracy (ACC) and Area Under the ROC Curve (AUC) are used as evaluation metrics to ensure a fair comparison with benchmark methods reported in the literature \cite{yang2023medmnist,liu2022feature,manzari2023medvit,zheng2023complex,doerrich2023unoranic,zhu2024bsda,zhemchuzhnikov2024ilpo}. To enhance the reliability of the results, we repeated the process three times with different random seeds and use the average value as the final performance of our model. 

\subsubsection{Overall Performance}
\begin{figure}[ht]
    \centering
    \includegraphics[width=1\textwidth]{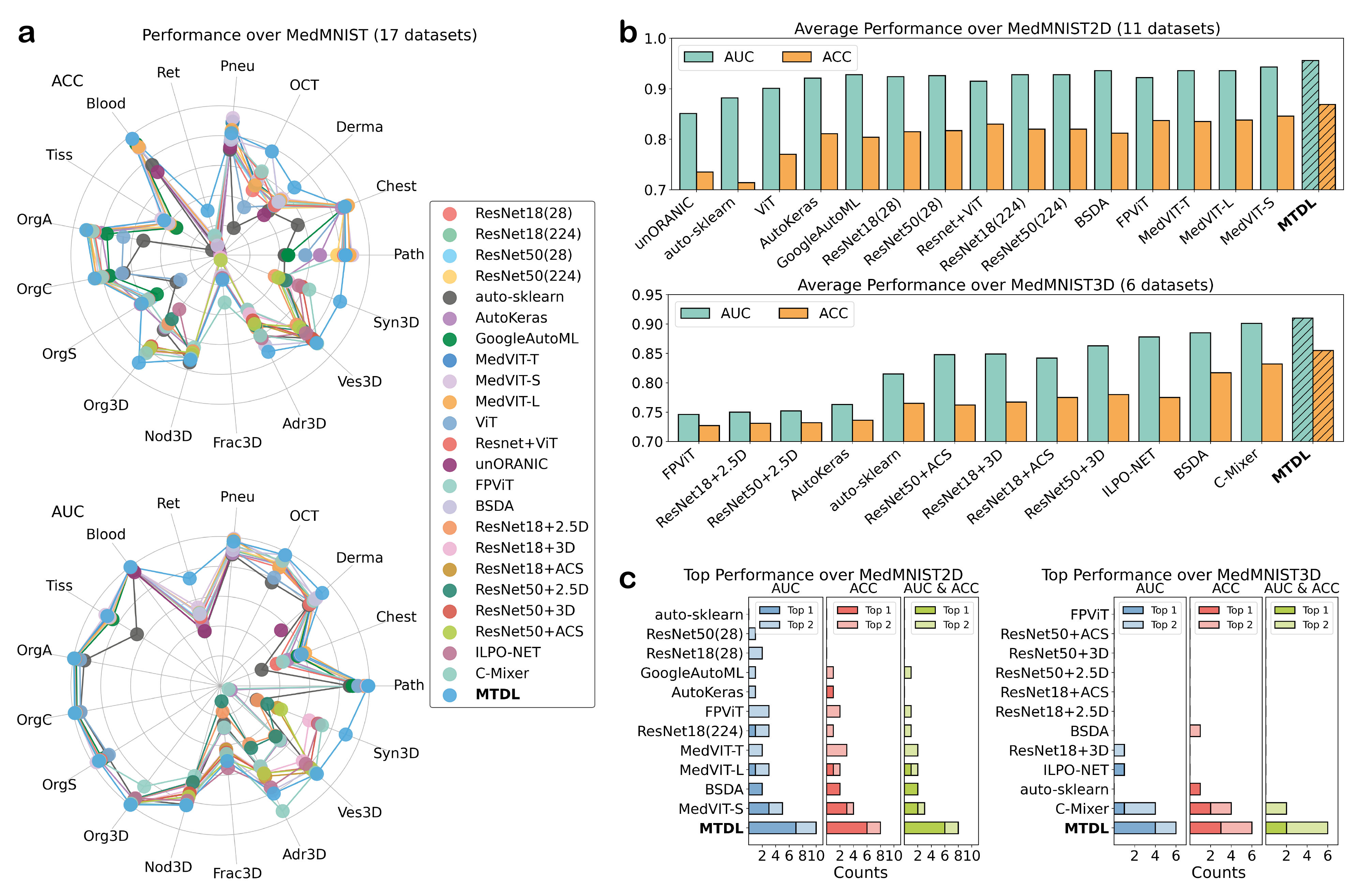}
    \caption{Performance comparison between MTDL model and other models on the MedMNIST v2 dataset. \textbf{a}: Comparison of model performance in terms of AUC and ACC across all 17 datasets of MedMNIST v2. The polygon representing the MTDL model covers the largest area, indicating its superior performance compared to the other models. \textbf{b}: Average performance of all models over 2D and 3D tasks. MTDL consistently achieves higher AUC and ACC values, outperforming all other models for both types of tasks. \textbf{c}: Frequency of top-ranking performance across 2D and 3D tasks. MTDL significantly surpasses all other models, demonstrating its consistent superiority in both 2D and 3D tasks.  }
    \label{fig:result1}
\end{figure}
Performance comparison of the proposed MTDL model with other state-of-the-art methods on the MedMNIST v2 dataset, in terms of AUC and ACC, is presented in Fig. \ref{fig:result1} (detailed values refer to Supplementary). Two radar charts are used to show the performance comparison among different models across all 17 datasets for ACC and AUC respectively. As shown in the figure, the polygon corresponding to MTDL model covers the largest area and is situated at the outermost edge of the region occupied by the polygons of all the models, demonstrating its superior overall performance for medical image classification (Fig. \ref{fig:result1}\textbf{a}). Notably, MTDL demonstrates significant improvements over the second-best models in specific datasets:
\begin{itemize}
    \item DermaMNIST: AUC improves from 0.937 to 0.962, and ACC improves from 0.780 to 0.836.
    \item RetinaMNIST: AUC improves from 0.773 to 0.874, and ACC improves from 0.568 to 0.655.
    \item OrganMNIST3D: AUC improves from 0.995 to 0.999, and ACC improves from 0.912 to 0.952.
    \item SynapseMNIST3D: AUC improves from 0.866 to 0.951, and ACC improves from 0.820 to 0.931.
\end{itemize} 

Additionally, we compute the average AUC and ACC separately for 2D and 3D datasets, MTDL consistently outperforms all other models for both 2D and 3D tasks (Fig. \ref{fig:result1}\textbf{b}). Specifically, for 2D datasets, MTDL achieves an average AUC of 0.956 and an ACC of 0.868, outperforming the second-best model, which achieves an average AUC of 0.943 and an ACC of 0.846. For 3D datasets, MTDL achieves an average AUC of 0.910 and an ACC of 0.855, compared to the second-best model's average AUC of 0.901 and ACC of 0.832.

Furthermore, we count the frequency of top performance for all models. As shown in the figure, MTDL can outperform all other models in AUC and ACC for both 2D and 3D tasks (Fig. \ref{fig:result1}\textbf{c}). Specifically, for 2D tasks, MTDL achieves the highest AUC and ACC on six tasks, including RetinaMNIST (1,600 samples), DermaMNIST (10,015 samples), BloodMNIST (17,092 samples), OrganCMNIST (23,660 samples), OrganAMNIST (58,850 samples), and OCTMNIST (109,309 samples). This demonstrates its ability to perform effectively on prediction tasks of varying data scales. When considering the top-2 models, MTDL ranks within the top 2 in a frequency of 10/11 for AUC, 8/11 for ACC, and 8/11 for both AUC and ACC, which is significantly better than the second-best model, which ranks within top 2 in 5/11 for AUC, 4/11 for ACC, and 3/11 for both AUC and ACC. For 3D tasks, MTDL ranks best in both AUC and ACC for 2 tasks while no other model achieves the top rank for both metrics on any dataset. Moreover, MTDL ranks in the top 2 with a frequency of 6/6 for AUC, 6/6 for ACC, and 6/6 for both ACU and ACC, compared with the second-best model's performance of 4/6 for AUC, 4/6 for ACC, and 2/6 for both AUC and ACC.  

These results highlight the overall superiority of MTDL in comparison to other state-of-the-art models, demonstrating its effectiveness in handling both 2D and 3D medical image classification tasks.

\subsubsection{Robustness Analysis Across Data Modality, Scale, and Task Type }
\begin{figure}[ht]
    \centering
    \includegraphics[width=0.95\textwidth]{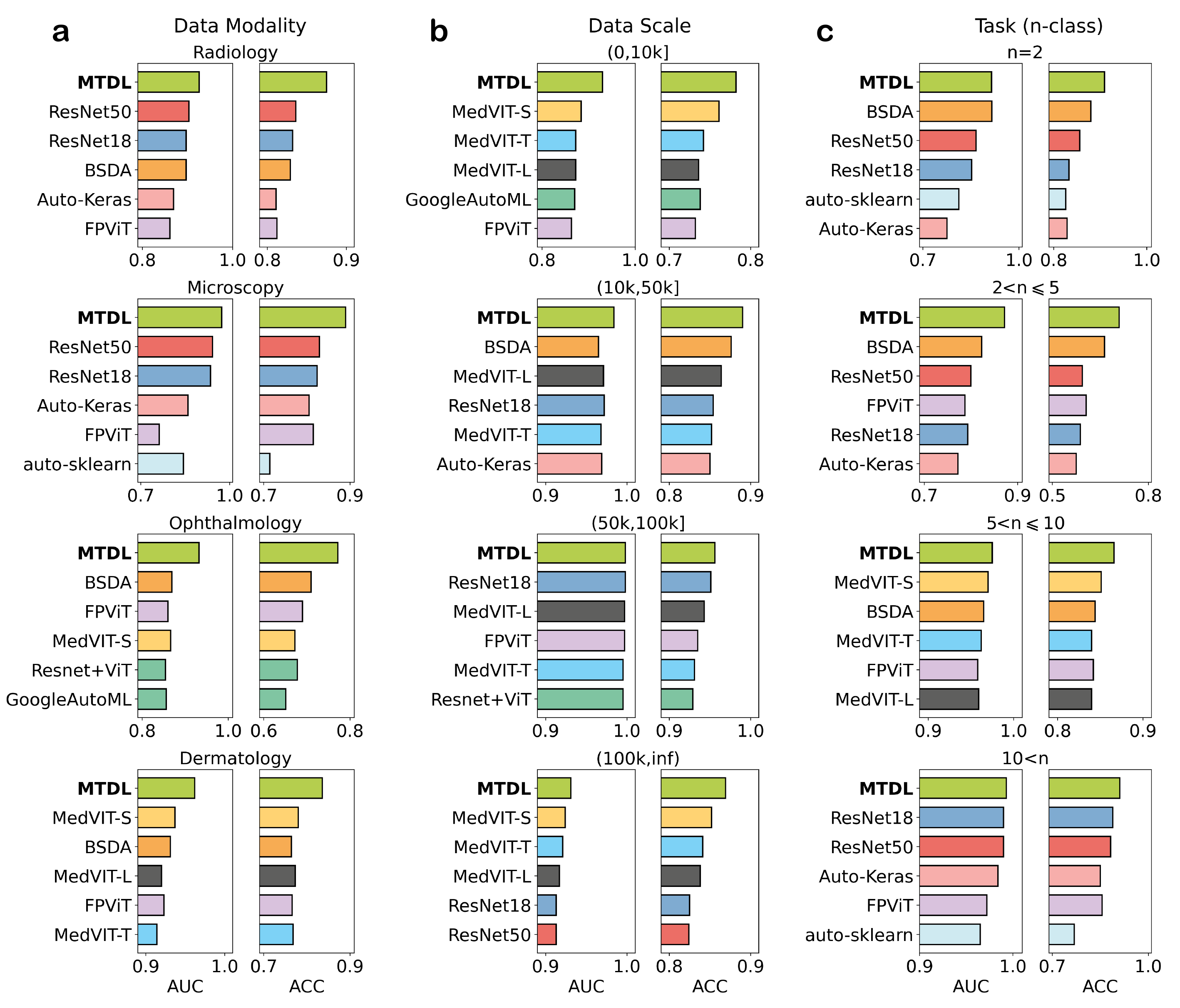}
    \caption{Performance comparison between MTDL and other models on different groups based on data modality, data scale, and task type. Here we only show the best six models for each group. \textbf{a}: Comparison on four data modality groups (Radiology, Microscopy, Ophthalmology, Dermatology). \textbf{b}: Comparison on four data scale groups ($n<10$K, $10{\rm K}\leqslant n<50{\rm K}$, $50{\rm K}\leqslant n<100{\rm K}$, $100{\rm K}<n$) where $n$ is the sample numbers of each dataset. \textbf{c}: Comparison on four task type groups ($n=2$, $2<n\leqslant5$, $5<n\leqslant10$, $10<n$) where $n$ is the class number of each dataset.}
    \label{fig:result2}
\end{figure}
To assess the robustness and generalizability of model performance, we divide the 17 datasets into groups based on data modality, data scale, and task type (refer to Supplementary Information), and then compare the average performance of all models within each group. To ensure a fair comparison, for each group, MTDL is evaluated only against models that reported results for all datasets in the respective group.

For data modality, we divide the datasets into four groups: Radiology (X-ray, CT, MRA), Microscopy (Pathology, Electron Microscope), Ophthalmology, and Dermatology. The performance comparison between MTDL and other models is shown in Fig. \ref{fig:result2}\textbf{a}. It can be seen that MTDL consistently outperforms all other models in both AUC and ACC across all groups. Specifically, 
MTDL achieves an average AUC (ACC) of 0.926 (0.875) for Radiology, compared to the second-best model's performance of 0.903 (0.836). For Microscopy, MTDL obtains an average AUC (ACC) of 0.973 (0.890) while the second-best model achieves a score of 0.942 (0.829). For Ophthalmology, MTDL gets an average AUC (ACC) of 0.932 (0.772), significantly surpassing the second-best model's score of 0.869 (0.710). For Dermatology, MTDL obtains an average AUC (ACC) of 0.962 (0.836),  compared to the second-best model's performance of 0.937 (0.780). Note that MTDL can maintain an AUC above 0.930 for all four groups and an ACC above 0.835 for groups except Ophthalmology, the ACC for Ophthalmology is slightly smaller than other groups. We attribute this to the RetinaMNIST dataset within the Ophthalmology group since this dataset only contains 1600 samples. Despite this, MTDL still significantly outperforms other models for this group. 

For data scale, we divide the datasets into four groups based on the sample size $n$ of each dataset: G1 ($n\leqslant$10K), G2 ($10{\rm K}<n\leqslant$50K), G3 ($50{\rm K}<n\leqslant$100K), and G4 ($100{\rm K}<n$). The performance in terms of AUC and ACC for all models is presented in Fig. \ref{fig:result2}\textbf{b}. MTDL also ranks best in both metrics for all groups. Specifically, MTDL achieves an average AUC (ACC) of 0.930 (0.782) for G1, compared to the second-best model's performance of 0.884 (0.761). For G2, MTDL scores 0.984 (0.890) compared to the second-best model's score of 0.965 (0.876). In G3, both MTDL and the second-best model achieve an average AUC of 0.998, but MTDL has a slightly higher ACC (0.956 vs. 0.951). For G4, MTDL obtains 0.931 (0.869) compared to the second-best model's 0.924 (0.852). MTDL can get an AUC exceeding 0.930 for all four groups and an ACC exceeding 0.860 for groups except G1, this is reasonable because bigger data usually leads to better performance.

For task type, we divide the datasets into four groups based on the number of classes $n$ for each classification task: G1 ($n$=2), G2 (2$<n\leqslant$5), G3 (5$<n\leqslant$10), and G4 (10$<n$). The performance for all models is shown in Fig. \ref{fig:result2}\textbf{c}. MTDL again achieves the best overall performance. Specifically, MTDL achieves an average AUC (ACC) of 0.914 (0.909) for G1 compared to the second-best model's performance of 0.915 (0.880). For G2, MTDL scores 0.872 (0.709), significantly surpassing the second-best model's score of 0.823 (0.663). For G3, MTDL obtains 0.975 (0.866) compared to the second-best model's 0.970 (0.851). In G4, MTDL attains an average AUC (ACC) of 0.993 (0.911) compared to the second-best model's value of 0.990 (0.882). MTDL achieves strong performance for G1, G3, and G4, with AUC exceeding 0.910 and ACC exceeding 0.880. We think the slightly lower performance on G2 is due to the inclusion of the small-sized RetinaMNIST dataset within this group.

These results demonstrate the superiority, robustness, and generalizability of MTDL across various data scales, data modalities, and task types, indicating its great potential for medical image analysis. It is noteworthy that MTDL has only 0.56M parameters for 2D tasks and 0.75M parameters for 3D tasks, which is significantly smaller than models such as ResNet, GoogleNet, Vision Transformer (ViT), and MedViT. Despite its lightweight architecture, MTDL demonstrates exceptional performance.

\subsubsection{Evaluation on Clinical Data}
In MedMNIST v2, most datasets are derived from clinical sources, that is, human subjects treated in hospitals and medical centers, such as the German National Center for Tumor Diseases \cite{kather_2018_1214456}, Zhongshan Hospital Affiliated to Fudan University \cite{yang2023medmnist}, and Guangzhou Women and Children’s Medical Center \cite{kermany2018large}, among others. The majority of source datasets are simply processed through center-cropping and resizing to uniform dimensions for inclusion in MedMNIST v2. Consequently, models based on MedMNIST v2 are generally reliable.

To better understand the clinical applicability of our model for medical image analysis, we need to check the effects of image resizing process on model performance. We utilized the HAM10000 dataset, the original clinical dataset for DermaMNIST, and evaluated the performance of MTDL on it. HAM10000 consists of 10015 dermatoscopic images from different populations, including a representative collection of all important diagnostic categories in the realm of pigmented lesions. Over 50\% of lesions in it have been confirmed by pathology, while the remaining cases are validated through either follow-up examinations, expert consensus, or in-vivo confocal microscopy \cite{tschandl2018ham10000}.

The images in HAM10000 have the same size of $3\times 600\times 450$. We center-crop the images to $3\times 450\times450$ and then resize then into five resolutions: $3\times 35\times 35$, $3\times 75\times 75$, $3\times 150\times 150$, $3\times 300\times 300$, $3\times 450\times 450$, with cubic spline interpolation. The performance of MTDL on these groups are shown in Table \ref{tab:clinical}. As seen in the table, MTDL achieves improved performance as the image resolution increases. This indicates that MTDL is capable of extracting more detailed features from higher-resolution inputs, which makes it well-suited for clinical applications where high-resolution images are prevalent. 
\begin{table}[h]
    \centering
    \caption{Performance of MTDL over different image reslutions}
    \begin{tabular}{c|c|c|c|c|c}
        \hline
         Resolution & $3\times 35\times 35$ & $3\times 75\times 75$ & $3\times 150\times 150$& $3\times 300\times 300$ & $3\times 450\times 450$ \\
         \hline
        AUC & 0.943 & 0.947 & 0.958 & 0.970 & 0.973\\
        ACC & 0.797 & 0.803 & 0.822 & 0.856 & 0.863\\
		\hline
	\end{tabular}
    \label{tab:clinical}
\end{table}
Notably, the best performance is achieved on the largest image resolution ($3\times 450\times 450$), surpassing the results obtained on the DermaMNIST dataset. Specifically, the AUC improves from 0.962 to 0.973 and the ACC improves from 0.836 to 0.863. Even at the lowest resolution ($3\times 35\times 35$), MTDL achieves an AUC (ACC) of 0.943 (0.797), outperforming the best existing models' performance of 0.937 (0.780). This highlights the robust lower-bound performance of MTDL across varying data resolutions, a critical attribute for addressing real-world clinical challenges.

\subsubsection{Ablation Study}
In our proposed MTDL model, the original image is decomposed into distinct orthogonal components, which are then concatenated to form a new composite image, serving as input to the CNN architecture.
To evaluate the importance of the Hodge decomposition method, we perform an ablation study by replacing the decomposed images with the original images and denote the resulting model as ImgCNN. We compare the performance of MTDL and ImgCNN on five 2D datasets spanning different data scales, including RetinaMNIST (1,600 samples), PneumoniaMNIST (5856 samples), DermaMNIST (10,015 samples), OrganAMNIST (58,830 samples), and PathMNIST (107,180 samples). Additionally, the comparison extends to two 3D datasets: VesselMNIST3D (binary classification) and FractureMNIST3D (three-class classification). The results are summarized in Table. \ref{tab:ablation}.
\begin{table}[h]
    \centering
    \caption{Performance Comparison between the original images and decomposition images, the best result is in bold.}
    \begin{tabular}{l|cc|cc}
        \hline
        Methods & \multicolumn{2}{c|}{ImgCNN} &  \multicolumn{2}{c}{\textbf{MTDL}} \\
        & AUC   & ACC   & AUC   & ACC  \\
        \hline
		RetinaMNIST   & 0.838  & 0.608 & \textbf{0.874} & \textbf{0.655} \\ 
	  PneumoniaMNIST & 0.978 & 0.885 & \textbf{0.986} & \textbf{0.910} \\
        DermaMNIST & 0.957 & 0.808 & \textbf{0.962} & \textbf{0.836} \\
        OrganAMNIST & \textbf{0.998} & 0.955 & \textbf{0.998} & \textbf{0.956}\\
        PathMNIST & 0.987 & 0.902 & \textbf{0.996} & \textbf{0.920}\\
        VesselMNIST3D &0.924 & 0.903 & \textbf{0.937} & \textbf{0.938} \\
        FractureMNIST3D &0.749 & 0.566 & \textbf{0.753} & \textbf{0.583} \\
		\hline
	\end{tabular}
    \label{tab:ablation}
\end{table}
As shown in the table, MTDL consistently outperforms ImgCNN across both 2D and 3D tasks. Notably:
\begin{itemize}
    \item For RetinaMNIST, AUC improves from 0.838 to 0.874, and ACC improves from 0.608 to 0.655.
    \item For DermaMNIST, AUC improves from 0.957 to 0.962, and ACC improves from 0.808 to 0.836.
    \item For VesselMNIST3D, AUC improves from 0.924 to 0.937, and ACC improves from 0.903 to 0.938.
\end{itemize}
These findings demonstrate the significant potential of the Hodge decomposition approach for enhancing medical image representation, enabling improved performance across diverse datasets and classification tasks.

\section{Discussion}\label{sec4}
TDL has achieved great success in applications involving point cloud and graph data. However, a dedicated TDL model for differentiable manifold data has not yet been developed, despite images being natural examples of such data. To bridge this gap, we introduce MTDL as a novel framework for extending TDL to differentiable manifold data. The systematic evaluation results demonstrate the efficiency, robustness, and generalizability of MTDL in medical image analysis. Additionally, our ablation studies highlight the significant potential of the Hodge decomposition approach in enhancing medical image representations.


In comparison to existing models on MedMNIST v2, MTDL is lightweight yet highly effective. For 2D datasets, the top three models in terms of average performance are MTDL, MedViT \cite{manzari2023medvit}, and FPViT \cite{liu2022feature}. Similarly, for 3D datasets, the leading models are MTDL, C-Mixer \cite{zheng2023complex}, and BSDA \cite{zhu2024bsda}. MedViT, which combines ViT with CNN, contains over 10M parameters. FPViT uses ResNet18 for feature extraction followed by shallow ViT layers for classification, its parameters also exceed 10 M since ResNet18 alone has more than 10 M parameters. C-Mixer, a model that integrates incentive learning, a C-Mixer network, and a self-supervised pretraining framework, does not report its parameter count or provide public code. Our rough estimate suggests it exceeds 1M parameters. BSDA is a Bayesian random semantic data augmentation techniques, which can be integrated with our model. In contrast, MTDL has only 0.56M parameters for 2D tasks and 0.75M parameters for 3D tasks, which is significantly fewer than other competing models. Despite its lightweight architecture, MTDL demonstrates exceptional performance.

For topological component of MTDL, the representation of images as vector fields plays a critical role in model performance, analogous to the importance of data representation in deep learning models. While this study adopts a specific method for generating vector fields in this study, we also present alternative methods in the Supplementary Information, which warrant further investigation. For the Hodge decomposition, we employ the standard three-component decomposition method. However, the five-component decomposition, which captures richer boundary and topological information of the image manifold, represents another promising direction for future research.

For deep learning component of MTDL, the key element is a modified Transformer encoder architecture by adding a maxpooling operation and replacing the multihead attention and feedforward layers by convolutions operations. 
Here we deliberately use this simple architecture to highlight the topological aspects of MTDL. 
In follow-up studies, we plan to integrate attention mechanisms for long-range inference on medical data tensors, enabling more complex clinical tasks such as lung CT screening and diagnosis \cite{niu2023specialty}. This will be explored in future work.

\section{Methods}\label{sec13}

\subsection{Topology-preserving Hodge Decomposition for Images}
Hodge decomposition is a fundamental result in differential geometry and algebraic topology, specifically for the analysis of differential forms on Riemannian manifolds. Recently, a discrete topology-preserving Hodge decomposition for manifolds with boundaries on Cartesian grids has been introduced \cite{su2024hodge}. This method is particularly well-suited for image analysis, as images can be naturally treated as discrete manifolds with boundaries embedded in Cartesian grids. 
\subsubsection{Hodge Decomposition in the Continuous Case}
Let $M$ be an $m$-dimensional smooth, orientable, compact manifold with boundary $\partial M$, $\Omega^k(M)$ represent the space of differential $k$-forms on $M$, and $d$ denote the differential (exterior derivative) from $k$-forms to $(k+1)$-forms. A differential $k$-form $\omega$ is called closed if $d\omega=0$ and exact if there exists a $(k-1)$-form $\zeta$ such that $d\zeta=\omega$.

Given a Riemannian metric $g$ on $M$, let $\star$ be the Hodge star operator that maps $k$-forms to $(m-k)$-forms and $(\cdot,\cdot)$ denote the induced Hodge $L^2$ inner product on $\Omega^k(M)$. The codifferential $\delta:\Omega^k(M)\rightarrow\Omega^{k-1}(M)$ is defined as
\begin{equation}
    \delta=(-1)^{m(k-1)+1}\star d \star.
\end{equation} 
A differential $k$-form $\omega$ is called coclosed if $\delta \omega=0$, and coexact if there exists a $(k+1)$-form $\zeta$ such that $\delta\zeta=\omega$. The operators $d$ and $\delta$ satisfy the following relationship
\begin{equation}\label{adjoint1}
    (d\omega,\eta)=(\omega,\delta\eta)+\int_{\partial M}\omega\wedge \star\eta,
\end{equation}
where $\omega$ is a $(k-1)$-form, $\eta$ is a $k$-form, and $\wedge$ is the wedge product on differential forms. This implies that $d$ and $\delta$ are adjoint if $M$ is a closed manifold, i.e., $\partial M=\emptyset$.

The Hodge Laplacian for differential forms is defined as 
\begin{equation}
    \Delta=d\delta+\delta d.
\end{equation}
The Laplacian operator maps $k$-forms to $k$-forms. The kernel of $\Delta$ is called the space of harmonic forms. We denote by $\mathcal{H}^k_\Delta(M)$ the space of harmonic $k$-forms and by $\mathcal{H}^k(M)$ the space of $k$-forms that are both closed and coclosed. We have $\mathcal{H}^k(M)\subset\mathcal{H}_\Delta^k(M)$.

When $M$ is a closed manifold, i.e., a compact manifold without boundary. The standard Hodge decomposition \cite{hodge1989theory} states that
\begin{equation}
    \Omega^k(M)=d\Omega^{k-1}(M)\oplus\delta\Omega^{k+1}(M)\oplus\mathcal{H}^k_\Delta(M),
\end{equation}
where the adjointness of $d$ and $\delta$ ensures that these three subspaces are orthogonal with respect to the Hodge $L^2$ inner product.

When $M$ is a manifold with non-empty boundary, the operators $d$ and $\delta$ are generally not adjoint, as noted in (\ref{adjoint1}). To ensure their adjointness and consequently achieve an orthogonal decomposition of differential forms, appropriate boundary conditions must be imposed.

Two most commonly used boundary conditions are the normal (Dirichlet) boundary condition and the tangential (Neumann) boundary condition. These conditions define the following subspaces,

\begin{equation}
    \Omega^k_n(M)=\{\omega\in\Omega^k(M)\;\vert \quad  \omega|_{\partial M}=0\},~~~
    \Omega^k_t(M)=\{\omega\in\Omega^k(M)\;\vert \quad \star \omega|_{\partial M}=0\}.
\end{equation}
The forms in $\Omega^k_n(M)$ and $\Omega^k_t(M)$ are called normal and tangential respectively. 

The Hodge-Morrey decomposition \cite{morrey1956variational} states that 
\begin{equation}\label{decomposition-with-boundary1}
    \Omega^k(M)=d\Omega^{k-1}_n(M)\oplus\delta\Omega^{k+1}_t(M)\oplus\mathcal{H}^k(M).
\end{equation}
The exterior derivative $d$ preserves the normal boundary condition and the codifferential $\delta$ preserves the tangential boundary condition. As a result, any $k$-form can be decomposed as the sum of an exact normal form, a coexact tangential form, and a harmonic form that is both closed and coclosed.
\begin{equation}\label{three-component1}
    \omega=d\alpha_n+\delta\gamma_t+\eta,
\end{equation}
where $\omega\in \Omega^k(M)$, $\alpha_n\in \Omega_n^{k-1}(M)$, $\gamma_t\in \Omega_t^{k+1}(M)$, $\eta\in\mathcal{H}^k(M)$. 
When we focus on the compact manifold in Euclidean spaces, the third term $\mathcal{H}^k(M)$ in \eqref{decomposition-with-boundary1} can be further decomposed into three orthogonal components \cite{shonkwiler2009poincare}, resulting in a five-component decomposition. A more detailed description of the Hodge decomposition can be found in the Supplementary Information.

\subsubsection{Discrete Topology-preserving Hodge Decomposition for Medical Images}\label{sec:discrete.hodge}

A medical image can be naturally seen as a level set function on a Cartesian grid, with its pixel values defining the scalar field. This makes discrete Hodge decomposition on Cartesian grids particularly suitable for the analysis of medical images.

Here we focus on 2D and 3D Cartesian grids, as medical images are typically in these dimensions. The discrete manifold $M$ on Cartesian grids can be given as a sublevel set of a level set function on the grid. We employ the strategy in \cite{ribando2024combinatorial} to determine the boundary of $M$ for two boundary conditions. For normal boundary condition, cells with at least one vertex inside $M$ are included, while for tangential boundary condition, cells with at least one vertex of their dual cells inside $M$ are included. The resulting sets of cells are referred to as the normal support for the normal boundary condition and the tangential support for the tangential boundary condition. These supports can be seen as discrete versions of the manifolds with boundary. The boundary of $M$ is typically detected using a projection matrix. The projection matrices $P_{k,n}$ and $P_{k,t}$ for normal and tangential boundary conditions are derived from the identity matrix by removing rows corresponding to cells outside the respective supports. 


On a Cartesian grid, vertices, edges, faces, and cubes are referred to as 0-cells, 1-cells, 2-cells and 3-cells. A differential $k$-form can be discretized as a $k$-cochain, which is a real-valued function on the $k$-cells. For instance, an image can be seen as a discrete 0-form since it is a 0-cochain on the Cartesian grid. The differential operators, including exterior derivative, Hodge star, codifferential, and Laplacian, can be discretized as matrices. Formally, let ${\rm I}_m$ be a Cartesian grid with cells oriented according to the coordinate axes, and $D_k$ denote the discrete exterior derivative on ${\rm I}_m$, then the discrete exterior derivative on $M$ for normal and tangential boundary conditions, denoted by $D_{k,n}$ and $D_{k,t}$ are 
\begin{equation}
    D_{k,n}=P_{k+1,n}D_kP_{k,n}^T,~~~
        D_{k,t}=P_{k+1,t}D_kP_{k,t}^T.
\end{equation}
Let $S_k$ denote the discrete Hodge star on ${\rm I}_m$, the discrete Hodge star on $M$ for normal and tangential boundary conditions are $S_{k,n}$ and $S_{k,t}$ respectively as follows
\begin{equation}
    S_{k,n}=P_{k,n}S_kP_{k,n}^T,~~~
    S_{k,t}=P_{k,t}S_kP_{k,t}^T.
\end{equation}
With the discrete Hodge star and discrete exterior derivative, the discrete codifferential can be expressed as $S_{k-1,n}^{-1}D_{k-1,n}^TS_{k,n}$ and $S_{k-1,t}^{-1}D_{k-1,t}^TS_{k,t}$ for normal and tangential boundary conditions respectively. The discrete Hodge Laplacian for normal and tangential boundary conditions $L_{k,n}$ and $L_{k,t}$ respectively are as follows
\begin{equation}
    \begin{split}
        L_{k,n}&=D_{k,n}^TS_{k+1,n}D_{k,n}+
        S_{k,n}D_{k-1,n}S_{k-1,n}^{-1}D_{k-1,n}^TS_{k,n},\\
        L_{k,t}&=D_{k,t}^TS_{k+1,t}D_{k,t}+
        S_{k,t}D_{k-1,t}S_{k-1,t}^{-1}D_{k-1,t}^TS_{k,t}.\\
    \end{split}
\end{equation}
As in the continuous case, the Kernels of these discrete Laplacians are fully determined by the topology of $M$.  Specifically, the dimension of ${\rm ker} L_{k,n}$ equals the Betti number $\beta_{m-k}$, while the dimension of ${\rm ker} L_{k,t}$ equals $\beta_k$. The Betti number $\beta_k$  quantifies the number of $k$-dimensional topological features in $M$: $\beta_0$ represents the number of connected components, $\beta_1$ the number of loops, and $\beta_2$ the number of voids.

\begin{figure}[ht]
    \centering
    \includegraphics[width=1\textwidth]{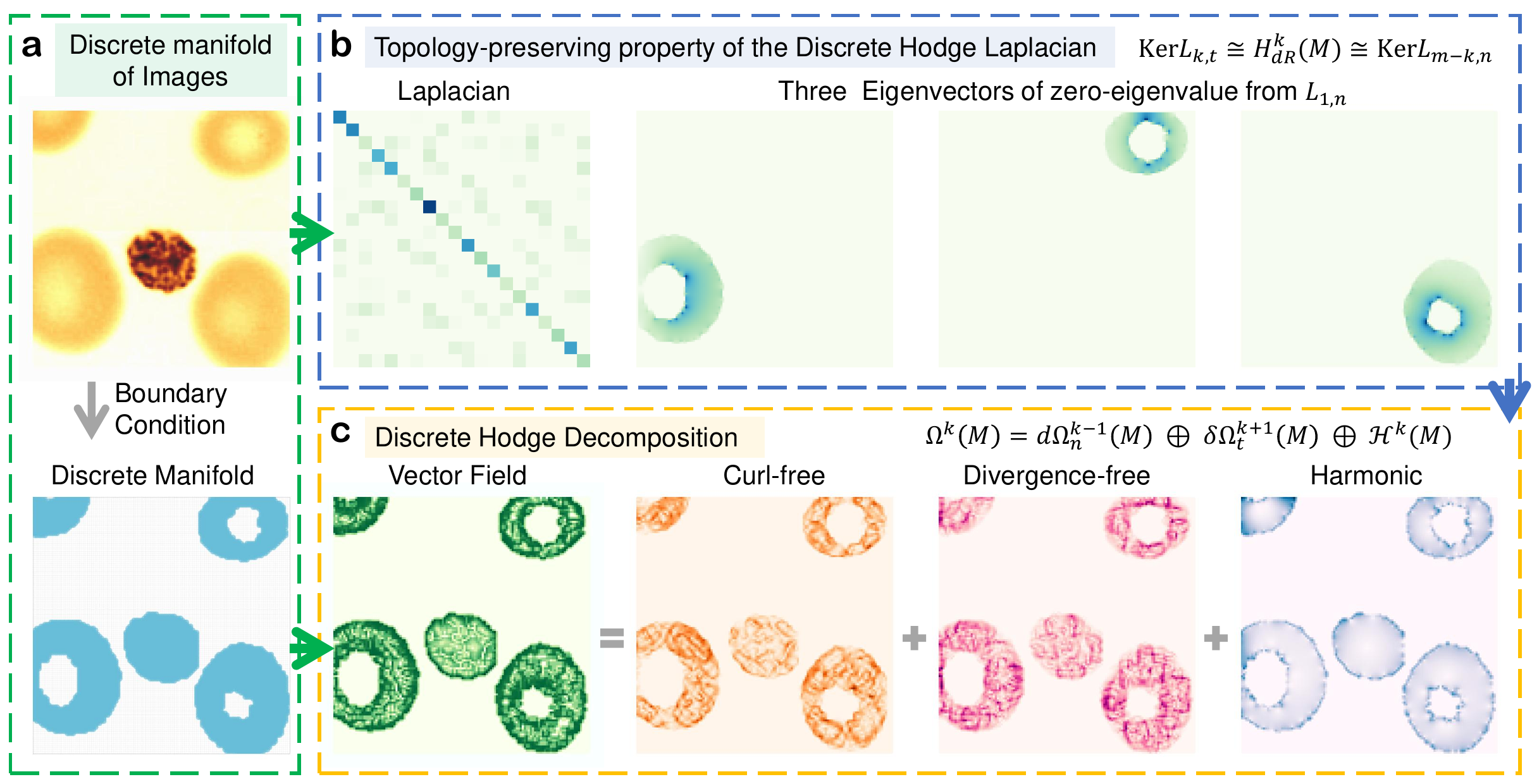}
    \caption{Illustration of the topology-preserving property of the discrete Hodge Laplacian and the Hodge decomposition for a medical image. In (\textbf{a}), the foreground of the image is represented as a manifold with boundary. The Laplacian $L_{1,n}$
    is computed and its eigenvectors corresponding to the zero eigenvalues are displayed, accurately capturing the three loops in the manifold (\textbf{b}). In (\textbf{c}), a vector field (1-form) is constructed from the image and decomposed into three orthogonal components: the curl-free, divergence-free, and harmonic parts. The harmonic component encapsulates the global topological information, while the other two components convey distinct aspects of local information. }
    \label{fig:hodge}
\end{figure}

Fig. \ref{fig:hodge} illustrates an example demonstrating the topology-preserving property of the discrete Laplacian. As shown in the figure, a blood cell image is represented as a discrete manifold under boundary conditions (Fig. \ref{fig:hodge}\textbf{a}). This manifold exhibits three distinct loop structures, resulting in a Betti number $\beta_1$ of 3. We compute the Laplacian $L_{1,n}$ under the normal boundary condition, and the eigenvectors corresponding to the three zero eigenvalues are displayed. These eigenvectors align precisely with the three loops present in the manifold (Fig. \ref{fig:hodge}\textbf{b}).

With the discrete versions of differential forms and operators established, the discrete Hodge decomposition is expressed as:
\begin{equation}
    V^k=D_{k-1,n}W_n+S_{k,t}^{-1}D_{k,t}^TS_{k+1,t}W_t+E,
\end{equation}
where $V^k$, $W_n$, $W_t$, and $E$ are the discrete version of $\omega$, $\alpha_n$, $\beta_t$, and $\eta$ in \eqref{three-component1} respectively. 

Fig. \ref{fig:hodge} illustrates an example of the Hodge decomposition applied to a blood cell image. As shown in Fig. \ref{fig:hodge}\textbf{c}, a vector field (1-form) on the manifold is first derived from the image using the flow-based method described in Supplementary Information. This vector field is subsequently decomposed into three orthogonal components: the curl-free, divergence-free, and harmonic parts. The harmonic component represents the global topological structure of the underlying manifold, whereas the normal and tangential components characterize distinct aspects of the local information. Specifically, the textures of the normal and tangential components exhibit an approximately perpendicular relationship, and the harmonic component appears smoother compared to the other two components.

\subsection{CNN Architecture}
The CNN we used is based on the Transformer encoder architecture by adding a maxpooling operation and replacing the multihead attention and feedforward layers by convolution operations (Fig. \ref{fig:model-architecture}\textbf{e}). 

As illustrated, the decomposed image $x$ is first processed through an initialization block, which consists of a convolutional layer, a batch normalization operation, and a nonlinear ReLU activation function. The initialized image is then passed through a sequence of Transformer-encoder-induced convolution layers to extract hierarchical features. Finally, the extracted features are spatially averaged and fed into a multilayer perceptron (MLP) for classification.

A Transformer-encoder-induced convolution layer is composed of two convolutional blocks followed by a pooling operation. Formally, for an input image 
$x$, the TransConv layer is defined as:
\begin{equation}
    \begin{split}
        x'&={\rm Norm}(x+{\rm MultiHeadConv}(x)),\\
        x''&={\rm Norm}(x'+{\rm FeedForwardConv(x')}),\\
        x'''&={\rm MaxPool}(x''),\\
    \end{split}
\end{equation}
where ${\rm Norm}$ is the batch normalization operation, MaxPool represents the max pooling operation, and $x'''$ is the output for $x$ after a TransConv layer. If the input 
$x$ is a 2D image of dimensions $(W,H,C)$ , where $W$, $H$ and 
$C$ correspond to the width, height, and number of channels, respectively, then the output $x'''$ will have dimensions $(\frac{W}{2},\frac{H}{2},C)$ due to the pooling operation.

The MultiHeadConv block is designed to mimic the multi-head attention mechanism in the Transformer encoder. It consists of a group convolution, a ReLU activation, and a $1\!\times\! 1$ convolution operation. Let ${\rm Conv}_{ C_{in},C_{out},k,g }$ denote a convolution operation with a kernel size $k$, group number $g$, input channel $C_{in}$, and output channel $C_{out}$. For an input image $x$ with $C$ channels, the MultiHeadConv block is expressed as
\begin{equation}\small
    \begin{split}
        x'&={\rm Conv}_{C,C\times h,3,h}(x)\\
        x''&={\rm ReLU}(x')\\
        x'''&={\rm Conv}_{C\times h,C,1,1}(x'')\\
    \end{split}
\end{equation}
where $x'''$ is the output of $x$ after the MultiHeadConv block, $h$ is a hyperparameter corresponding to the number of heads in the multi-head attention mechanism. The first convolution emulates the multi-head attention operation, while the second $1\times 1$ convolution serves as a linear layer for feature fusion. Importantly, the MultiHeadConv block preserves the input image dimensions.

The FeedForwardConv block imitates the feedforward neural network layers typically found in the Transformer's encoder. It consists of two group convolutions separated by a ReLU activation function. Formally, for an input image $x$ with 
$C$ channels, the FeedForwardConv block is defined as:
\begin{equation}
    \begin{split}
        x'&={\rm Conv}_{C,2\times C,1,g}(x),\\
        x''&={\rm ReLU}(x'),\\
        x'''&={\rm Conv}_{2\times C,C,1,g}(x''),\\
    \end{split}
\end{equation}
where the two $1\times 1$ convolutions mimic the linear layers in a standard feedforward neural network. Similar to the MultiHeadConv block, the FeedForwardConv block maintains the input image dimensions.

\subsection{Model Implementation Detail}

\subsubsection{Decomposed Image Generation}
\label{sec:decomposed-img-construction}
In our implementation, each image is considered as a scalar field on the vertices of a standard Cartesian grid. The discrete manifold is generated by a segmentation, which involves extracting the foreground pixels by applying a threshold to remove background pixels from the images. We use the grid vertices, edges, faces, and cubes to construct the differential operators and projection operators in Sec.~\ref{sec:discrete.hodge}. 

Instead of taking the differential operator directly on the scalar field to construct the 1-form $\omega.$ We instead follow a 2-step procedure to provide noise resilience. 
First, we use the discrete gradient operation to get a vector field stored on the vertices. Formally, For a 3D image $\mathcal{I}$, where $\mathcal{I}(i,j,k)$ represents the pixel value at position $(i,j,k)$, a vector $(x_{i,j,k},y_{i,j,k},z_{i,j,k})$ for the pixel at $(i,j,k)$ is constructed by the following centered finite differences
\begin{equation}\label{eq:vector-field-construction}
    \begin{split}
        x_{i,j,k} &= \frac{\mathcal{I}(i+1,j,k)-\mathcal{I}(i-1,j,k)}{2},\\
        y_{i,j,k} &= \frac{\mathcal{I}(i,j+1,k)-\mathcal{I}(i,j-1,k)}{2},\\
        z_{i,j,k} &= \frac{\mathcal{I}(i,j,k+1)-\mathcal{I}(i,j,k-1)}{2}.\\
    \end{split}
\end{equation}
Second, this vector field is averaged into a 1-form $\omega$ on the edges. Let $e^x_{i,j,k}$ denote the edge connecting the vertices at $(i,j,k)$ and $(i\!+\!1,j,k)$, $e^y_{i,j,k}$ denote the edge connecting the vertices at $(i,j,k)$ and $(i,j\!+\!1,k)$, and $e^z_{i,j,k}$ denote the edge connecting the vertices at $(i,j,k)$ and $(i,j,k\!+\!1)$. The 1-form $\omega$ is defined as
\begin{equation}\label{eq:vector-field-to-1-form}
    \begin{split}
        \omega(e^x_{i,j,k}) &= \frac{x_{i,j,k}+x_{i+1,j,k}}{2},\\
        \omega(e^y_{i,j,k}) &= \frac{y_{i,j,k}+y_{i,j+1,k}}{2},\\
        \omega(e^z_{i,j,k}) &= \frac{z_{i,j,k}+z_{i,j,k+1}}{2}.\\
    \end{split}
\end{equation}

Finally, following the decomposition described in (\ref{three-component1}), the 1-form $\omega$ is decomposed into three orthogonal components
\begin{equation}\label{img-three-component}
    \omega=\omega_1+\omega_2+\omega_3.
\end{equation}
Here the decomposition is performed by the BIG Laplacian. For each component 1-form $\eta$ resulting from this decomposition, it is represented as a vector field stored on grid cubes. For the cube with the lowest indexed corner at position $(i,j,k)$, the corresponding vector $(\eta^x_{i,j,k},\eta^y_{i,j,k},\eta^z_{i,j,k})$ is given by averaging its projection on an axis direction along 4 edges in that direction:
\begin{equation}\label{eq:1-form-to-2-form}
    \begin{split}
        \eta^x_{i,j,k}&=[\eta(e^x_{i,j,k})+\eta(e^x_{i,j+1,k})+\eta(e^x_{i,j,k+1})+\\ &\qquad\eta(e^x_{i,j+1,k+1})]/4\\
        \eta^y_{i,j,k}&=[\eta(e^y_{i,j,k})+\eta(e^y_{i+1,j,k})+\eta(e^y_{i,j,k+1})+\\ &\qquad\eta(e^y_{i+1,j,k+1})]/4\\
        \eta^z_{i,j,k}&=[\eta(e^z_{i,j,k})+\eta(e^z_{i+1,j,k})+\eta(e^z_{i,j+1,k})+\\ &\qquad\eta(e^z_{i+1,j+1,k})]/4\\
    \end{split}
\end{equation}
The resulting vector field $\eta$ can be interpreted as a three-channel image, with each channel corresponding to one of the $x$, $y$ and $z$-axes. Finally, we concatenate the three-channel images derived from the three components in (\ref{img-three-component}) to construct a nine-channel image, which serves as the final decomposed representation. For 2D images, a similar procedure is applied, using only the $x$ and $y$-components in equations (\ref{eq:vector-field-construction}), (\ref{eq:vector-field-to-1-form}), (\ref{img-three-component}), and (\ref{eq:1-form-to-2-form}) to obtain the decomposed representations.

\subsubsection{Model details}
The proposed MTDL model is implemented using PyTorch \cite{paszke2019pytorch} and evaluated on an NVIDIA Tesla V100S GPU. For 2D datasets, the batch size and learning rate are set to 64 and $10^{-3}$, respectively, across all tasks. The training process spans 30 epochs for tasks with a sample size smaller than 100,000 and 10 epochs for tasks with a sample size exceeding this threshold. The number of layers in the model is adapted based on the data distribution. For the majority of tasks, a 5-layer structure is employed, with detailed configurations provided in the Supplementary Information.
The hidden channel dimension $C$ and head number $h$ are set to 72 and 4, with the group number $g$ configured as 1 for grayscale images and 3 for colored images.

For 3D datasets, the batch size and learning rate are set to 16 and $10^{-3}$, respectively, for all tasks. The training process involves 10 epochs for the FractureMNIST dataset and 20 epochs for the remaining datasets. Similar to the 2D case, the number of layers is determined by the data distribution, with further details available in the Supplementary Information. The hidden channel dimension $C$ and head number $h$ are set to 64 and 4, with a group number $g$ of 1 since all the 3D images are grayscale.

The model is optimized using the AdamW optimizer \cite{loshchilov2017fixing} with a weight decay of $10^{-5}$, and a one-cycle learning rate scheduler employed \cite{smith2019super}. For the ChestMNIST2D task, a multi-label classification problem, the Binary Cross-Entropy with Logits is used as the loss function, while Cross-Entropy Loss is applied for all other tasks.

\section*{Data and Code Availability}
The data used in this study can be found on the MedMNIST official website \href{https://medmnist.com/}{medmnist.com}. 

\section*{Acknowledgment}
This work was supported in part by NIH grants R01GM126189, R01AI164266, and R35GM148196, National Science Foundation grants DMS2052983 and IIS-1900473, Michigan State University Research Foundation, and  Bristol-Myers Squibb 65109. The work of YS and GW was supported in part by NIH grants R01EB032716, 
R01EB031102, and 
R01HL151561.



\newpage
\section*{Supporting Information}

\setcounter{section}{0}
\section{Hodge theory}\label{secA1}
\subsection{Hodge Theory in the Continuous Case}
\label{sec:appendix-continuous-hodge}
Here, we provide a comprehensive review of the Hodge decomposition in the continuous setting, as this work represents the first endeavor to apply the Hodge decomposition within the domain of medical image analysis.
\subsubsection{Hodge Laplacian}
Let $M$ be an $m$-dimensional smooth, orientable, compact manifold with boundary, and let $\Omega^k(M)$ denote the space of differential $k$-forms on $M$, i.e., the space of all smooth, antisymmetric covariant tensor fields of degree $k$ on $M$. A differential $k$-form can be integrated over any orientable $k$-dimensional submanifold of $M$. For any orientable $(k+1)$-dimensional submanifold $S\subset M$ with boundary $\partial S$, Stokes' theorem states that the integral of a differential $k$-form $\omega$ over the boundary $\partial S$ is equal to the integral of its differential over the manifold $S$. Explicitly, this is expressed as
\begin{equation}
    \int_Sd\omega=\int_{\partial S}\omega,
\end{equation}
where the differential $d$ (exterior derivative) is the unique $\mathbb{R}$-linear mapping from $\Omega^k(M)$ to $\Omega^{k+1}(M)$ that satisfies the Leibniz rule with respect to the wedge product $\wedge$ and the property $dd=0$. A differential $k$-form $\omega$ is called closed if $d\omega=0$ and exact if there exists a $(k-1)$-form $\psi$ such that $d\psi=\omega$. The pair $(\Omega^*(M),d)$ forms a cochain complex known as the de Rham complex, and its $k$-th cohomology, denoted by $H_{DR}^k(M)$, is called the $k$-th de Rham cohomology of $M$.

Let $g$ be a Riemannian metric on $M$ and let $<\cdot,\cdot>$ denote the inner product on $\Omega^k(M)$ induced by $g$. The Hodge star operator $\star$ is an isomorphism from the space of $k$-forms $\Omega^k(M)$ to the space of $(m-k)$-forms $\Omega^{m-k}(M)$ satisfying the property
\begin{equation}\label{hodge-star}
    \omega\wedge\star\eta=<\omega,\eta>_g\mu_g,
\end{equation}
where $\omega$ and $\eta$ are $k$-forms, and $\mu_g$ is the volume form induced by $g$ on $M$. By taking the integral of formula (\ref{hodge-star}), we obtain the Hodge $L^2$-inner product on the space of differential forms $\Omega^k(M)$
\begin{equation}\label{inner-product}
    (\omega,\eta)=\int_M\omega\wedge\star\eta.
\end{equation}
The codifferential $\delta:\Omega^k(M)\rightarrow\Omega^{k-1}(M)$ is defined as
\begin{equation}
    \delta=(-1)^{m(k-1)+1}\star d \star.
\end{equation}
A differential $k$-form $w$ is called coclosed if $\delta w=0$, and coexact if there exists a $(k+1)$-form $\psi$ such that $\delta\psi=w$. The differential $d$ and the codifferential $\delta$ satisfy the following relationship based on integration by parts
\begin{equation}\label{adjoint}
    (d\omega,\eta)=(\omega,\delta\eta)+\int_{\partial M}\omega\wedge \star\eta,
\end{equation}
where $\omega$ is a $(k-1)$-form and $\eta$ is a $k$-form. This shows that $d$ and $\delta$ are adjoint if $M$ is a closed manifold, i.e., $\partial M=\emptyset$.

The Hodge Laplacian for differential forms is defined as 
\begin{equation}
    \Delta=d\delta+\delta d.
\end{equation}
The Laplacian operator maps $k$-forms to $k$-forms. The kernel of $\Delta$ is called the space of harmonic forms. We denote by $\mathcal{H}^k_\Delta(M)$ the space of harmonic $k$-forms and by $\mathcal{H}^k(M)$ the space of $k$-forms that are both closed and coclosed.
\subsubsection{Hodge Decomposition for Closed Manifolds}
Assume $M$ is a closed manifold, i.e., a compact manifold without boundary. The standard Hodge decomposition \cite{hodge1989theory} states that
\begin{equation}
    \Omega^k(M)=d\Omega^{k-1}(M)\oplus\delta\Omega^{k+1}(M)\oplus\mathcal{H}^k_\Delta(M),
\end{equation}
where the adjointness of $d$ and $\delta$ ensures that these three subspaces are orthogonal with respect to the inner product defined in  (\ref{inner-product}). Consequently, any $k$-form can be uniquely decomposed as the sum of an exact form, a coexact form, and a harmonic form,
\begin{equation}
    \omega=d\alpha+\delta\beta+h,
\end{equation}
where $\omega\in\Omega^k(M)$, $\alpha\in\Omega^{k-1}(M)$, $\beta\in\Omega^{k+1}(M)$, $h\in \mathcal{H}^k_\Delta(M)$. The Hodge isomorphism theorem asserts that the space of harmonic $k$-forms is isomorphic to the $k$-th de Rham cohomology $H_{DR}^k(M)$ of $M$, this implies that the dimension of the harmonic space $\mathcal{H}_\Delta^k(M)$ is a topological invariant of the manifold, determined entirely by its topology.

\subsubsection{Hodge Decomposition for Manifolds with Boundary}
\begin{figure}[ht]
    \centering
    \includegraphics[width=1\linewidth]{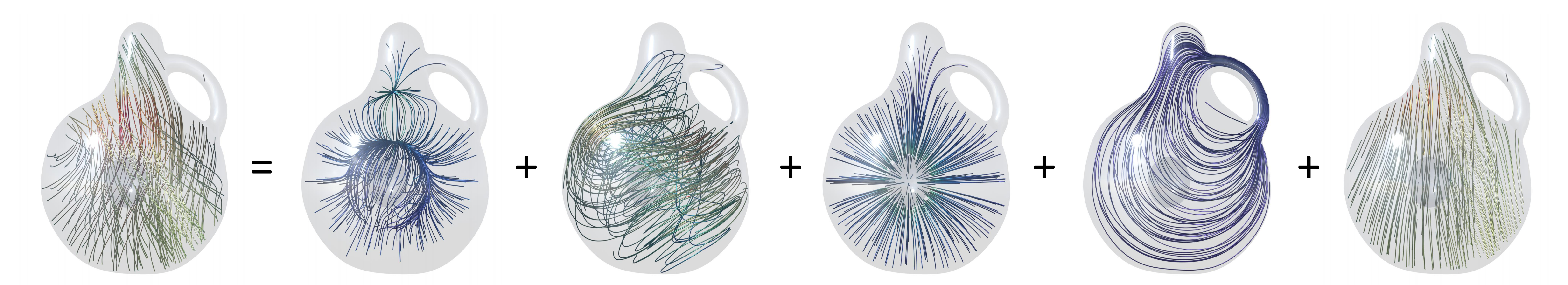}
    \caption{Illustration of the 3D Hodge decomposition on a pear with a tunnel model. From left to right: the original vector field, the curl-free field, the divergence-free field, the normal harmonic field, the tangential harmonic field, and the curl gradient field.}
    \label{fig:five-decomposition}
\end{figure}

When $M$ is a manifold with non-empty boundary, the operators $d$ and $\delta$ are generally not adjoint, as noted in (\ref{adjoint}). To ensure their adjointness and consequently achieve an orthogonal decomposition of differential forms, appropriate boundary conditions must be imposed.

The two most commonly used boundary conditions that ensure the adjointness of $d$ and $\delta$ are the normal (Dirichlet) boundary condition and the tangential (Neumann) boundary condition. A form is called normal if it vanishes when applied to tangential vectors of the boundary and tangential if its dual vanishes when applied to tangential vectors of the boundary. These conditions define the following subspaces,
\begin{equation}
        \Omega^k_n(M)=\{\omega\in\Omega^k(M) \;\vert \quad\omega|_{\partial M}=0\},~~~
        \Omega^k_t(M)=\{\omega\in\Omega^k(M)\;\vert \quad\star \omega|_{\partial M}=0\}.  
\end{equation}
The Hodge star $\star$ provides an isomorphism between $\Omega^k_n(M)$ and $\Omega^{m-k}_t(M)$.

The Hodge-Morrey decomposition \cite{morrey1956variational} states that 
\begin{equation}\label{decomposition-with-boundary}
    \Omega^k(M)=d\Omega^{k-1}_n(M)\oplus\delta\Omega^{k+1}_t(M)\oplus\mathcal{H}^k(M),
\end{equation}
where the boundary conditions ensure the adjointness of $d$ and 
$\delta$, guaranteeing the orthogonality of the decomposition. The exterior derivative $d$ preserves the normal boundary condition and the codifferential $\delta$ preserves the tangential boundary condition. As a result, any $k$-form can be decomposed as the sum of an exact normal form, a coexact tangential form, and a form that is both closed and coclosed.
\begin{equation}\label{three-component}
    \omega=d\alpha_n+\delta\gamma_t+\eta,
\end{equation}
where $\omega\in \Omega^k(M)$, $\alpha_n\in \Omega_n^{k-1}(M)$, $\gamma_t\in \Omega_t^{k+1}(M)$, $\eta\in\mathcal{H}^k(M)$. 
To compute the components of this decomposition, we can firstly determine the potentials $\alpha_n$ and $\gamma_t$, and then compute $\eta$ as $\eta=\omega-d\alpha_n-\delta\gamma_t$. However, the potentials $\alpha_n$ and $\gamma_t$ are not unique, as $\alpha_n+d\eta$ and $\gamma_t+\delta\gamma$, with any $\eta\in \Omega^{k-2}_n(M)$ and $\gamma\in \Omega_t^{k+2}(M)$, can serve as valid potentials for the first two terms. This issue can be solved by imposing gauge conditions. Specifically, we restrict 
\begin{equation}
    \alpha_n\in {\rm ker} \delta\cap\Omega_n^{k-1}(M),~\gamma_t\in {\rm ker}d\cap\Omega_t^{k+1}(M).
\end{equation}
Under these conditions, the potentials satisfy the following equations,
\begin{equation}\label{gauge-condition}
        \delta\omega=\delta d\alpha_n=(\delta d+d\delta)\alpha_n=\Delta\alpha_n, ~~~
        d\gamma_t =d\delta\gamma_t=(d\delta+\delta d)\gamma_t=\Delta \gamma_t.
\end{equation}
Up to a difference of harmonic forms in the kernel of $\Delta$, the potentials $\alpha_n$ and $\gamma_t$ can be uniquely determined by the equations in (\ref{gauge-condition}) by enforcing the boundary conditions $\delta\alpha_n|_{\partial M}=0$, and $\star d\gamma_t|_{\partial M}=0$, i.e., a nonsingular linear system by resolving the rank deficiency of $\Delta$. 

When we focus on the compact manifold in Euclidean spaces, the third term $\mathcal{H}^k(M)$ in (\ref{decomposition-with-boundary}) can be further decomposed into three orthogonal components \cite{shonkwiler2009poincare},
\begin{equation}
    \mathcal{H}^k(M)=\mathcal{H}_n^k(M)\oplus\mathcal{H}_t^k(M)\oplus(d\Omega^{k-1}(M)\cap\delta\Omega^{k+1}(M)). 
\end{equation}
As a result, a five-component decomposition is obtained
\begin{equation}\label{eq:5-decomposition}
        \Omega^k(M)=d\Omega^{k-1}_n(M)\oplus\delta\Omega_t^{k+1}(M)\oplus\mathcal{H}^k_n(M)\oplus\mathcal{H}_t^k(M)\oplus(d\Omega^{k-1}(M)\cap\delta\Omega^{k+1}(M)),
\end{equation}
where all five components are mutually orthogonal with respect to the inner product defined in (\ref{inner-product}). Fig. \ref{fig:five-decomposition} gives an example of the five-component decomposition of a 3D pear with a tunnel model. The five components in the figure correspond to the five components in Equation \ref{eq:5-decomposition}. Particularly, the third term corresponds to a two-dimensional homology which is a void, and the forth term corresponds to a one-dimensional homology which is a loop.

\subsection{Discrete Topology-preserving Hodge Theory on Cartesian Grids }
 \begin{figure}[ht]
    \centering
    \includegraphics[width=0.8\linewidth]{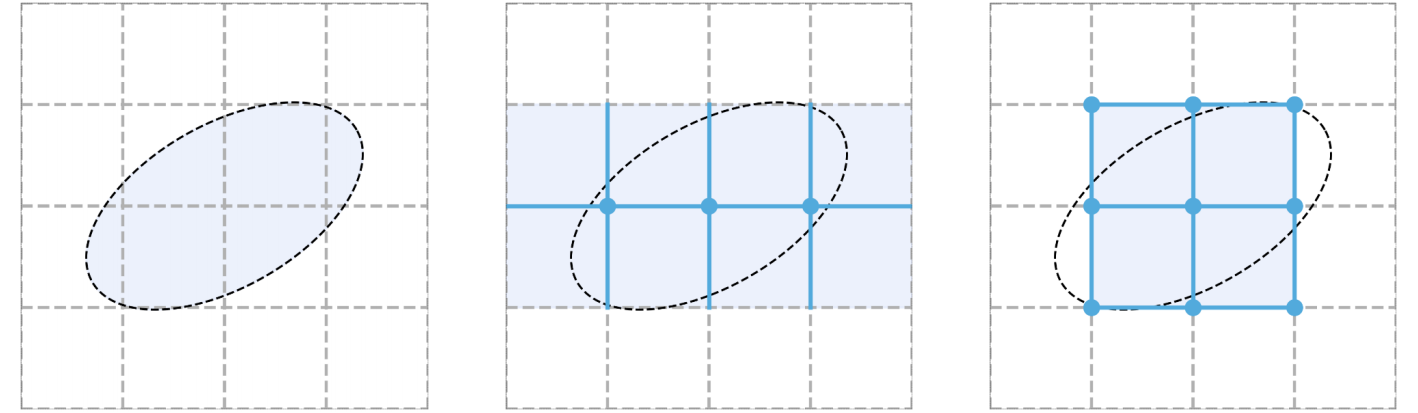}
    \caption{Illustration of discrete manifold representation for an image under normal and tangential boundary conditions. From left to right: the original image, the discrete manifold under normal boundary condition, and the discrete manifold on tangential boundary condition.}
    \label{fig:manifold}
\end{figure}
To obtain the discrete Hodge decomposition, it suffices to construct discrete versions of differential forms and differential operators, then replace the continuous forms and operators in \eqref{three-component} and \eqref{gauge-condition} with their discrete counterparts. 
Here we focus on 2D/3D domains bounded by level set surfaces on Cartesian grids. The manifold is given as a sublevel set of a level set function defined on a Cartesian grid. Note that a medical image can be naturally seen as a level set function on a Cartesian grid, with its pixel values defining the scalar field. Therefore, the discrete Hodge decomposition on Cartesian grids can be directly used for medical image analysis. 

\subsubsection{Discrete Manifolds with Boundary}
The discrete manifold $M$ in the Cartesian grid can be given as a sublevel set of a level set function on the grid. The boundary of $M$ is typically detected using a projection matrix. Note that in the grid representation, the boundary of $M$ often intersects with boundary cells instead of being fully contained within them. To address this issue, we restrict computations to relevant cells by employing the inclusion or exclusion strategy proposed in \cite{ribando2024combinatorial}. For normal boundary condition, cells with at least one vertex inside $M$ are included, while for tangential boundary condition, cells with at least one vertex of their dual cells inside $M$ are included. The resulting set of cells is referred to as the normal support for the normal boundary condition and the tangential support for the tangential boundary condition. These supports can be seen as discrete versions of the manifolds with boundary. The projection matrices $P_{k,n}$ and $P_{k,t}$ for these boundary conditions are derived from the identity matrix by removing rows corresponding to cells outside the respective supports. Fig. \ref{fig:manifold}
shows an example for the discrete manifold representation of an image under normal and tangential boundary conditions. It can be seen that the normal and tangential supports are different, and neither is a subset of the other.

\subsubsection{Discrete Differential Forms}
The discretization of a differential form can be achieved by the de Rham map, which maps a form to a cochain by integrating the form over cells \cite{desbrun2006discrete}. For a Cartesian grid ${\rm I}_m$ with cells oriented according to the coordinate axes, let $\omega$ be a differential $k$-form on $I_m$, the discretization assigns to each $k$-cell $\sigma_k$ the value $\int_{\sigma_k}\omega$, creating a cochain.

\subsubsection{Discrete Exterior Derivative Operator}
The discrete exterior derivative operator $d$ on discrete $k$-forms can be derived by Stokes' theorem
\begin{equation}
    \int_\sigma d\omega=\int_{\partial\sigma}\omega.
\end{equation}
In matrix form, $d$ corresponds to the transpose of the boundary matrix from $(k+1)$-cells to $k$-cells. Let $D_k$ denote the discrete exterior derivative on the entire grid, the discrete exterior derivative on the manifold $M$ for normal and tangential boundary conditions, denoted by $D_{k,n}$ and $D_{k,t}$, are
\begin{equation}
        D_{k,n} = P_{k+1,n}D_kP_{k,n}^T,~~~
        D_{k,t} = P_{k+1,t}D_kP_{k,t}^T.
\end{equation}
We still have $D_{k+1,n}D_{k,n}=0$ and $D_{k+1,t}D_{k,t}=0$.

\subsubsection{Discrete Hodge Star Operator}
A dual grid with respect to the primal grid ${\rm I}_m$ can be constructed by treating the centers of $m$-cells of ${\rm I}_m$ as grid points of the dual grid. The discretization of Hodge star operator can be obtained by the following relationship in the continuous case
\begin{equation}
    \frac{1}{|\sigma_k|}\int_{\sigma_k}\omega\approx\frac{1}{|\star\sigma_k|}\int_{\star\sigma_k}\star\omega,
\end{equation}
where $|\sigma_k|$ is the volume of primal $k$-cell $\sigma_k$, $\star\sigma_k$ is the dual $(m-k)$-cell of $\sigma_k$, and $\omega$ is a $k$-form. This gives a one-to-one correspondence between discrete $k$-forms on the primal grids and discrete $(m-k)$-forms on its dual grids. And the correspondence leads to the discrete Hodge star as a diagonal matrix whose diagonal entries are the ratios of the volumes of dual $(m-k)$-cells to primal $k$-cells. Let $S_k$ denote the discrete Hodge star matrix on the entire grid, the discrete Hodge star on the manifold $M$ for normal and tangential boundary conditions are $S_{k,n}$ and $S_{k,t}$ respectively as follows
\begin{equation}
        S_{k,n}=P_{k,n}S_kP_{k,n}^T,~~~
        S_{k,t}=P_{k,t}S_kP_{k,t}^T.
\end{equation}
The discrete Hodge $L^2$ inner product of two discrete $k$-forms $V_k$ and $W_k$ on ${\rm I}_m$ is
\begin{equation}
    (V_k,W_k)=(V_k)^TS_kW_k.
\end{equation}

\subsubsection{Discrete Hodge Laplacian}\label{sec:appendix-bigL}
Having the discrete Hodge star and discrete exterior derivative, the discrete codifferential can be expressed as $S_{k-1,n}^{-1}D_{k-1,n}^TS_{k,n}$ and $S_{k-1,t}^{-1}D_{k-1,t}^TS_{k,t}$ for normal and tangential boundary conditions respectively. Using these discrete differential and codifferential operators, the Laplacian $\Delta=\delta d+d\delta$ is not symmetric. Therefore the symmetric  $\star\Delta$ matrix as used to define the discrete Laplacian. The discrete Laplacian for normal and tangential boundary conditions $L_{k,n}$ and $L_{k,t}$ are respectively defined as follows
\begin{equation}
    \begin{split}
        L_{k,n}&=D_{k,n}^TS_{k+1,n}D_{k,n}+
        S_{k,n}D_{k-1,n}S_{k-1,n}^{-1}D_{k-1,n}^TS_{k,n},\\
        L_{k,t}&=D_{k,t}^TS_{k+1,t}D_{k,t}+
        S_{k,t}D_{k-1,t}S_{k-1,t}^{-1}D_{k-1,t}^TS_{k,t}.
    \end{split}
\end{equation}
As in the continuous case, the Kernels of these discrete Laplacians are fully determined by the topology of $M$. Specifically, the dimension of ${\rm ker} L_{k,n}$ equals the Betti number $\beta_{m-k}$, while the dimension of ${\rm ker} L_{k,t}$ equals $\beta_k$.

When the Hodge star matrix is replaced by the identity matrix, the discrete Laplacians reduce to the Boundary-Induced Graph (BIG) Laplacians,
\begin{equation}
    \begin{split}
        L_{k,n}^B&=D_{k,n}^TD_{k,n}+D_{k-1,n}D_{k-1,n}^T,\\
        L_{k,t}^B&=D_{k,t}^TD_{k,t}+D_{k-1,t}D_{k-1,t}^T.
    \end{split}
\end{equation}
The BIG Laplacians preserve the differential calculus properties of the Hodge Laplacian while retaining the combinatorial nature of the discrete Laplacian \cite{ribando2024combinatorial}. 

Note that the discrete Hodge Laplacian differs from the combinatorial Laplacian. For instance, when performing the spectral decomposition of a vector field on a point cloud, the use of combinatorial Laplacians defined on commonly employed simplicial complexes does not yield the same curl-free and divergence-free components as those obtained through the spectral decomposition of a vector field using discretized Hodge Laplacians. The latter are defined either on a point cloud with a boundary in the Eulerian representation or on a regular mesh in the Eulerian representation. A detailed comparison between the Hodge Laplacian and the combinatorial Laplacian can be found in the referenced literature \cite{ribando2024combinatorial}.

\subsubsection{Discrete Hodge Decomposition}
With the discrete version of differential forms and differential operators, the discrete Hodge decomposition can be expressed as 
\begin{equation}
    V^k=D_{k-1,n}W_n+S_{k,t}^{-1}D_{k,t}^TS_{k+1,t}W_t+E,
\end{equation}
where $V^k$, $W_n$, $W_t$, and $E$ are the discrete version of $\omega$, $\alpha_n$, $\beta_t$, and $\eta$ in (\ref{three-component}) respectively. As in the continuous case, we can first find $W_n$ and $W_t$, then compute $E$ as $E=V^k-D_{k-1,n}W_n-S_{k,t}^{-1}D_{k,t}^TS_{k+1,t}W_t$. And $W_n$ and $W_t$ can be uniquely determined by the discrete version of equation (\ref{gauge-condition})
\begin{equation}
        L_{k-1,n}W_n=D_{k-1,n}^TS_{k,n}V^k_n,~~~
        L_{k+1,t}W_t=S_{k+1,t}D_{k,t}V^k_t.
\end{equation}
where $V^k_n$ and $V^k_t$ are the vectors of $V^k$ under normal and tangential supports respectively.
\begin{figure}[ht]
    \centering
    \includegraphics[width=0.8\linewidth]{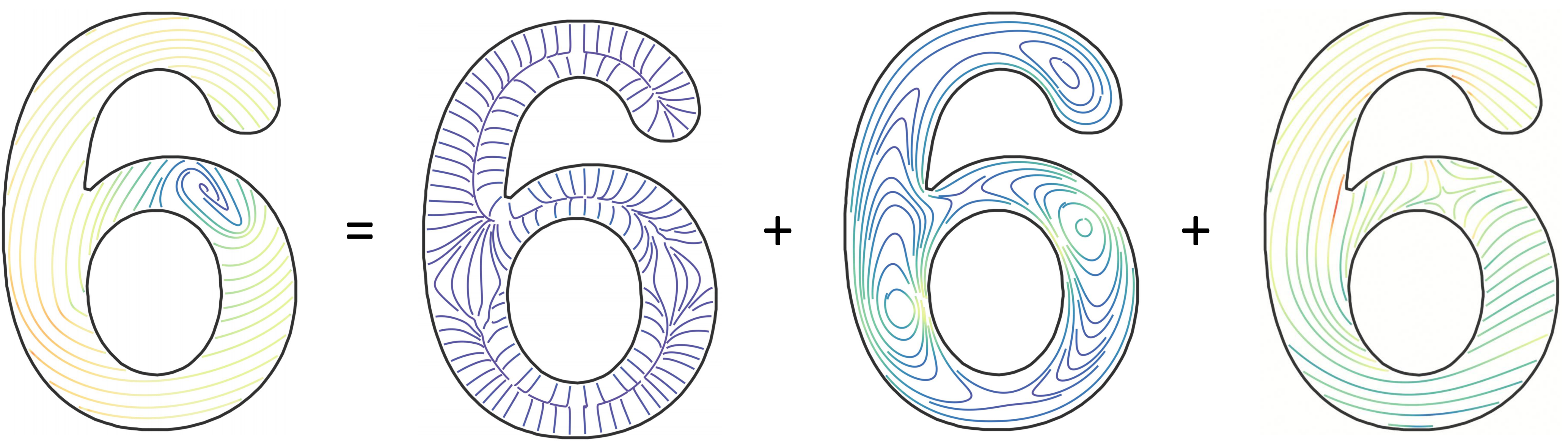}
    \caption{Illustration of the three-component Hodge decomposition on a 2D 6 domain. From left to right: original vector field, the curl-free field, the divergence-free field, and the harmonic field.}
    \label{fig:three-decomposition}
\end{figure}
Fig. \ref{fig:three-decomposition} gives an example for the three-component Hodge decomposition on a 2D 6 domain.

\section{Vector Field Generation}
In this section, we introduce several methods for generating noise-resilient vector fields (1-forms) from images. Without loss of generality, we focus on 3D images, and these methods can be easily adapted to 2D images.
\subsection{Gradient-based Method}
The gradient-based method uses the discrete gradient operation to construct a vector field from the images. Formally, for a 3D image $\mathcal{I}$, where $\mathcal{I}(i,j,k)$ represents the pixel value at position $(i,j,k)$, a vector $(x_{i,j,k},y_{i,j,k},z_{i,j,k})$ is constructed for each pixel at $(i,j,k)$ as follows
\begin{equation}\label{eq:gradient-method}
    \begin{split}
        x_{i,j,k} &= \frac{\mathcal{I}(i+s,j,k)-\mathcal{I}(i-t,j,k)}{2}\\
        y_{i,j,k} &= \frac{\mathcal{I}(i,j+s,k)-\mathcal{I}(i,j-t,k)}{2}\\
        z_{i,j,k} &= \frac{\mathcal{I}(i,j,k+s)-\mathcal{I}(i,j,k-t)}{2}\\
    \end{split}
\end{equation}
where $s,t$ are parameters to control the forward and backward steps. The resulting vector field is called the gradient-based $(s,t)$-step vector field. This method is a generalization of the standard finite difference method for computing gradient. It allows different forward and backward steps for computing the difference of a point. The method described in section ``Methods" is a special case of this gradient-based approach, corresponding to the standard case where $s=t=1$. 

\subsection{Flow-based Method}
The flow-based method constructs the vector field by analyzing the flow of pixel values, akin to unwind scheme. Formally, for a 3D image $\mathcal{I}$, where $\mathcal{I}(i,j,k)$ represents the pixel value at position $(i,j,k)$, a vector $(x_{i,j,k},y_{i,j,k},z_{i,j,k})$ for the pixel at $(i,j,k)$ is constructed by the following steps
\begin{enumerate}
    \item For the 26 voxel values adjacent to position $(i,j,k)$, let $S$ be the set of positions whose pixel values are smaller than $\mathcal{I}(i,j,k)$.
    \item If $S=\emptyset$, set $(x_{i,j,k},y_{i,j,k},z_{i,j,k})$=(0,0,0)
    \item If $S\neq \emptyset$, Identify the positions in $S$ with the smallest pixel value. If there is a unique position, let $(x_{i,j,k},y_{i,j,k},z_{i,j,k})$ be the vector pointing from $(i,j,k)$ to this position, with magnitude $\mathcal{I}(i,j,k)$.
    If multiple positions have the same smallest value, compute the average direction from $(i,j,k)$ to these positions and assign the resulting vector a magnitude of $\mathcal{I}(i,j,k)$.
\end{enumerate}
This vector field captures the flow of pixel values from regions of higher intensity to those of lower intensity. Alternatively, one can construct a vector field that represents the reverse flow, from lower-intensity to higher-intensity regions.

\subsection{Other Methods}
For color images, vector fields can be derived by selecting pairs of color channels. For instance, a $(r, g, b)$-channel image can yield three distinct vector fields based on the $(r,g)$, $(r,b)$ and $(g,b)$ channel pairs.

For images with large dimensions, the image can be divided into smaller patches, and vector fields can be computed for these patches. For example, consider an image of size 1024$\times$1024. By dividing it into $16\!\times\! 16$ patches, a new 64$\times$64 image can be formed, where each ``pixel" represents a patch. Topological indices of the patches, such as $(\beta_0,\beta_1)$, can then be used to define vectors for the corresponding pixels in the new image.

\subsection{Evaluation of Different Methods}
We conducted a comparative analysis of three distinct vector field generation methods: gradient-based, flow-based, and RGB-based, on the OrganSMNIST dataset.  For gradient-based method, we consider three cases: $s=t=1$, $s=t=2$, and $s=t=3$. For RGB-based method, we converted grayscale images into color images with three channels, then use the pairs (r,b), (r,b), and (g,b) to form vector fields. The results of these methods are shown in Table \ref{tab:vector-field-test}.
\begin{table}[h]
    \centering
    \caption{Performance of MTDL using different vector field generation methods on the OrganSMNIST dataset.}
    \begin{tabular}{c|c|c|c|c|c}
        \hline
         Methods &Gradient (1,1) & Gradient (2,2) & Gradient (3,3) & Flow & RGB \\
         \hline
        AUC & 0.978 & 0.977 & 0.980 & 0.976 & 0.981\\
        ACC & 0.809 & 0.806 & 0.817 & 0.792 & 0.818\\
		\hline
	\end{tabular}
    \label{tab:vector-field-test}
\end{table} 
The gradient-based method with $s=t=1$ is the model we used in the work.  It can be seen that the model performance can be further improved by  considering the RGB-based method or gradient-based method with $s=t=3$. This could be further explored in the future study.

\section{Dataset Details}

The MedMNIST v2 dataset is a standardized, MNIST-like collection of biomedical images. All images are preprocessed into a uniform size and labeled, eliminating the need for domain knowledge from users. The dataset includes twelve 2D datasets and six 3D datasets, covering a range of data modalities, data scales, and task types. In total, it includes 708,069 2D images and 9,998 3D images, with standard train-validation-test splits provided for all datasets. The detailed data scale, data modality, and task type information for each dataset are shown in Table \ref{tab:medmnist}.
\begin{table}[ht]
\centering
\scriptsize
\caption{Overview of MedMNIST v2 dataset}
\label{tab:medmnist}
\begin{tabular}{l l l r r}
    \hline
    \textbf{MedMNIST2D} & \textbf{Data Modality} & \textbf{Task (\# Classes / Labels)} & \textbf{\# Samples} & \textbf{\# Training / Validation / Test} \\
    \hline
    PathMNIST  & Colon Pathology   & Multi-Class (9)             & 107,180    & 89,996 / 10,004 / 7,180 \\
    ChestMNIST & Chest X-Ray       & Multi-Label (14) Binary-Class (2)   & 112,120   & 78,468 / 11,219 / 22,433 \\
    DermaMNIST & Dermatoscope      & Multi-Class (7)             & 10,015     & 7,007 / 1,003 / 2,005 \\
    OCTMNIST   & Retinal OCT       & Multi-Class (4)             & 109,309    & 97,477 / 10,832 / 1,000 \\
    PneumoniaMNIST & Chest X-Ray   & Binary-Class (2)            & 5,856      & 4,708 / 524 / 624 \\
    RetinaMNIST   & Fundus Camera  & Ordinal Regression (5)        & 1,600 & 1,080 / 120 / 400 \\
    BloodMNIST & Blood Cell Microscope  & Multi-Class (8)             & 17,092     & 11,959 / 1,712 / 3,421 \\
    TissueMNIST& Kidney Cortex Microscope & Multi-Class (8)        & 236,386    & 165,466 / 23,640 / 47,280 \\
    OrganAMNIST& Abdominal CT      & Multi-Class (11)             & 58,850     & 34,581 / 6,491 / 17,778 \\
    OrganCMNIST& Abdominal CT      & Multi-Class (11)             & 23,660     & 13,000 / 2,392 / 8,268 \\
    OrganSMNIST& Abdominal CT      & Multi-Class (11)             & 25,221     & 13,940 / 2,452 / 8,829 \\
    \hline
    \hline
    \textbf{MedMNIST3D} & \textbf{Data Modality} & \textbf{Task (\# Classes / Labels)} & \textbf{\# Samples} & \textbf{\# Training / Validation / Test} \\
    \hline
    OrganMNIST3D	&Abdominal CT	&Multi-Class (11)	&1,742	&971 / 161 / 610\\
    NoduleMNIST3D	&Chest CT	    &Binary-Class (2)	&1,633	&1,158 / 165 / 310\\
    AdrenalMNIST3D	&Shape from Abdominal CT&Binary-Class (2)&	1,584   &1,188 / 98 / 298\\
    FractureMNIST3D	&Chest CT	   &Multi-Class (3)	&1,370	&1,027 / 103 / 240\\
    VesselMNIST3D	&Shape from Brain MRA&Binary-Class (2)	&1,908	&1,335 / 191 / 382\\
    SynapseMNIST3D	&Electron Microscope	&Binary-Class (2)	&1,759&	1,230 / 177 / 352\\
    \hline
\end{tabular}
\end{table}
The available data resolutions are 28$\times$28, 64$\times$64, 128$\times$128, 224$\times$224 for 2D datasets, and 28$\times$28$\times$28, 64$\times$64$\times$64 for 3D datasets. Here we also give a brief introduction of all the 17 datasets. The example illustration of 2D and 3D datasets are shown in Fig. \ref{fig:2d-data} and Fig. \ref{fig:3d-data} respectively.
\subsection{2D Datasets}
\begin{itemize}
    \item PathMNIST: the PathMNIST is based on a prior study \cite{kather2019predicting,kather_2018_1214456} for predicting survival from colorectal cancer histology slides. The dataset is comprised of 9 types of tissues, resulting in a multi-class classification task. The labels are adipose, background, debris, lymphocytes, mucus, smooth muscle, normal colon mucosa, cancer-associated stroma,  and colorectal adenocarcinoma epithelium. These images were manually extracted from N=86 H\&E stained human cancer tissue slides from formalin-fixed paraffin-embedded (FFPE) samples from the NCT Biobank (National Center for Tumor Diseases, Heidelberg, Germany) and the UMM pathology archive (University Medical Center Mannheim, Mannheim, Germany). Example images are shown in Fig. \ref{fig:path-data}.
    \begin{figure}
        \centering
        \includegraphics[width=1\linewidth]{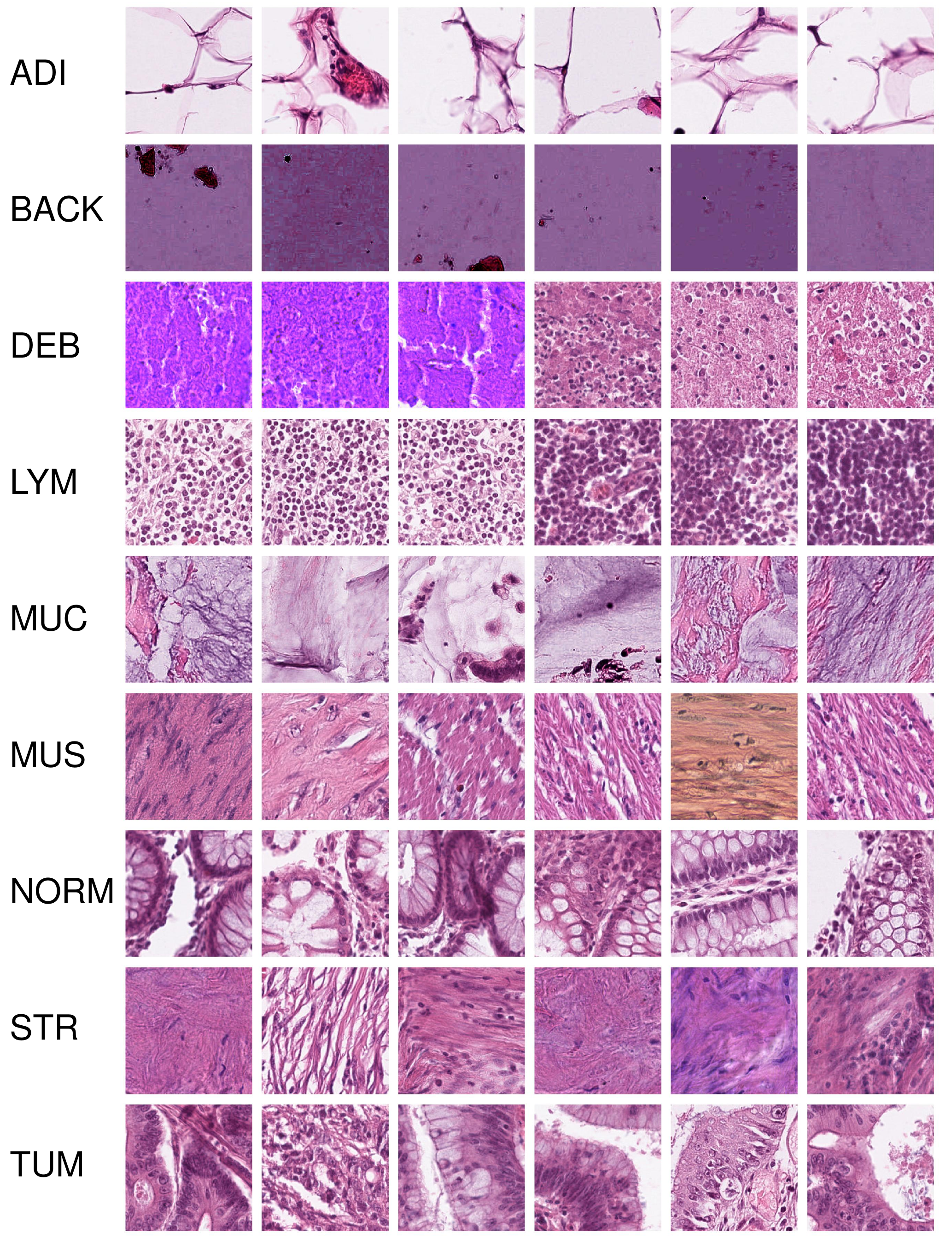}
        \caption{Example images of the nine classes in the PathMNIST dataset. ADI: adipose tissue; BACK: background; DEB: debris; LYM: lymphocytes; MUC: mucus; MUS: smooth muscle; NORM: normal colon mucosa; STR: cancer-associated stroma; TUM: colorectal adenocarcinoma epithelium.}
        \label{fig:path-data}
    \end{figure}
    \item ChestMNIST: the ChestMNIST is based on the NIH-ChestXray14 dataset \cite{wang2017chestx}, consisting of frontal-view X-Ray images of 30,805 unique patients with the text-mined 14 disease labels, leads to a multi-label binary-class classification task. The labels are atelectasis, cardiomegaly, effusion, infiltration, mass, nodule, pneumonia, pneumothorax, consolidation, edema, emphysema, fibrosis, pleural, and hernia. Example images are shown in Fig. \ref{fig:chest-data1} and Fig. \ref{fig:chest-data2}. 
    \begin{figure}
        \centering
        \includegraphics[width=1\linewidth]{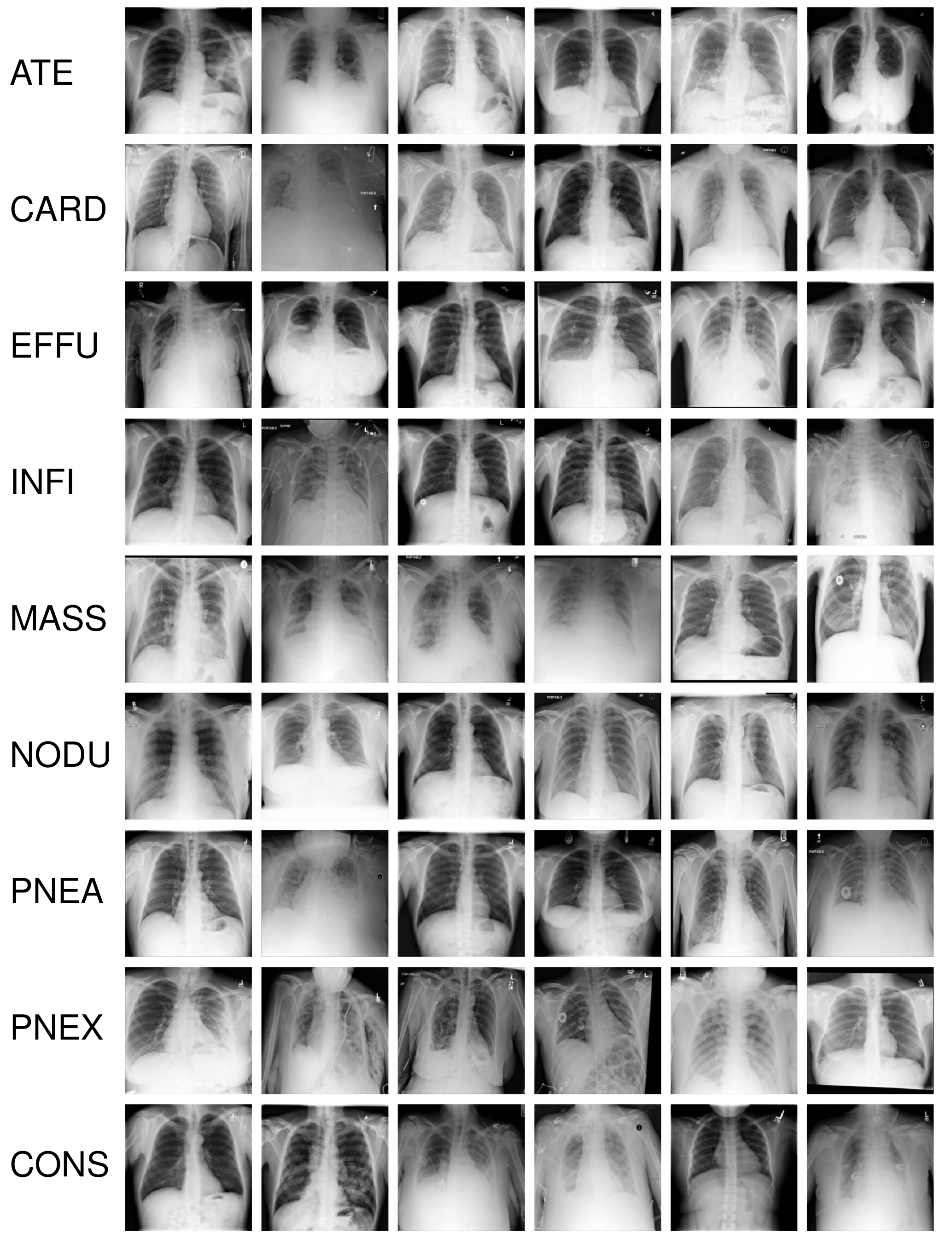}
        \caption{Example images of the first nine classes in the ChestMNIST dataset. ATE: atelectasis; CARD: cardiomegaly; EFFU: effusion; INFI: infiltration; MASS: mass; NODU: nodule; PNEA: pneumonia; PNEX: pneumothorax; CONS: consolidation.}
        \label{fig:chest-data1}
    \end{figure}
    \begin{figure}
        \centering
        \includegraphics[width=1\linewidth]{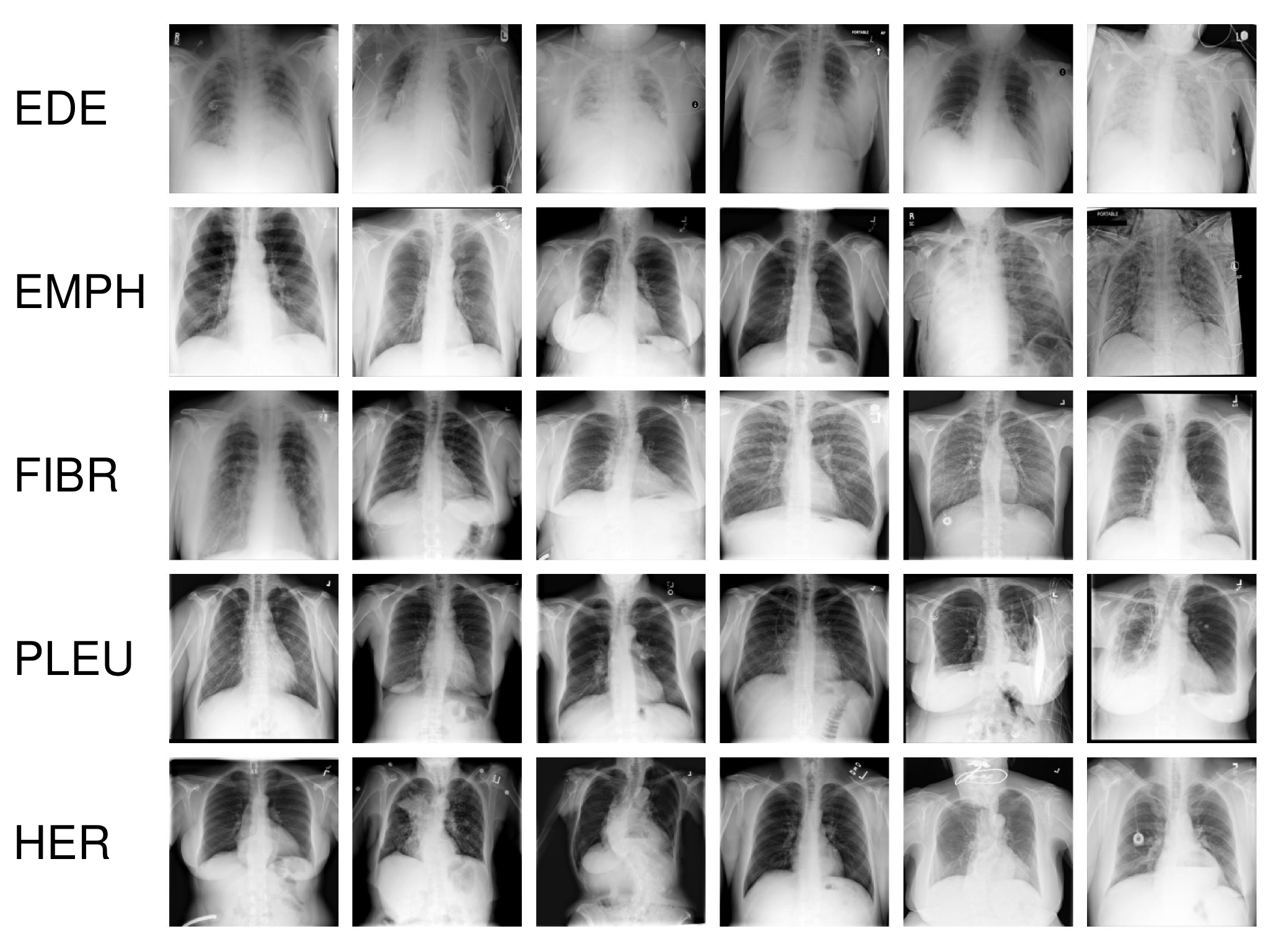}
        \caption{Example images of the rest five classes in the ChestMNIST dataset. EDE: edema; EMPH: emphysema; FIBR: fibrosis; PLEU: pleural; HER: hernia.}
        \label{fig:chest-data2}
    \end{figure}
    \item DermaMNIST: the DermaMNIST is based on the HAM10000 dataset \cite{tschandl2018ham10000,codella2019skin}, a collection of multi-source dermatoscopic images of common pigmented skin lesions, consisting of 7 diseases, formalizing as a multi-class classification task. The labels are actinic keratoses and intraepithelial carcinoma, basal cell carcinoma, benign keratosis-like lesions, dermatofibroma, melanoma, melanocytic nevi, and vascular lesions. Example images are shown in Fig. \ref{fig:derma-data}.
    \begin{figure}
        \centering
        \includegraphics[width=1\linewidth]{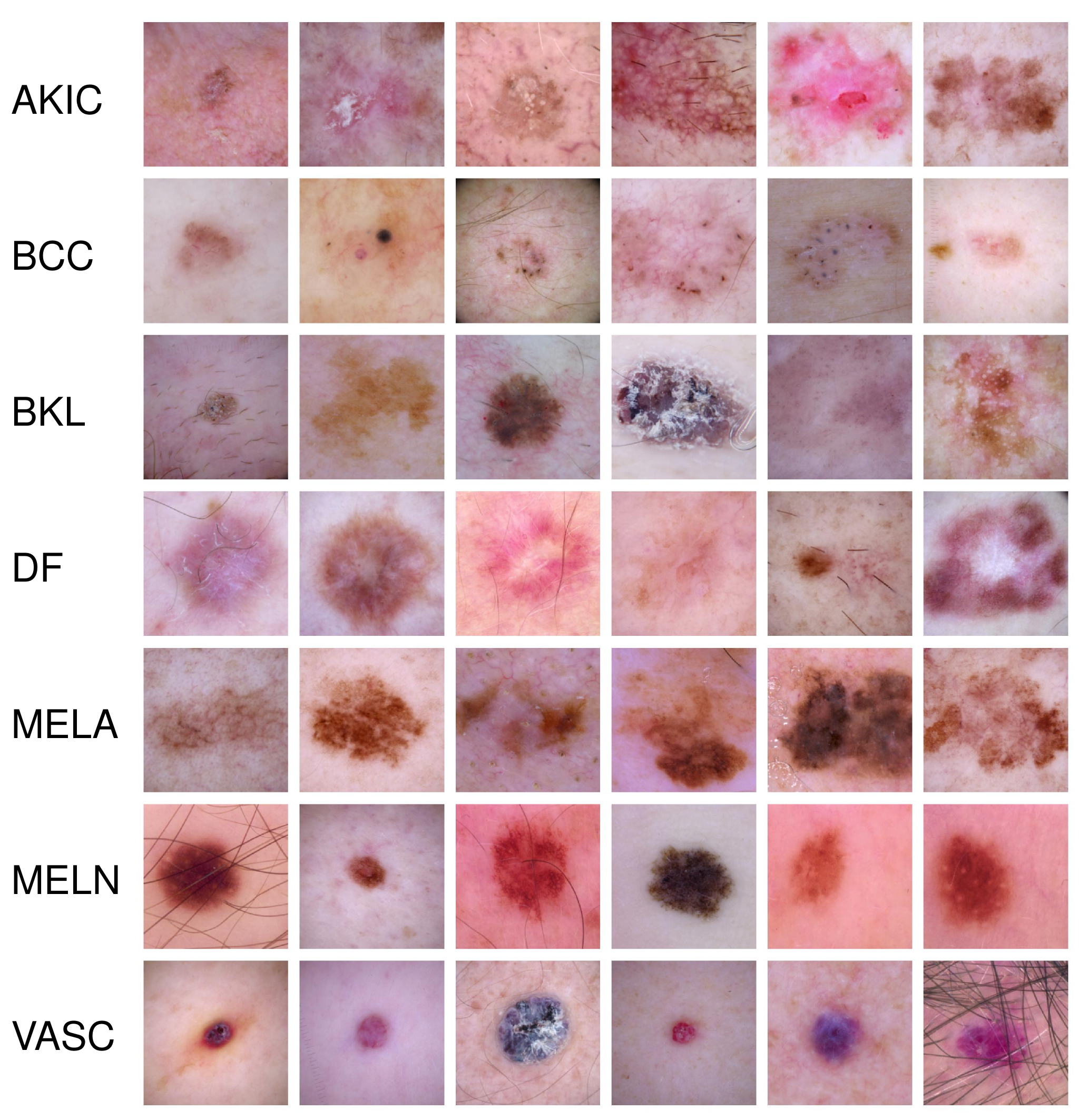}
        \caption{Example images of the seven classes in the DermaMNIST dataset. AKIC: actinic keratoses and intraepithelial carcinoma; BCC: basal cell carcinoma; BKL: benign keratosis-like lesions; DF: dermatofibroma; MELA: melanoma; MELN: melanocytic nevi; VASC: vascular lesions.}
        \label{fig:derma-data}
    \end{figure}
    \item OCTMNIST: the OCTMNIST is from a dataset \cite{kermany2018identifying} of valid optical coherence tomography (OCT) images for retinal diseases, consisting of 4 categories, leading to a multi-class classification task. The labels are choroidal neovascularization, diabetic macular edema, drusen, and normal. Example images are shown in Fig. \ref{fig:oct-data}.
    \begin{figure}
        \centering
        \includegraphics[width=1\linewidth]{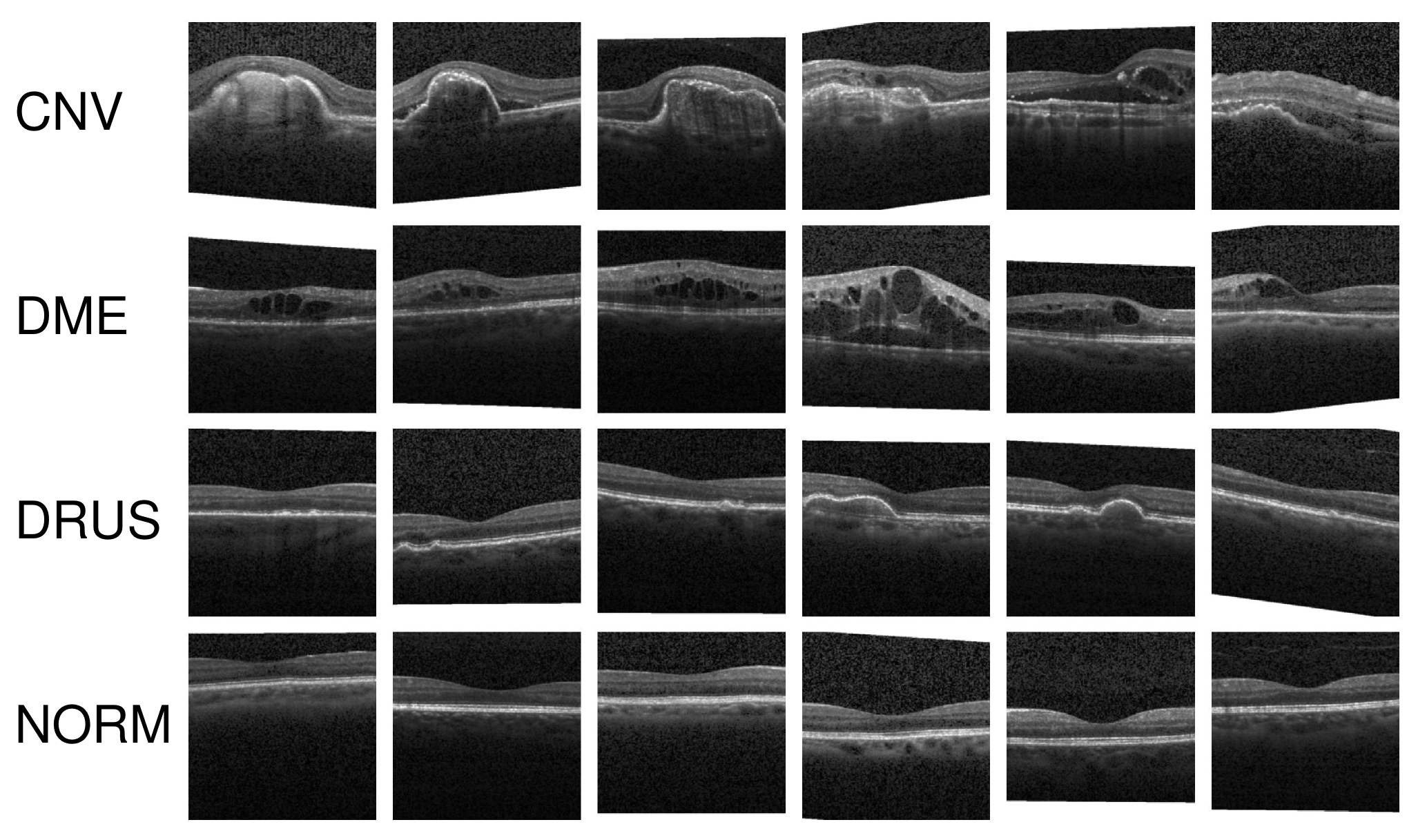}
        \caption{Example images of the four classes in the OCTMNIST dataset. CNV: choroidal neovascularization; DME: diabetic macular edema; DRUS: drusen; NORM: normal.}
        \label{fig:oct-data}
    \end{figure}
    \item PneumoniaMNIST: the PneumoniaMNIST is from a dataset \cite{kermany2018large} of 5856 pediatric chest X-Ray images, the task is binary-class classification of pneumonia against normal. These images were selected from retrospective cohorts of pediatric patients of one to five years old from Guangzhou Women and Children’s Medical Center, Guangzhou. All chest X-ray imaging was performed as part of patients’ routine clinical care. Example images are shown in Fig. \ref{fig:pneumonia-data}.
    \begin{figure}
        \centering
        \includegraphics[width=1\linewidth]{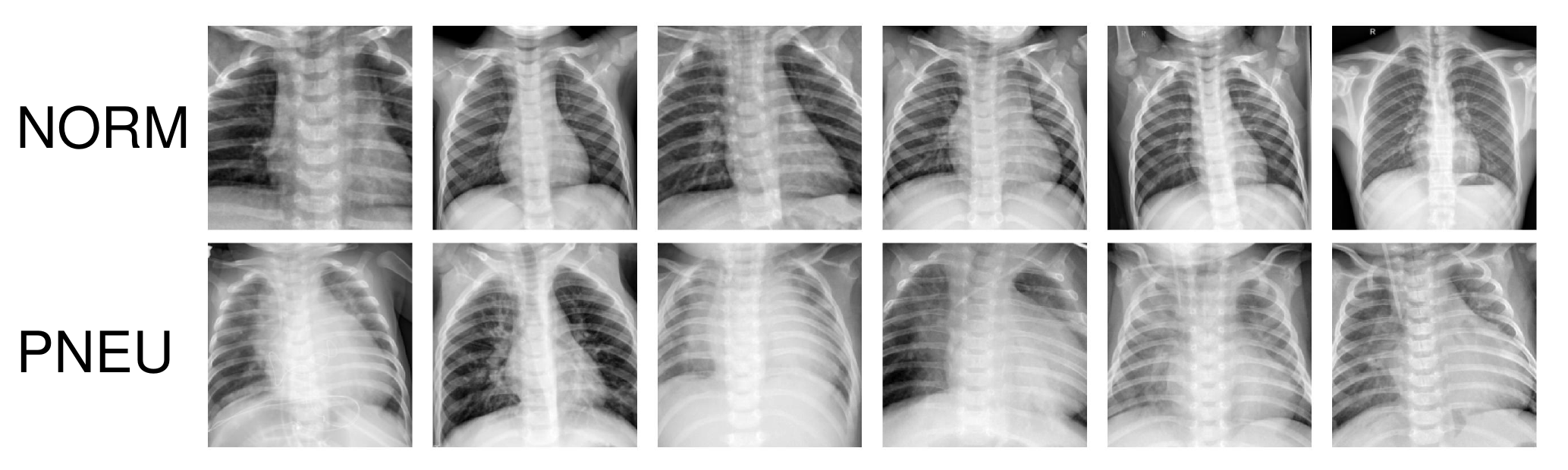}
        \caption{Example images of the two classes in the PneumoniaMNIST dataset. NORM: normal; PNEU: pneumonia.}
        \label{fig:pneumonia-data}
    \end{figure}
    \item RetinaMNIST: the RetinaMNIST is based on the DeepDRiD24 challenge \cite{liu2022deepdrid}, which provides a collection of 1600 retina fundus images. The task is ordinal regression of 5-level grading of diabetic retinopathy severity. An internationally accepted method of grading the DR levels classifies DR into non-proliferative DR (NPDR) and proliferative DR (PDR) \cite{world2006prevention}. NPDR is the early stage of DR and is characterized by the presence of microaneurysms, whereas PDR is an advanced stage of DR and can lead to severe vision loss. Example images are shown in Fig. \ref{fig:retina-data}.
    \begin{figure}
        \centering
        \includegraphics[width=1\linewidth]{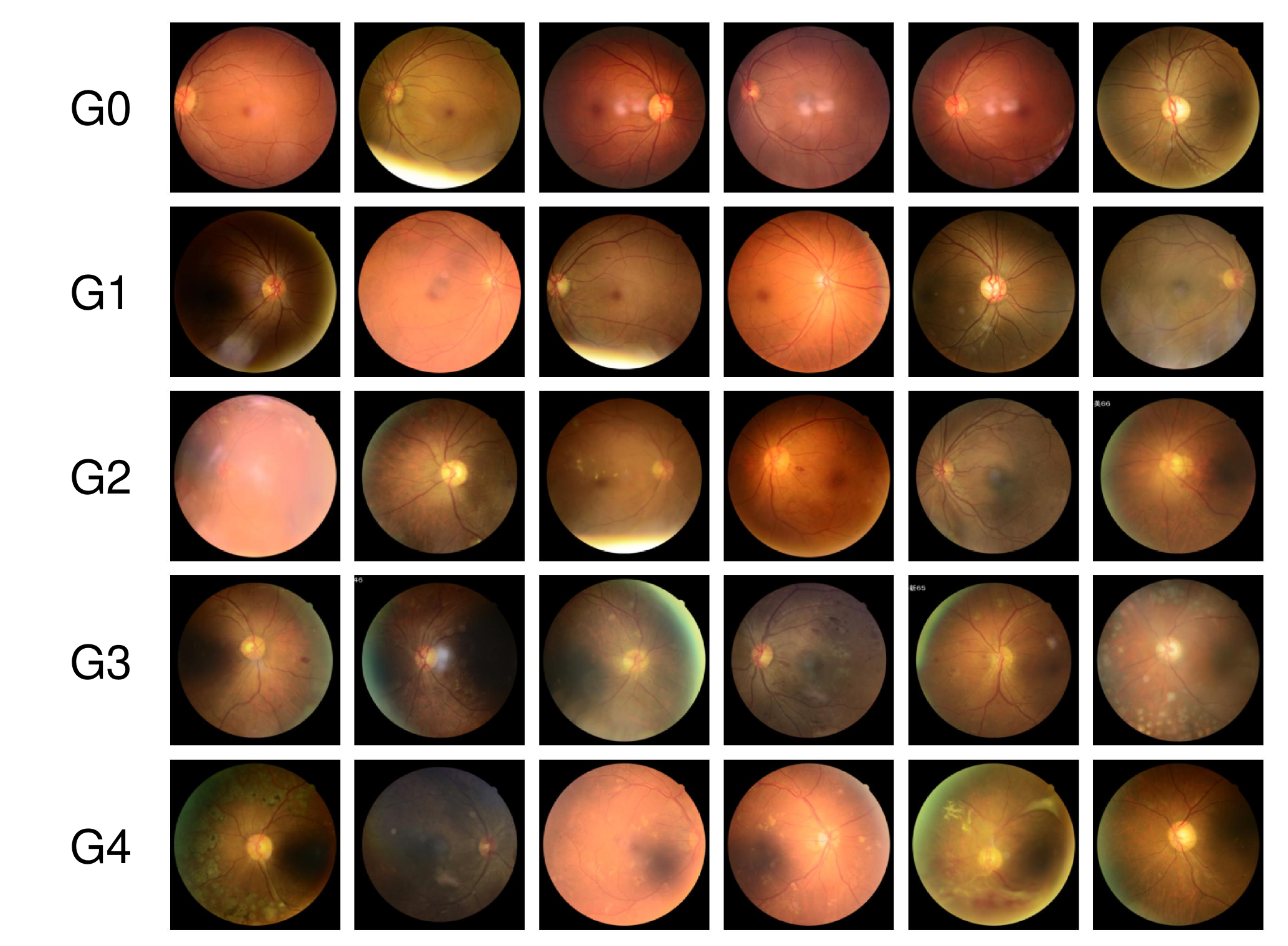}
        \caption{Example images of the five classes in the RetinaMNIST dataset. G0: no apparent retinopathy; G1: mild NPDR; G2: moderate NPDR; G3: severe NPDR; G4: PDR.}
        \label{fig:retina-data}
    \end{figure}
    \item BloodMNIST: the BloodMNIST is based on a dataset of individual normal cells \cite{acevedo2020dataset}, captured from individuals without infection, hematologic or oncologic disease and free of any pharmacologic treatment at the moment of blood collection, consisting of 8 classes. The labels are basophil, eosinophil, erythroblast, immature granulocytes(myelocytes, metamyelocytes and promyelocytes), lymphocyte, monocyte, neutrophil, and platelet. The images were acquired using the analyzer CellaVision DM96 in the Core Laboratory at the Hospital Clinic of Barcelona. Example images are shown in Fig. \ref{fig:blood-data}.
    \begin{figure}
        \centering
        \includegraphics[width=1\linewidth]{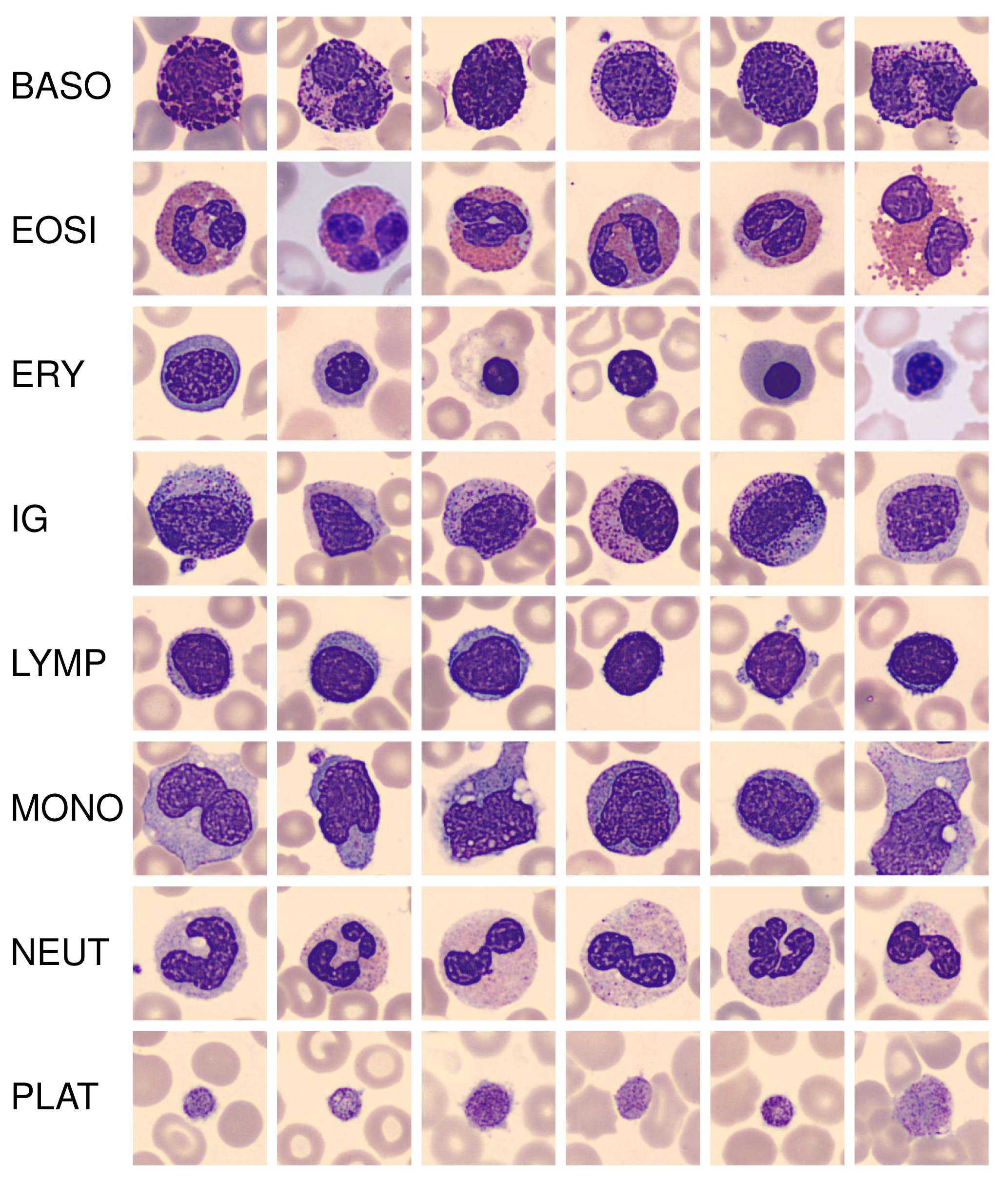}
        \caption{Example images of the eight classes in the BloodMNIST dataset. BASO: basophil; EOSI: eosinophil; ERY: erythroblast; IG: immature granulocytes (myelocytes, metamyelocytes and promyelocytes); LYMP: lymphocyte; MONO: monocyte; NEUT: neutrophil; PLAT: platelet.}
        \label{fig:blood-data}
    \end{figure}
    \item TissueMNIST: the TissueMNIST is from the BBBC051 \cite{woloshuk2021situ}, available from the Broad Bioimage Benchmark Collection \cite{ljosa2012annotated}. It contains human kidney cortex cells, segmented from 3 reference tissue specimens and organized into 8 categories. The labels are Collecting Duct, Connecting Tubule, Distal Convoluted Tubule,
    Glomerular endothelial cells, Interstitial endothelial cells, Leukocytes, Podocytes, Proximal Tubule Segments, and Thick Ascending Limb. Example images are shown in Fig. \ref{fig:tissue-data}.
    \begin{figure}
        \centering
        \includegraphics[width=1\linewidth]{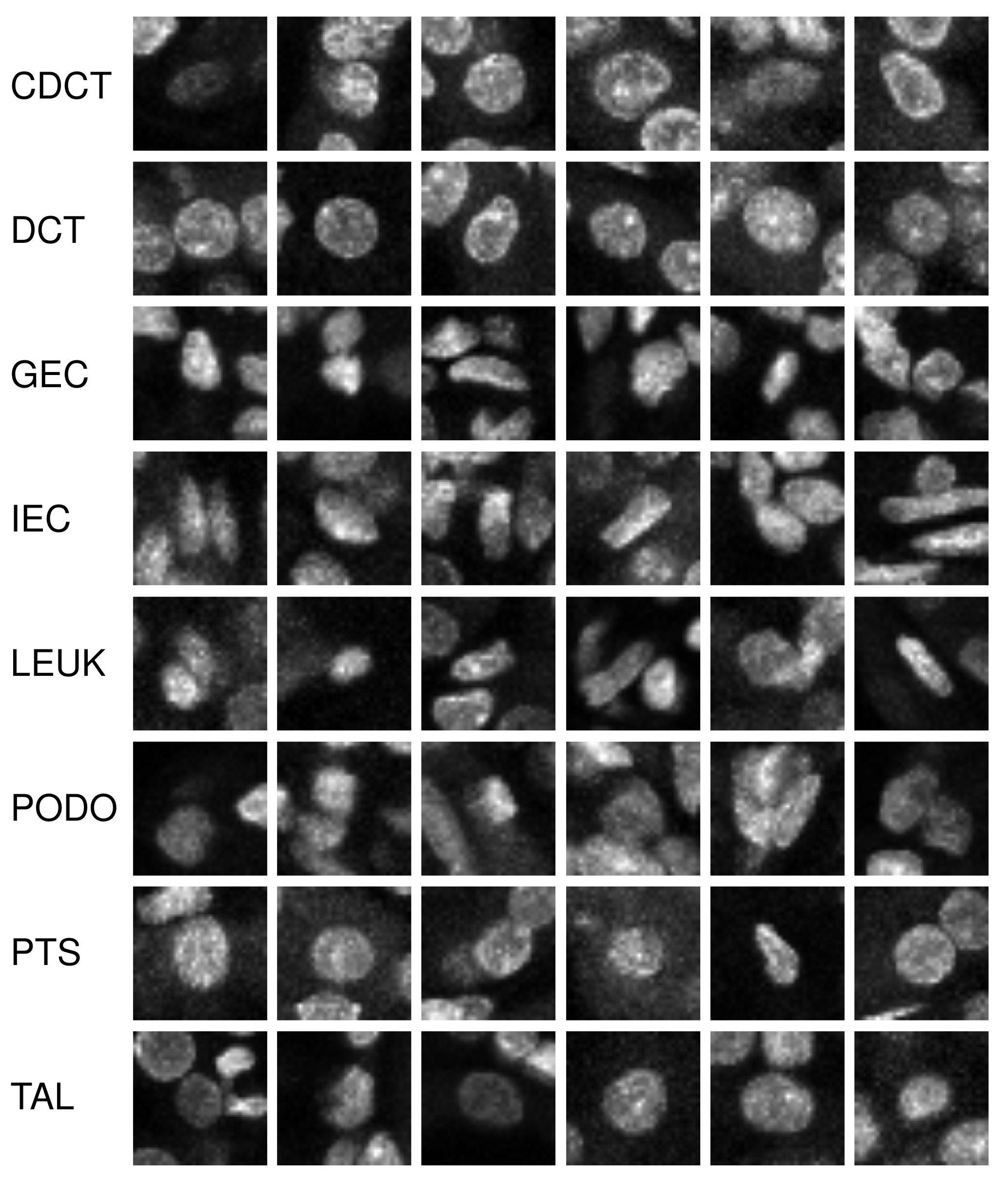}
        \caption{Example images of the eight classes in the TissueMNIST dataset. CDCT: Collecting Duct, Connecting Tubule; DCT: Distal Convoluted Tubule; GEC: Glomerular Endothelial Cells; IEC: Interstitial Endothelial Cells; LEUK: Leukocytes; PODO: Podocytes; PTS: Proximal Tubule Segments; TAL: Thick Ascending Limb.}
        \label{fig:tissue-data}
    \end{figure}
    \item Organ\{A,C,S\}MNIST: the Organ\{A,C,S\} datasets is based on the 3D computed tomography (CT) images from Liver Tumor Segmentation Benchmark \cite{bilic2023liver}, they are from the center slices of the 3D bounding boxes in axial/coronal/sagittal views respectively. The tasks are all multi-class classification with 11-classes. The labels are bladder, femur-left, femur-right, heart, kidney-left, kidney-right, liver, lung-left, lung-right, pancreas, and spleen. Example images for OrganAMNIST are shown in Fig. \ref{fig:organa-data1} and Fig. \ref{fig:organa-data2}, for OrganCMNIST are shown in Fig. \ref{fig:organc-data1} and Fig. \ref{fig:organc-data2}, and for OrganSMNIST are shown in Fig. \ref{fig:organs-data1} and Fig. \ref{fig:organs-data2}.
    \begin{figure}
        \centering
        \includegraphics[width=1\linewidth]{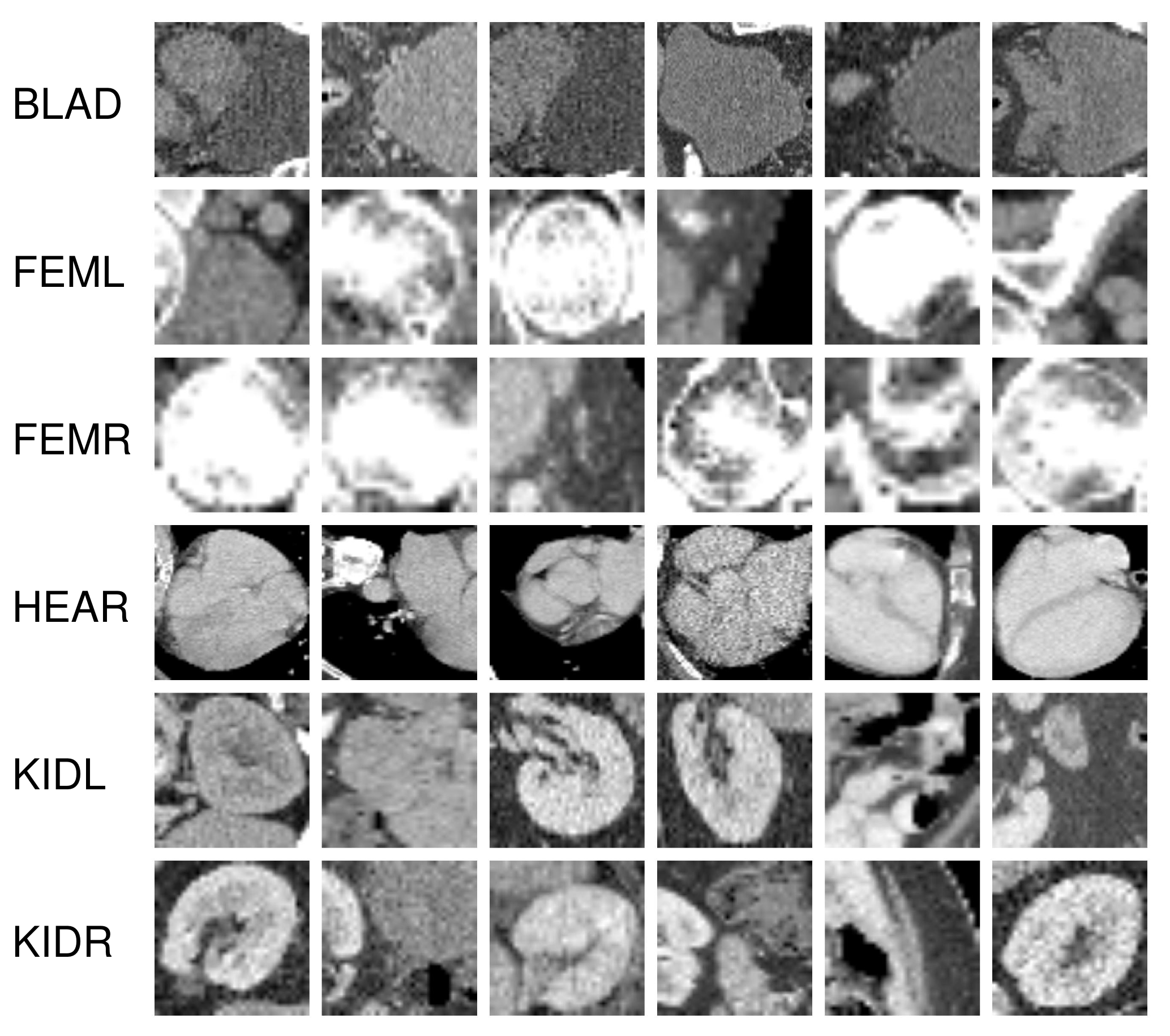}
        \caption{Example images of the first six classes in the OrganAMNIST dataset. BLAD: bladder; FEML: femur-left; FEMR: femur-right; HEAR: heart; KIDL: kidney-left; KIDR: kidney-right. }
        \label{fig:organa-data1}
    \end{figure}
    \begin{figure}
        \centering
        \includegraphics[width=1\linewidth]{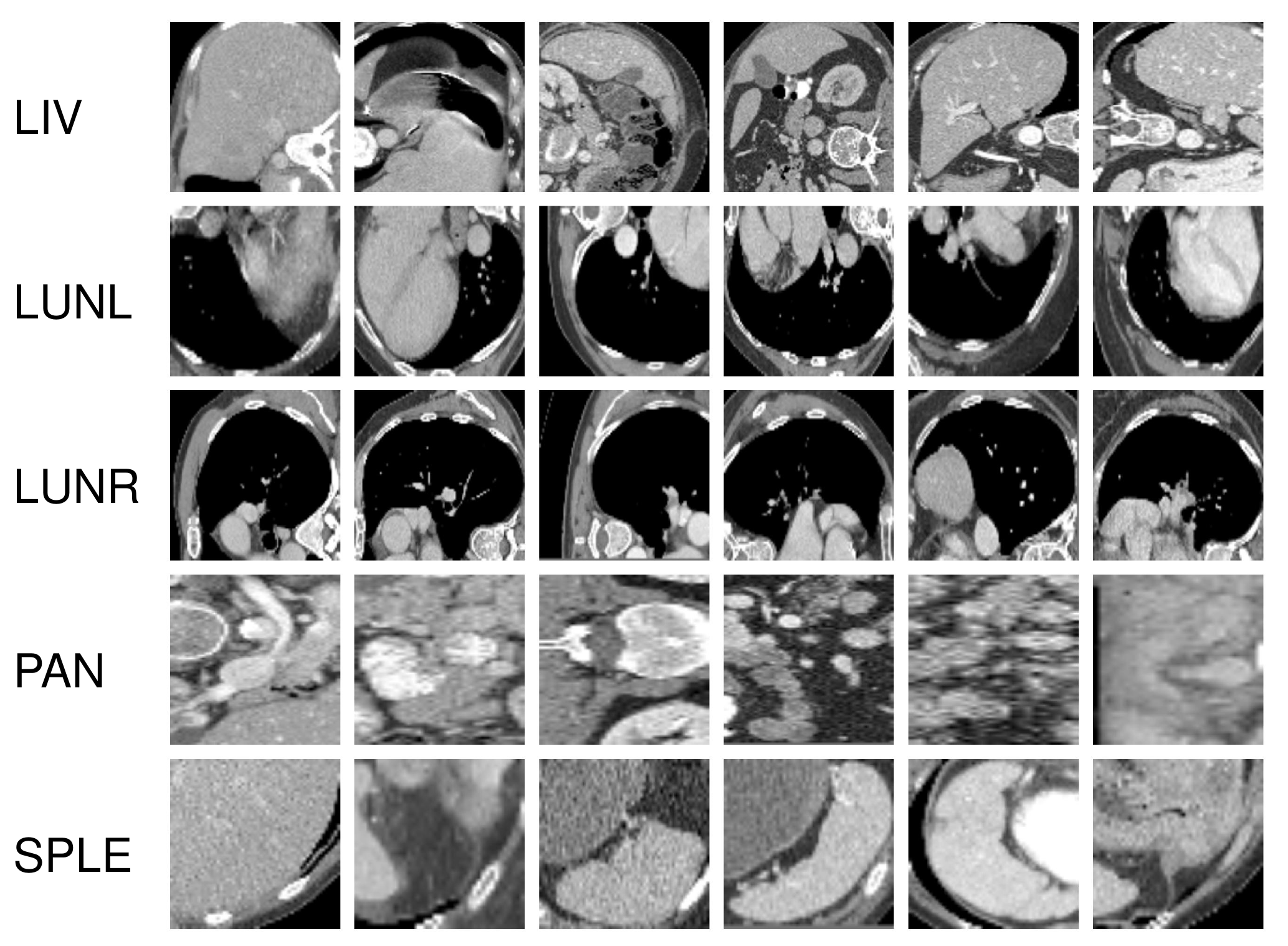}
        \caption{Example images of the rest five classes in the OrganAMNIST dataset. LIV: liver; LUNL: lung-left; LUNR: lung-right; PAN: pancreas; SPLE: spleen.}
        \label{fig:organa-data2}
    \end{figure}
    \begin{figure}
        \centering
        \includegraphics[width=1\linewidth]{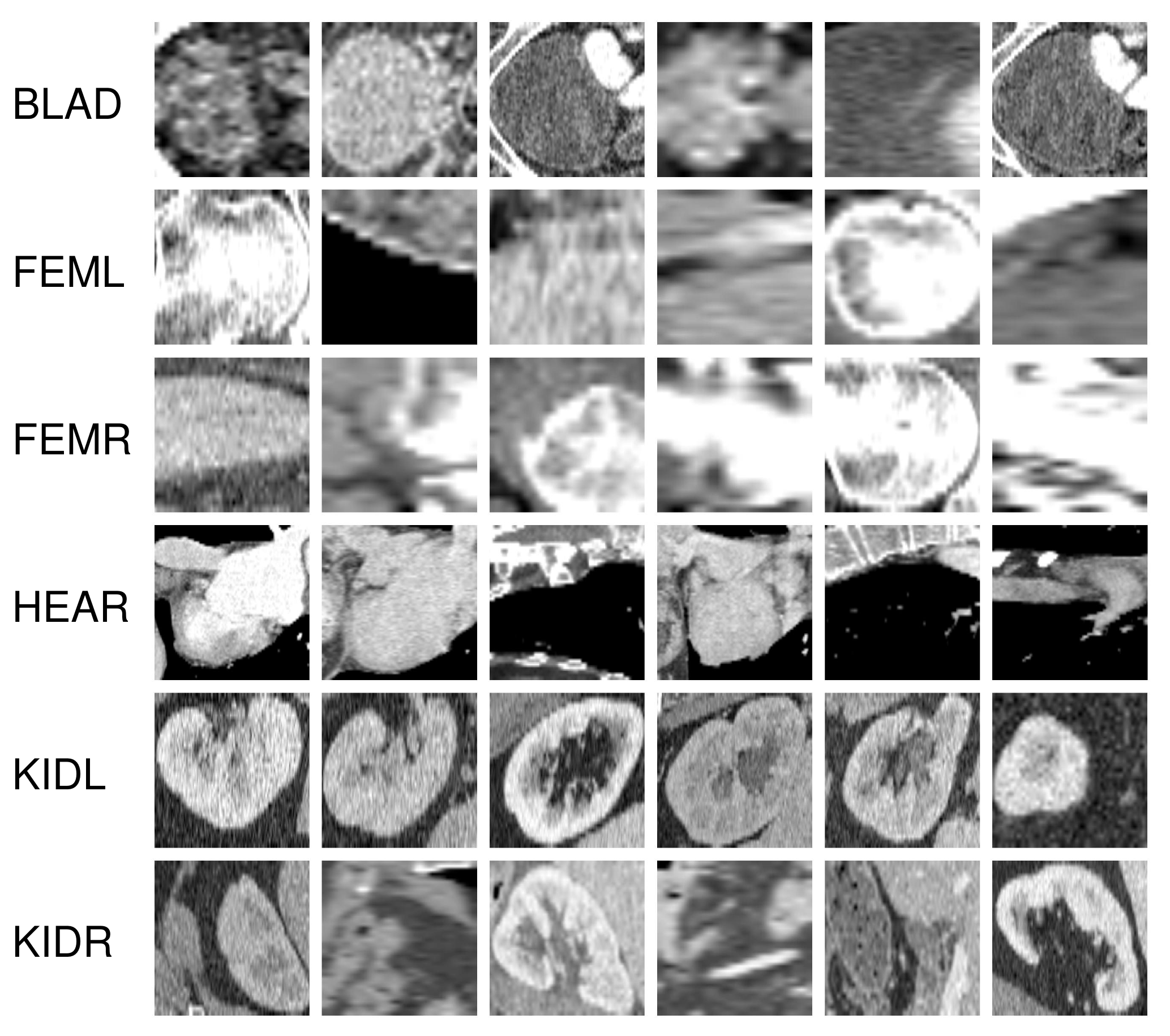}
        \caption{Example images of the first six classes in the OrganCMNIST dataset. BLAD: bladder; FEML: femur-left; FEMR: femur-right; HEAR: heart; KIDL: kidney-left; KIDR: kidney-right. }
        \label{fig:organc-data1}
    \end{figure}
    \begin{figure}
        \centering
        \includegraphics[width=1\linewidth]{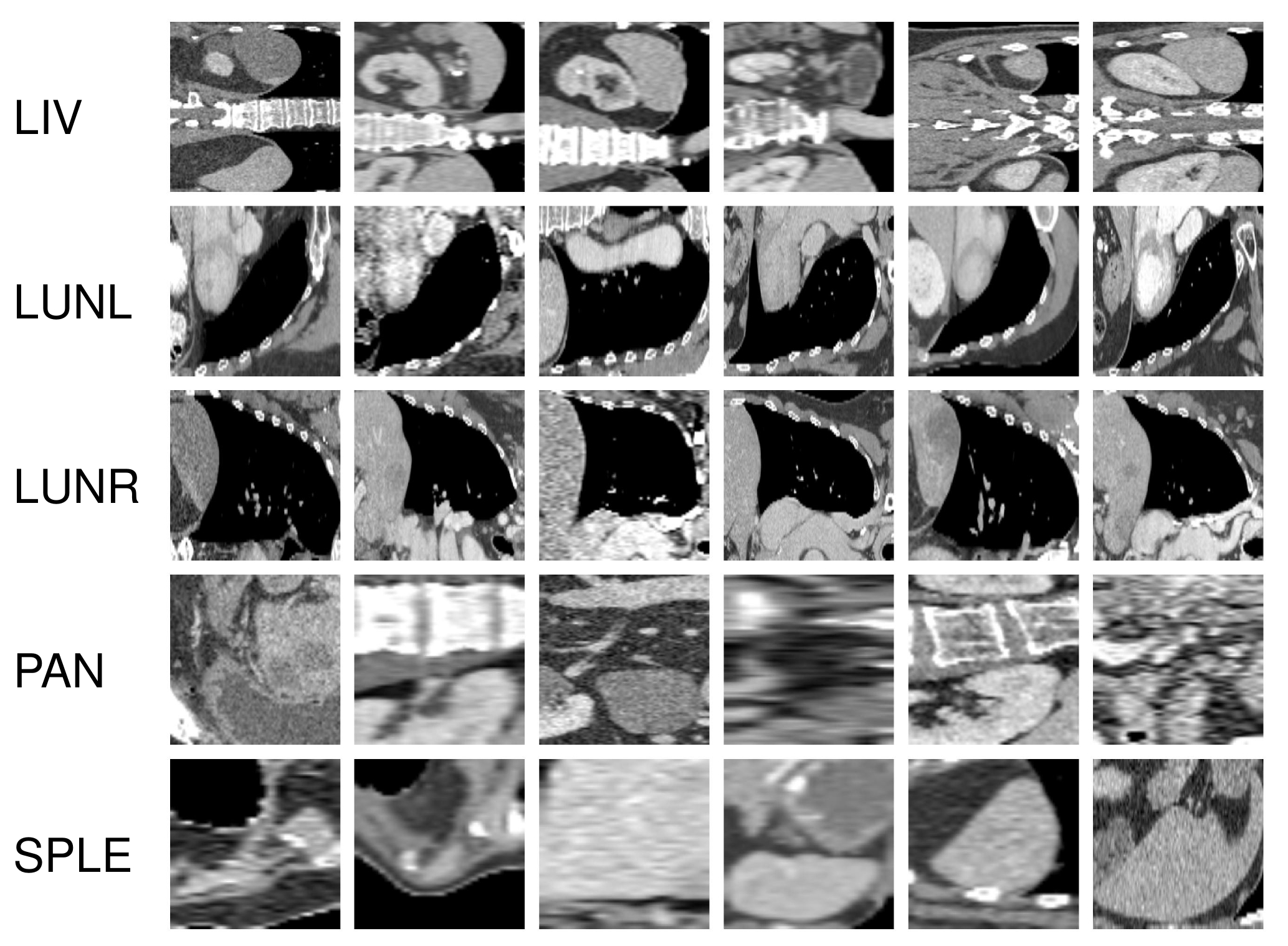}
        \caption{Example images of the rest five classes in the OrganCMNIST dataset. LIV: liver; LUNL: lung-left; LUNR: lung-right; PAN: pancreas; SPLE: spleen.}
        \label{fig:organc-data2}
    \end{figure}
    \begin{figure}
        \centering
        \includegraphics[width=1\linewidth]{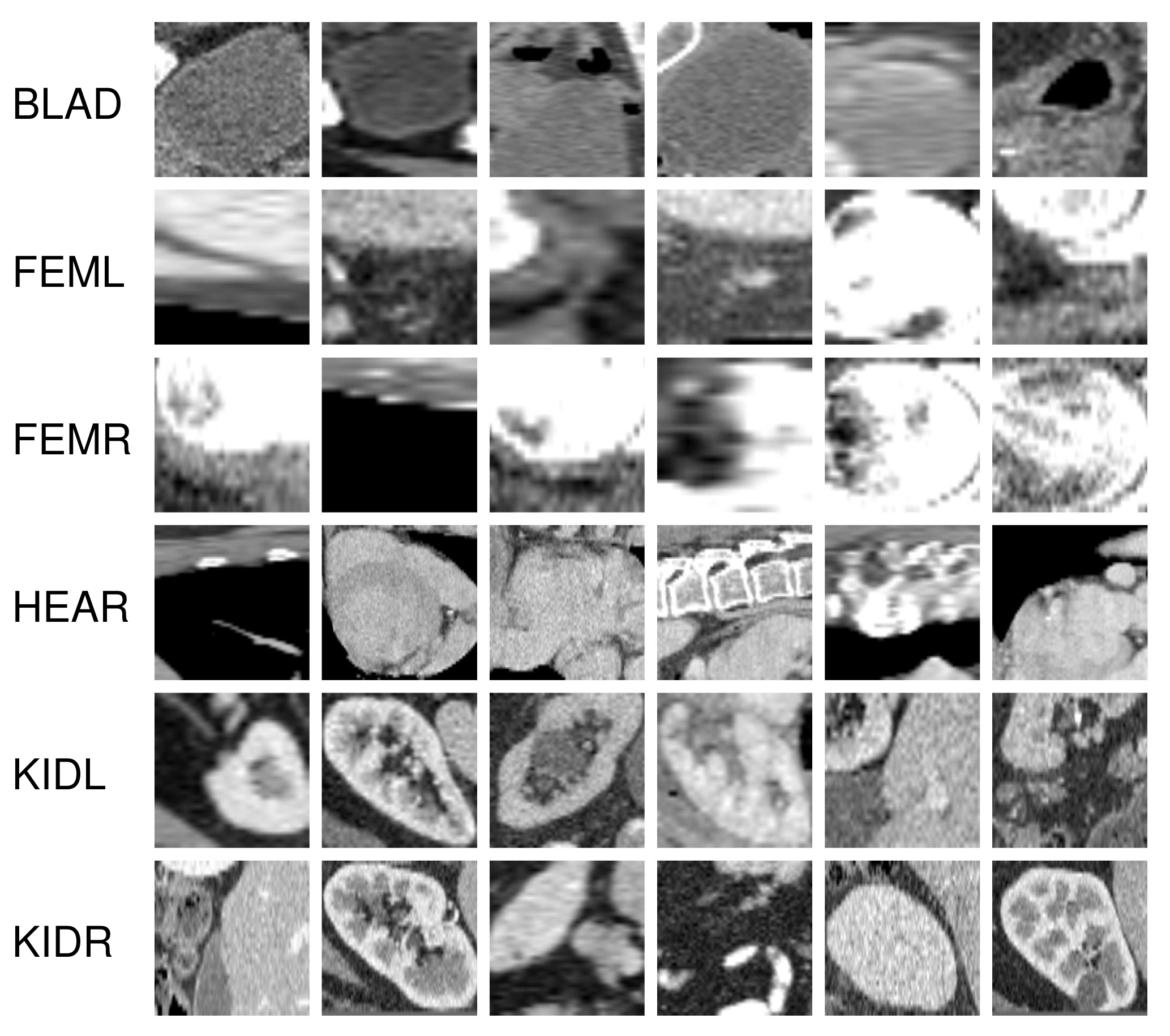}
        \caption{Example images of the first six classes in the OrganSMNIST dataset. BLAD: bladder; FEML: femur-left; FEMR: femur-right; HEAR: heart; KIDL: kidney-left; KIDR: kidney-right. }
        \label{fig:organs-data1}
    \end{figure}
    \begin{figure}
        \centering
        \includegraphics[width=1\linewidth]{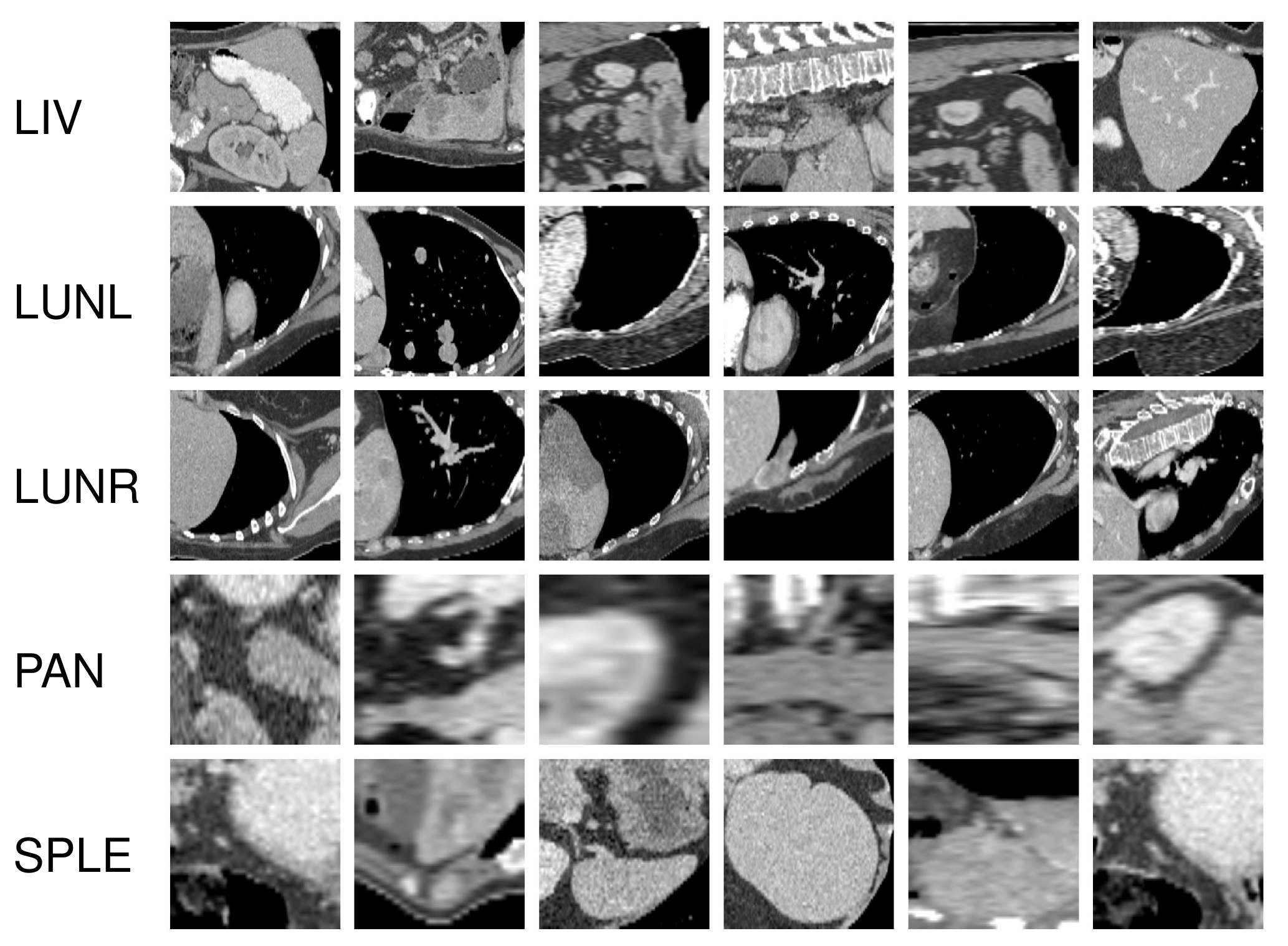}
        \caption{Example images of the rest five classes in the OrganSMNIST dataset. LIV: liver; LUNL: lung-left; LUNR: lung-right; PAN: pancreas; SPLE: spleen.}
        \label{fig:organs-data2}
    \end{figure}
\end{itemize}

\subsection{3D Datasets}
\begin{itemize}
    \item OrganMNIST3D: the OrganMNIST3D is from the same source with 2D Organ datasets. The tasks are multi-class classification with 11-classes. The labels are bladder, femur-left, femur-right, heart, kidney-left, kidney-right, liver, lung-left, lung-right, pancreas, and spleen. Example images are shown in Fig. \ref{fig:organ3d-data1} and Fig. \ref{fig:organ3d-data2}.
    \begin{figure}
        \centering
        \includegraphics[width=1\linewidth]{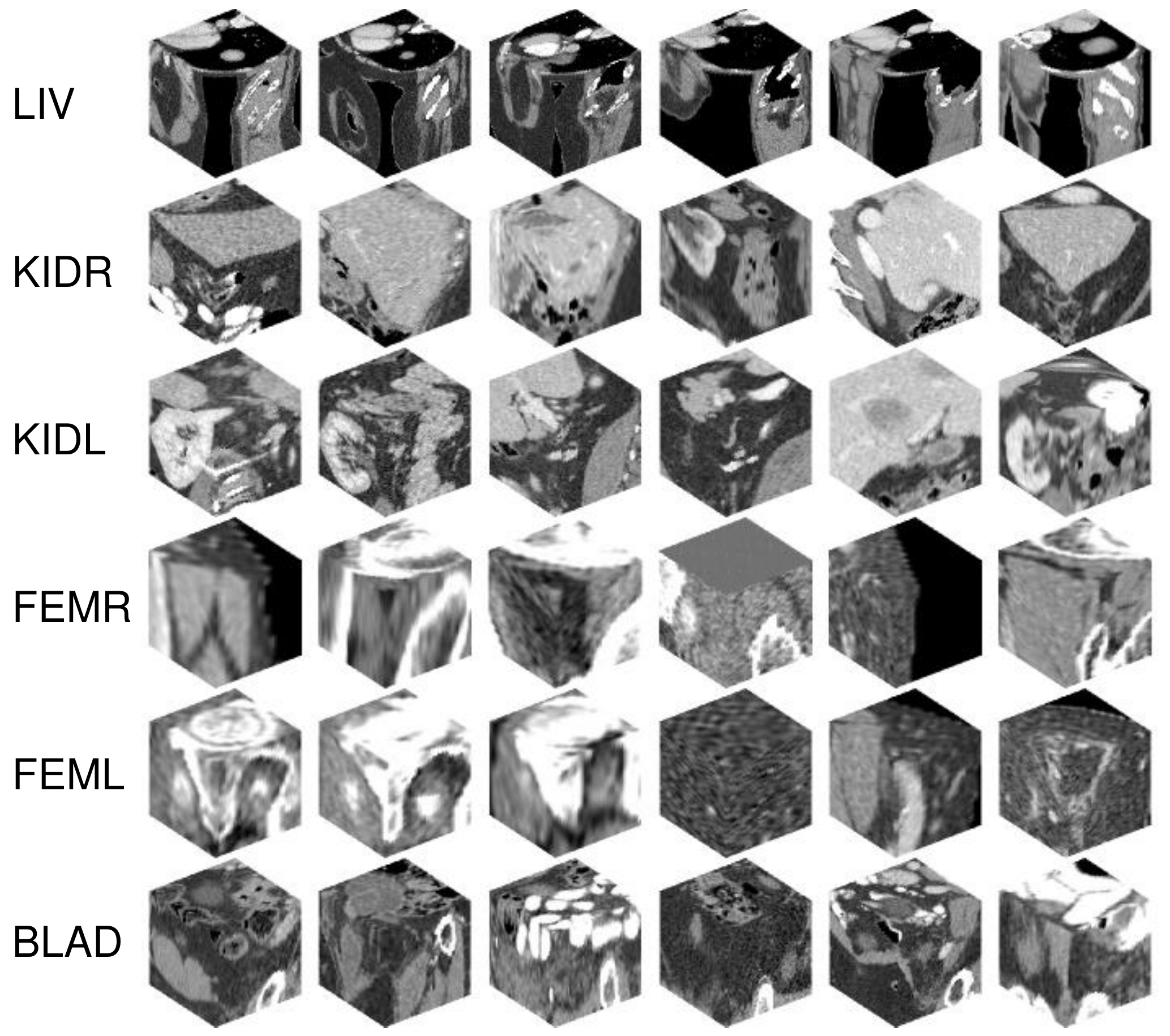}
        \caption{Example images of the first six classes in the OrganMNIST3D dataset. LIV: liver; KIDR: kidney-right; KIDL: kidney-left; FEMR: femur-right; FEML: femur-left; BLAD: bladder.}
        \label{fig:organ3d-data1}
    \end{figure}
     \begin{figure}
        \centering
        \includegraphics[width=1\linewidth]{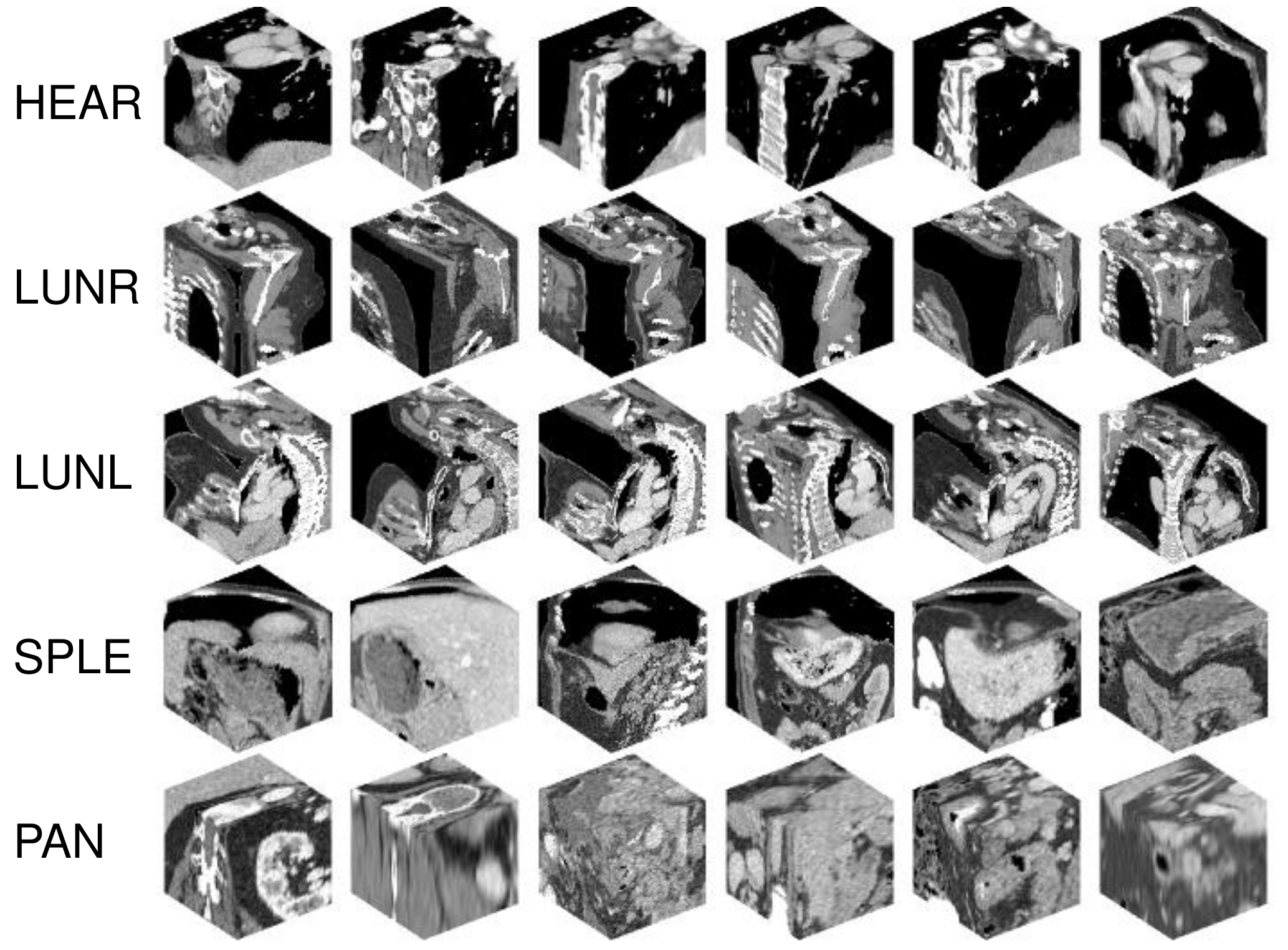}
        \caption{Example images of the rest five classes in the OrganMNIST3D dataset. HEAR: heart; LUNR: lung-right; LUNL: lung-left; SPLE: spleen; PAN: pancreas.}
        \label{fig:organ3d-data2}
    \end{figure}
    \item NoduleMNIST3D: the NoduleMNIST3D is based on the LIDC-IDRI dataset \cite{armato2011lung} that contains images from thoracic CT scans. It contains 1018 cases, each of which includes images from a clinical thoracic CT scan and an associated XML file that records the results of a two-phase image annotation process performed by four experienced thoracic radiologists. The task is binary classification of benign against malignant.
    Example images are shown in Fig. \ref{fig:nodule3d-data}. 
    \begin{figure}
        \centering
        \includegraphics[width=1\linewidth]{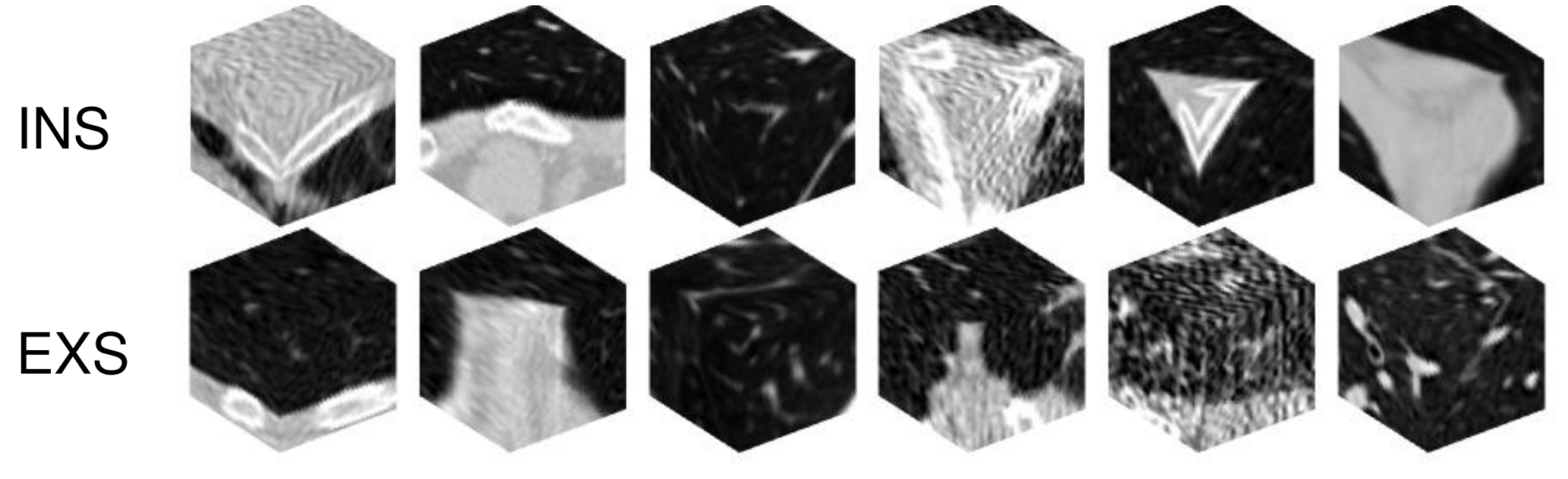}
        \caption{Example images of the two classes in the NoduleMNIST3D dataset. BENI: benign; MALI: maligant.}
        \label{fig:nodule3d-data}
    \end{figure}
    \item AdrenalMNIST3D: the AdrenalMNIST3D dataset consists of shape masks from 1,584 left and right adrenal glands, collected from Zhongshan Hospital Affliated to Fudan University, each 3D shape of adrenal gland is annotated by an expert endocrinologist using abdominal computed tomography (CT), together with a binary classifcation label of normal adrenal gland or adrenal mass. The task is binary classification of normal against mass. 
    Example images are shown in Fig. \ref{fig:adrenal3d-data}. 
    \begin{figure}
        \centering
        \includegraphics[width=1\linewidth]{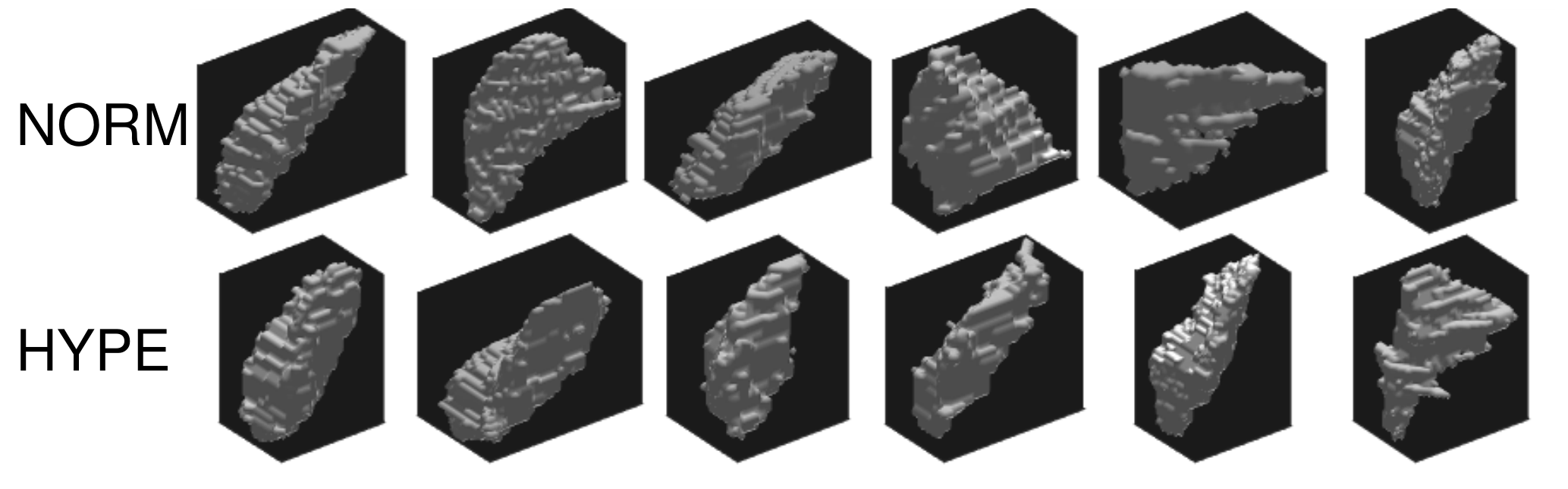}
        \caption{Example images of the two classes in the adrenalMNIST3D dataset. NORM: norm; HYPE: hyperplasia.}
        \label{fig:adrenal3d-data}
    \end{figure}
    \item FractureMNIST3D: the FractureMNIST3D is based on the RibFrac Dataset3 \cite{jin2020deep}, containing around 5,000 rib fractures from 660 computed tomography (CT) scans, which were annotated with a human-in-the-loop labeling procedure. The task is a 3-class classification. The labels are buckle rib fracture, nondisplaced rib fracture, and displaced rib fracture. Example images are shown in Fig. \ref{fig:fracture3d-data}. 
    \begin{figure}
        \centering
        \includegraphics[width=1\linewidth]{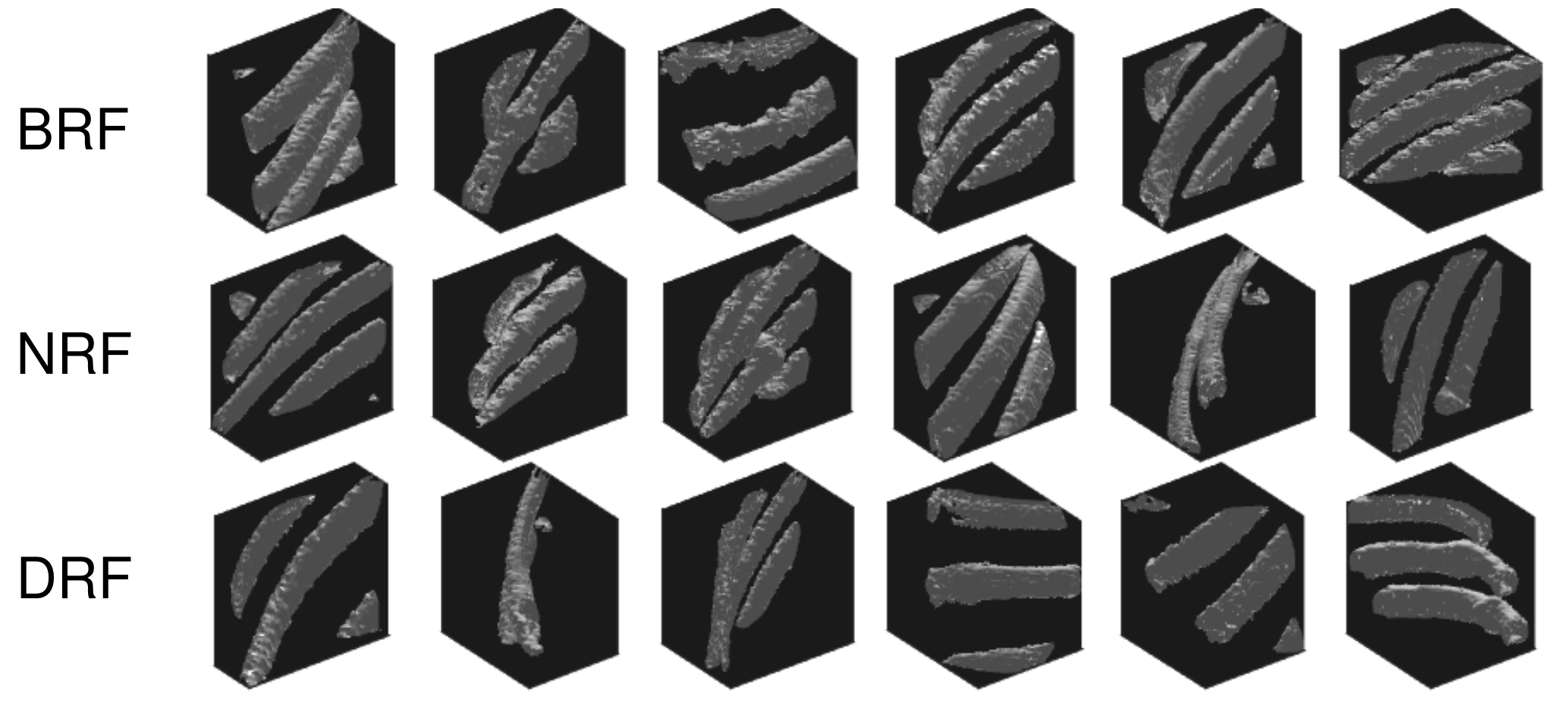}
        \caption{Example images of the three classes in the fractureMNIST3D dataset. BRF: buckle rib fracture; NRF: nondisplaced rib fracture; DRF: displaced rib fracture.}
        \label{fig:fracture3d-data}
    \end{figure}
    \item VesselMNIST3D: the VesselMNIST3D is based on an open-access 3D intracranial aneurysm dataset, IntrA3 \cite{yang2020intra}, containing 103 3D models (meshes) of entire brain vessels collected by reconstructing MRA images. The task is a binary classification of vessel and aneurysm. Example images are shown in Fig. \ref{fig:vessel3d-data}.
    \begin{figure}
        \centering
        \includegraphics[width=1\linewidth]{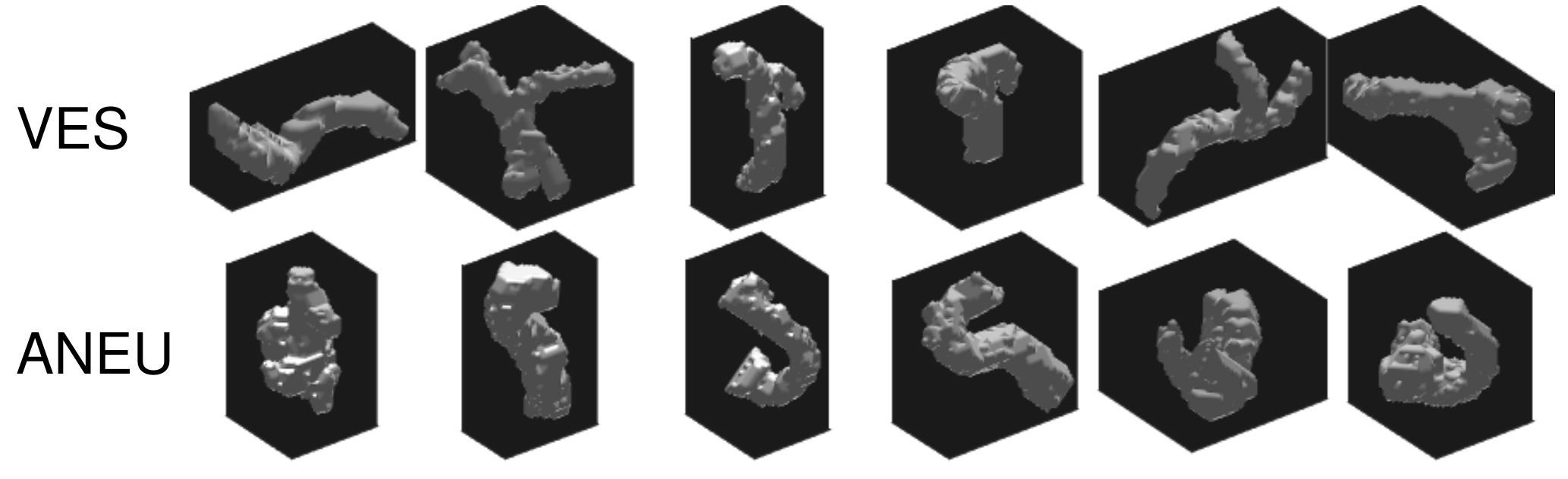}
        \caption{Example images of the two classes in the VesselMNIST3D dataset. VES: vessel; ANEU: aneurysm.}
        \label{fig:vessel3d-data}
    \end{figure}
    \item SynapseMNIST3D: the SynapseMNIST3D dataset is to classify whether a synapse is excitatory or inhibitory. It uses a 3D image volume of an adult rat acquired by a multi-beam scanning electron microscope. Three neuroscience experts segment a pyramidal neuron within the whole volume and proofread all the synapses on this neuron with excitatory/inhibitory labels. Example images are shown in Fig. \ref{fig:synapse3d-data}.
    \begin{figure}
        \centering
        \includegraphics[width=1\linewidth]{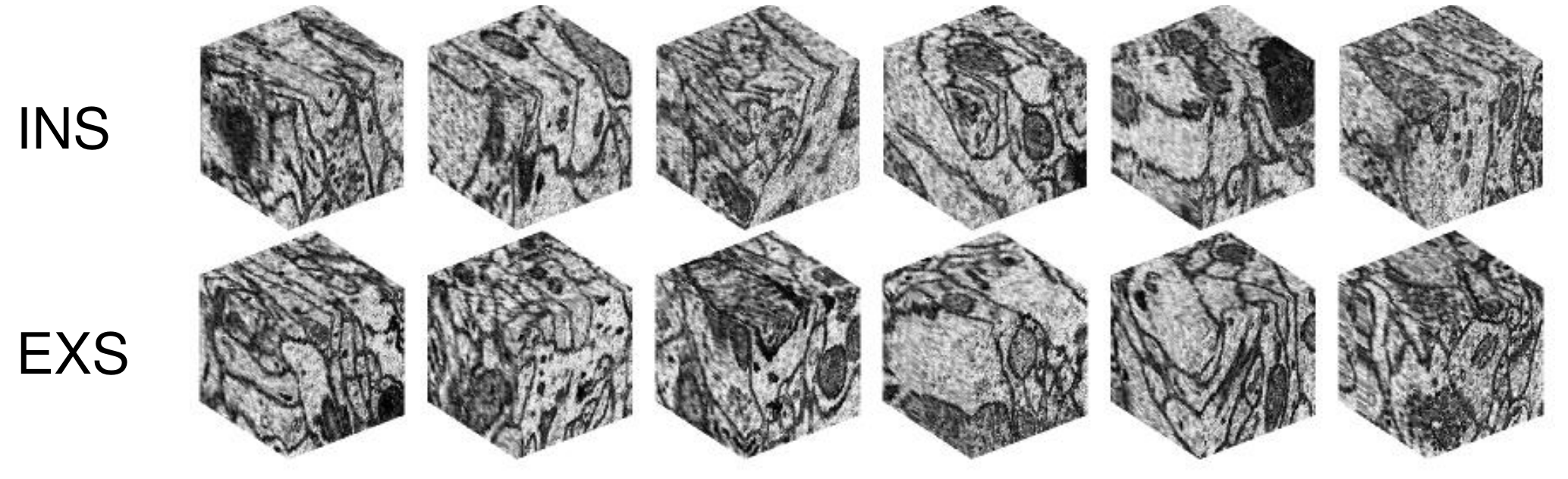}
        \caption{Example images of the two classes in the SynapseMNIST3D dataset. INS: inhibitory synapse; EXS: excitatory synapse.}
        \label{fig:synapse3d-data}
    \end{figure}
\end{itemize}

\section{Data Separation by Task Type, Data Scale, and Data Modality}
Here we give the detailed breakdown of the 17 datasets into different groups based on data modality, data scale, and task type.

\subsection{Data Modality}
For data modality, we have four groups: Radiology, Microscopy, Ophthalmology, and Dermatology. 
\begin{itemize}
    \item Radiology: ChestMNIST (Chest X-Ray), PneumoniaMNIST (Chest X-Ray), OrganAMNIST (Abdominal CT), OrganCMNIST (Abdominal CT), OrganSMNIST (Abdominal CT), OrganMNIST3D (Abdominal CT), NoduleMNIST3D (Chest CT), FractureMNIST3D (Chest CT), AdrenalMNIST3D (Abdominal CT), and VesselMNIST3D (Brain MRA).
    \item Microscopy: PathMNIST (Colon Pathology), BloodMNIST (Blood Cell Microscope), TissueMNIST (Kidney Cortex Microscope), and SynapseMNIST3D (Electron Microscope).
    \item Ophthalmology: OCTMNIST (Retinal OCT), and RetinaMNIST (Fundus Camera).
    \item Dermatology: DermaMNIST (Dermatoscope).
\end{itemize}
\subsection{Data Scale}
For data scale, we have four groups based on the sample size $n$ of each dataset: G1 ($n\leqslant$10K), G2 ($10{\rm K}<n\leqslant$50K), G3 ($50{\rm K}<n\leqslant$100K), and G4 ($100{\rm K}<n$). Here we only use 2D datasets since all the six 3D datasets has a sample size about 1500. 
\begin{itemize}
    \item G1 ($n\leqslant$10K): RetinaMNIST (1600 samples), PneumoniaMNIST (5856 samples)
    \item G2 ($10{\rm K}<n\leqslant$50K): DermaMNIST (10015 samples), BloodMNIST (17092 samples), OrganCMNIST (23583 samples), OrganSMNIST (25211 samples).
    \item G3 ($50{\rm K}<n\leqslant$100K): OrganAMNIST (58830 samples).
    \item G4 ($100{\rm K}<n$): PathMNIST (107180 samples), OCTMNIST (109309 samples), ChestMNIST (112120 samples), TissueMNIST (236386 samples).
\end{itemize}

\subsection{Task Type}
For task type, we have four groups based on the number of class $n$ for each classification task: G1 ($n$=2), G2 (2$<n\leqslant$5), G3 (5$<n\leqslant$10), and G4 (10$<n$).
\begin{itemize}
    \item G1 ($n$=2): ChestMNIST (Multi-Label (14) Binary-Class (2)), PneumoniaMNIST (Binary-Class ), NoduleMNIST3D (Binary-Class), AdrenalMNIST3D (Binary-Class), VesselMNIST3D (Binary-Class), SynapseMNIST3D (Binary-Class).
    \item G2 (2$<n\leqslant$5): OCTMNIST (4-Class), RetinaMNIST (5-Class), FractureMNIST3D (3-Class).
    \item G3 (5$<n\leqslant$10): PathMNIST (9-Class), DermaMNIST (7-Class), BloodMNIST (8-Class), TissueMNIST (8-Class).
    \item G4 (10$<n$): OrganAMNIST (11-Class), OrganCMNIST (11-Class), OrganSMNIST (11-Class), OrganMNIST3D (11-Class).
\end{itemize}

\section{Data-distribution-based Model Layer Design}
The design of model layers is determined based on the data distribution for specific tasks. Formally, for a $k$-class classification dataset, we compute the average pixel values of all images in the training set for each class and normalize these values to the range [0, 1]. This results in $k$ values, and the variance of these values serves as an index for determining the model architecture. The variance and the corresponding model layer configurations for eleven 2D datasets and six 3D datasets are presented in Table. \ref{tab:model-layer}.
\begin{table}[h]
    \centering
    \caption{Data distribution and model layer for different datasets.}
    \begin{tabular}{l|c|c}
        \hline
        \textbf{MedMNIST2D} & \textbf{Variance} & \textbf{Layer} \\
        \hline
         RetinaMNIST& 0.11 & 5 \\
         PneumoniaMNIST & 0.25 & 4 \\
         DermaMNIST & 0.13 & 5 \\
         BloodMNIST & 0.10 & 5 \\
         OrganAMNIST & 0.09 & 5 \\
         OrganCMNIST & 0.09 & 5 \\
         OrganSMNIST & 0.09 & 5 \\
         PathMNIST & 0.07 & 6 \\
         OCTMNIST & 0.15 & 5 \\
         ChestMNIST & 0.25 & 4 \\
         TissueMNIST & 0.10 & 5 \\
         \hline
         \hline
         \textbf{MedMNIST3D} & \textbf{Variance} & \textbf{Layer} \\
         \hline
         OrganMNIST3D & 0.09 & 6\\
         NoduleMNIST3D & 0.25 & 5\\
         FractureMNIST3D & 0.18 & 5\\
         AdrenalMNIST3D & 0.25 & 5\\
         VesselMNIST3D & 0.25 & 5 \\
         SynapseMNIST3D & 0.25 & 5\\
         \hline
    \end{tabular}
    \label{tab:model-layer}
\end{table}
The variance reflects the degree of dispersion among samples from different classes within the dataset. A larger variance indicates greater separability between classes, making the classification task comparatively easier. Consequently, datasets with larger variances require fewer model layers, while those with smaller variances require deeper models.

For 2D datasets, a 5-layer architecture serves as the baseline for datasets with variances around 0.10. For datasets with significantly larger variances, such as PneumoniaMNIST2D (0.25) and ChestMNIST2D (0.25), the number of layers is reduced to 4. Conversely, for datasets with smaller variances, such as PathMNIST2D (0.07), the number of layers is increased to 6.

Similarly, for 3D datasets, a 5-layer architecture is used as the baseline for datasets with variances around 0.25. For datasets with greatly smaller variances, such as OrganMNIST3D (0.09), the model depth is increased to 6 to accommodate the greater complexity of the classification task.

\section{Loss Function and Evaluation Metrics}
\subsection{Loss Function}
There are mainly two different tasks in MedMNIST v2 (\{multi-label,binary-class\},\{binary/multi-class,ordinal regression\}), we use different loss functions $\mathcal{L}$ for them respectively. For multi-label, binary-class, we employ the BCEWithLogitsLoss $\mathcal{L}=\{L_1,L_2,\cdots,L_n\}$ where $n$ is the label number and $L_i$ is as follows
\begin{equation}
    \mathcal{L}=-\frac{1}{N}\sum_{n=1}^N\sum_{k=1}^C(z_{n,k}\times \log{\sigma(y_{n,k})}+(1-z_{n,k})\times \log{(1-\sigma(y_{n,k}))}
\end{equation}
where $N$ is the number of samples in the current batch, $C$ is the label number, $z_{n,k}$ is the binary label for sample $n$ and label $k$, $y_{n,k}$ is the logit for sample $n$ and class label $k$, $\sigma$ is the sigmoid function. 
For binary/multi-class, ordinal regression, we use the CrossEntropyLoss
\begin{equation}
    \mathcal{L}=-\frac{1}{N}\sum_{n=1}^N\log{\frac{\exp{(y_{n,l_n})}}{\sum_{t=1}^C\exp{(y_{n,t})}}}
\end{equation}
where $N$ is the number of samples in the current batch, $C$ is the total number of classes, $y_{n,t}$ is the logit for class $c$ of the $n$-th sample, and $l_n$ is the logit of the target class of the $n$-th sample.

\subsection{Evaluation Metrics}
The evaluation metrics employed are the Area under the ROC curve (AUC) and Accuracy (ACC). AUC evaluates the continuous prediction scores without relying on a threshold, whereas ACC assesses the discrete prediction labels based on a given threshold. AUC is less susceptible to class imbalance compared to ACC. There is no severe class imbalance in the MeMNIST v2 dataset, so ACC can effectively serve as a reliable metric. While numerous other metrics exist, we simply use AUC and ACC have ensure a fair comparison with existing benchmark methods.

\section{Detailed Model Performance}
The detailed model performance between MTDL and other models over MedMNIST2D and MedMNIST3D are shown in Table \ref{tab:result-2d} and Table \ref{tab:result-3d} respectively. The average performance of all models over MedMNIST2D and MedMNIST3D are illustrated in Table \ref{tab:ave-result-2d} and Table \ref{tab:ave-result-3d} respectively.

\begin{table}
    \centering
    \scriptsize
    \caption{Performance comparison between our model and other benchmarks in terms of Accuracy (ACC) and Area Under the ROC Curve (AUC) on the MedMNIST2D dataset. The benchmark results are taken from the original papers \cite{liu2022feature,yang2023medmnist,manzari2023medvit,doerrich2023unoranic}. The best results are in bold.
    }
    \begin{tabular}{l|cc|cc|cc|cc|cc|cc}
    \hline
    Methods & \multicolumn{2}{c|}{Path} & \multicolumn{2}{c|}{Chest} & \multicolumn{2}{c|}{Derma} & \multicolumn{2}{c|}{OCT} & \multicolumn{2}{c|}{Pneumonia} & \multicolumn{2}{c}{Retina} \\
    & AUC   & ACC   & AUC   & ACC   & AUC   & ACC   & AUC   & ACC   & AUC   & ACC   & AUC   & ACC   \\
    \hline
    ResNet-18 (28) & 0.983 & 0.907 & 0.768 & 0.947 & 0.917 & 0.735 & 0.943 & 0.743 & 0.944 & 0.854 & 0.717 & 0.524 \\
    ResNet-18 (224) & 0.989 & 0.909 & 0.773 & 0.947 & 0.920 & 0.754 & 0.958 & 0.763 & 0.956 & 0.864 & 0.710 & 0.493 \\
    ResNet-50 (28) & 0.990 & 0.911 & 0.769 & 0.947 & 0.913 & 0.735 & 0.952 & 0.762 & 0.948 & 0.854 & 0.726 & 0.528 \\
    ResNet-50 (224) & 0.989 & 0.892 & 0.773 & 0.948 & 0.912 & 0.731 & 0.958 & 0.776 & 0.962 & 0.884 & 0.716 & 0.511 \\
    auto-sklearn & 0.934 & 0.716 & 0.649 & 0.779 & 0.902 & 0.719 & 0.887 & 0.601 & 0.942 & 0.855 & 0.690 & 0.515 \\
    AutoKeras & 0.959 & 0.834 & 0.742 & 0.937 & 0.915 & 0.749 & 0.955 & 0.763 & 0.947 & 0.878 & 0.719 & 0.503 \\
    Google AutoML & 0.944 & 0.728 & 0.778 & 0.948 & 0.914 & 0.768 & 0.963 & 0.771 & 0.991 & 0.946 & 0.750 & 0.531 \\
    MedVIT-T (224)& {0.994} & {0.938} & 0.786 & {0.956} & 0.914 & 0.768 & 0.961 & 0.767 & {0.993} & {0.949} & 0.752 & 0.534 \\
    MedVIT-S (224)& 0.993 & \textbf{0.942} & 0.791 & 0.954 & {0.937} & {0.780} & 0.960 & 0.782 & \textbf{0.995} & \textbf{0.961} & {0.773} & 0.561 \\
    MedVIT-L (224)& 0.984 & 0.933 & \textbf{0.805} & \textbf{0.959} & 0.920 & 0.773 & 0.945 & 0.761 & 0.991 & 0.921 & 0.754 & 0.552 \\
    ViT           & 0.962 & 0.785 & 0.724 & 0.947 & 0.914 & 0.745 & 0.905 & 0.679 & 0.958 & 0.885 & 0.750 & 0.565\\
    Resnet+ViT    & 0.991 & 0.915 & 0.703 & 0.947 & 0.906 & 0.748 & {0.968} & 0.807 & 0.972 & 0.897 & 0.740 & 0.548\\
    unORANIC & - & - & - & -& 0.776 & 0.699 & - & - &0.961 & 0.862 & 0.691 & 0.530 \\
    FPViT         & {0.994} & 0.918 & 0.725 & 0.948 & 0.923 & 0.766 & {0.968} & {0.813} & 0.973 & 0.896 & 0.753 & {0.568}\\
    BSDA & 0.992 & 0.919 & - & - & 0.931 & 0.764 & \textbf{0.989} & \textbf{0.888} & 0.957 & 0.888 & 0.750 & 0.533\\
    \textbf{MTDL}     & \textbf{0.996} & 0.920  &   {0.793}   & 0.948    & \textbf{0.962} & \textbf{0.836} & \textbf{0.989} & \textbf{0.888} & 0.986 & 0.910 & \textbf{0.874} & \textbf{0.655} \\
    \hline
    \hline
    Methods & \multicolumn{2}{c|}{Blood} & \multicolumn{2}{c|}{Tissue} & \multicolumn{2}{c|}{OrganA} & \multicolumn{2}{c|}{OrganC} & \multicolumn{2}{c}{OrganS} & \\
    & AUC   & ACC   & AUC   & ACC   & AUC   & ACC   & AUC   & ACC   & AUC   & ACC & &  \\
    \hline
    ResNet-18 (28) & {0.998} & 0.958 & 0.930 & 0.676 & {0.997} & 0.935 & 0.992 & 0.900 & 0.972 & 0.782 & &\\
    ResNet-18 (224) & {0.998} & 0.963 & 0.933 & 0.681 & \textbf{0.998} & {0.951} & {0.994} & 0.920 & 0.974 & 0.778 & &\\
    ResNet-50 (28) & 0.997 & 0.956 & 0.931 & 0.680 & {0.997} & 0.935 & 0.992 & 0.905 & 0.972 & 0.770 & &\\
    ResNet-50 (224) & 0.997 & 0.950 & 0.932 & 0.680 & \textbf{0.998} & 0.947 & 0.993 & 0.911 & 0.975 & 0.785 & &\\
    auto-sklearn & 0.984 & 0.878 & 0.828 & 0.532 & 0.963 & 0.762 & 0.976 & 0.829 & 0.945 & 0.672 & &\\
    AutoKeras & {0.998} & 0.961 & 0.941 & 0.703 & 0.994 & 0.905 & 0.990 & 0.879 & 0.974 & \textbf{0.813} & &\\
    Google AutoML & {0.998} & {0.966} & 0.924 & 0.673 & 0.990 & 0.886 & 0.988 & 0.877 & 0.964 & 0.749 & &\\
    MedVIT-T (224)& 0.996 & 0.950 & 0.943 & 0.703 & 0.995 & 0.931 & 0.991 & 0.901 & 0.972 & 0.789 \\
    MedVIT-S (224)& 0.997 & 0.951 & {0.952} & \textbf{0.731} & 0.996 & 0.928 & 0.993 & 0.916 & \textbf{0.987} & 0.805 \\
    MedVIT-L (224)& 0.996 & 0.954 & 0.935 & 0.699 & {0.997} & 0.943 & {0.994} & {0.922} & 0.973 & 0.806 \\
    ViT           &  -    &   -   &  -    &  -    & 0.978 & 0.830 & 0.976 & 0.835 & 0.939 & 0.657  \\
    Resnet+ViT    &  -    &   -   &  -    &  -    & 0.995 & 0.929 & 0.991 & 0.900 & 0.971 & 0.783\\
    unORANIC & 0.977 & 0.848 & -&-&-&-&-&-&-&-\\
    FPViT         & -     &   -   &  -    &  -    & {0.997} & 0.935 & 0.993 & 0.903 & 0.976 & 0.785\\
    BSDA & \textbf{0.999} & \textbf{0.988} & 0.937 & 0.704 & - & - & - & - & - & - \\
    \textbf{MTDL}     & \textbf{0.999} & \textbf{0.988} & \textbf{0.945} & {0.721} & \textbf{0.998} & \textbf{0.956} & \textbf{0.996} & \textbf{0.928} & {0.978} & {0.809}\\
    \end{tabular}
    \label{tab:result-2d}
\end{table}

\begin{table}
    \centering
    \scriptsize
    \caption{Performance comparison between our model and other benchmarks in terms of Accuracy (ACC) and Area Under the ROC Curve (AUC) on the MedMNIST3D dataset. The benchmark results are taken from the original papers \cite{liu2022feature,zheng2023complex,yang2023medmnist,zhu2024bsda,zhemchuzhnikov2024ilpo}. The results are in bold.
    }
    \begin{tabular}{l|cc|cc|cc|cc|cc|cc}
    \hline
    Methods & \multicolumn{2}{c|}{Organ} & \multicolumn{2}{c|}{Nodule} & \multicolumn{2}{c|}{Fracture} & \multicolumn{2}{c|}{Adrenal} & \multicolumn{2}{c|}{Vessel} & \multicolumn{2}{c}{Synapse} \\
    & AUC   & ACC   & AUC   & ACC   & AUC   & ACC   & AUC   & ACC   & AUC   & ACC   & AUC   & ACC   \\
    \hline
    ResNet-18 + 2.5D & 0.977 & 0.788 & 0.838 & 0.835 &0.587 &0.451 &0.718& 0.772& 0.748 &0.846 &0.634 &0.696 \\
    ResNet-18 + 3D & {0.996} &0.907 &0.863 &0.844 &0.712 &0.508 &0.827 &0.721 &0.874 &0.877 &0.820 &0.745 \\
    ResNet-18 + ACS & 0.994 &0.900 &0.873 &0.847 &0.714 &0.497 &0.839& 0.754 &0.930 &0.928 &0.705 &0.722 \\
    ResNet-50 + 2.5D & 0.974 &0.769 &0.835 &0.848 &0.552 &0.397 &0.732& 0.763 &0.751 &0.877 &0.669 &0.735 \\
    ResNet-50 + 3D & 0.994 &0.883 &0.875 &0.847 &0.725 &0.494 &0.828& 0.745 &0.907 &0.918 &0.851 &0.795 \\
    ResNet-50 + ACS & 0.994 &0.889 &0.886 &0.841 &0.750 &0.517 &0.828& 0.758 &0.912 &0.858 &0.719 &0.709 \\
    auto-sklearn & 0.977 &0.814 & 0.914 &\textbf{0.874} &0.628 &0.453 &0.828& 0.802&0.910 &0.915 &0.631 &0.730 \\
    AutoKeras& 0.979 &0.804 &0.844 &0.834 &0.642 &0.458 &0.804 &0.705& 0.773 &0.894 &0.538 &0.724 \\
    FPViT (224)& 0.923 &0.800 &0.814 &0.822 &0.640 &0.438 &0.801 &0.704& 0.770 &0.888 &0.530 &0.712 \\
    BSDA    & 0.994 & 0.887 & 0.892 & 0.861 & 0.731 & 0.569 &0.892 & {0.838}& 0.917 & 0.932 & - & -\\
    ILPO-NET (average) & 0.972 &0.728 &0.900 &0.861 &\textbf{0.776} &0.577 &0.880 &0.811 &0.888 &0.888&0.854&0.782\\
    C-Mixer  & 0.995 &{0.912} &{0.915} &0.860 &0.729 &\textbf{0.660} &\textbf{0.969} &0.801& {0.932} &\textbf{0.940} &{0.866} &{0.820} \\
    \textbf{MTDL}     & \textbf{0.999} & \textbf{0.952}  &   \textbf{0.916}   &  {0.865}    & {0.753} & {0.583} & {0.903} & \textbf{0.862} & \textbf{0.938} & {0.937} & \textbf{0.951} & \textbf{0.931} \\
    \hline
    \end{tabular}
    \label{tab:result-3d}
\end{table}

\begin{table}
    \centering
    \caption{Average Performance Comparison in ACC and AUC over all datasets in MedMNIST2D. The best result is in bold. }
    \begin{tabular}{l|cc}
        \hline
        Methods & \multicolumn{2}{c}{Average}  \\
        & AUC   & ACC     \\
        \hline
		ResNet-18 (28)   & 0.924 &0.815 \\ 
		ResNet-18 (224)  & 0.928 &0.820\\ 
        ResNet-50 (28)   & 0.926 &0.817\\
        ResNet-50 (224)  & 0.928 &0.820\\
        auto-sklearn     & 0.882 &0.714\\
        AutoKeras        & 0.921 &0.811\\
        Google AutoML    & 0.928 &0.804\\
        MedVIT-T         & 0.936 &0.835\\
        MedVIT-S         & 0.943 &0.846\\
        MedVIT-L         & 0.936 &0.838\\
        ViT              & 0.901 &0.770\\
        Resnet+ViT       & 0.915 &0.830\\
        unORANIC         & 0.851 &0.735\\
        FPViT            & 0.922 &0.837\\
        BSDA             & 0.936 & 0.812\\
        \textbf{MTDL}   & \textbf{0.956} &\textbf{0.868}\\
        \hline
	\end{tabular}
	\label{tab:ave-result-2d}
\end{table}

\begin{table}
    \centering
    \caption{Average Performance Comparison in ACC and AUC over all datasets in MedMNIST3D, the best result is in bold.}
    \begin{tabular}{l|cc}
        \hline
        Methods & \multicolumn{2}{c}{Average}  \\
        & AUC   & ACC     \\
        \hline
		ResNet-18 + 2.5D   & 0.750 &0.731 \\ 
		ResNet-18 + 3D  & 0.849 &0.767\\ 
        ResNet-18 + ACS   & 0.842 &0.775\\
        ResNet-50 + 2.5D  & 0.752 &0.732\\
        ResNet-50 + 3D    & 0.863 &0.780\\
        ResNet-50 + ACS   & 0.848 &0.762\\
        auto-sklearn     & 0.815 &0.765\\
        AutoKeras         & 0.763 &0.737\\
        FPVT             & 0.746 &0.727\\
        BSDA             & 0.885 & 0.817\\
        ILPO-NET(average)& 0.878 &0.775\\
        C-Mixer         & 0.901 &0.832\\
        \textbf{MTDL}  &\textbf{0.910} &\textbf{0.855}\\
        \hline
	\end{tabular}
	\label{tab:ave-result-3d}
\end{table}


\begin{thebibliography}{48}
\providecommand{\natexlab}[1]{#1}
\providecommand{\url}[1]{\texttt{#1}}
\expandafter\ifx\csname urlstyle\endcsname\relax
  \providecommand{\doi}[1]{doi: #1}\else
  \providecommand{\doi}{doi: \begingroup \urlstyle{rm}\Url}\fi

\bibitem[Hajij et~al.(2022)Hajij, Zamzmi, Papamarkou, Miolane, Guzm{\'a}n-S{\'a}enz, Ramamurthy, Birdal, Dey, Mukherjee, Samaga, et~al.]{hajij2022topological}
Mustafa Hajij, Ghada Zamzmi, Theodore Papamarkou, Nina Miolane, Aldo Guzm{\'a}n-S{\'a}enz, Karthikeyan~Natesan Ramamurthy, Tolga Birdal, Tamal~K Dey, Soham Mukherjee, Shreyas~N Samaga, et~al.
\newblock Topological deep learning: Going beyond graph data.
\newblock \emph{arXiv preprint arXiv:2206.00606}, 2022.

\bibitem[Cang and Wei(2017)]{cang2017topologynet}
Zixuan Cang and Guo-Wei Wei.
\newblock Topologynet: Topology based deep convolutional and multi-task neural networks for biomolecular property predictions.
\newblock \emph{PLoS computational biology}, 13\penalty0 (7):\penalty0 e1005690, 2017.

\bibitem[Nguyen et~al.(2019)Nguyen, Cang, Wu, Wang, Cao, and Wei]{nguyen2019mathematical}
Duc~Duy Nguyen, Zixuan Cang, Kedi Wu, Menglun Wang, Yin Cao, and Guo-Wei Wei.
\newblock Mathematical deep learning for pose and binding affinity prediction and ranking in d3r grand challenges.
\newblock \emph{Journal of computer-aided molecular design}, 33:\penalty0 71--82, 2019.

\bibitem[Papamarkou et~al.(2024)Papamarkou, Birdal, Bronstein, Carlsson, Curry, Gao, Hajij, Kwitt, Lio, Di~Lorenzo, et~al.]{papamarkou2024position}
Theodore Papamarkou, Tolga Birdal, Michael~M Bronstein, Gunnar~E Carlsson, Justin Curry, Yue Gao, Mustafa Hajij, Roland Kwitt, Pietro Lio, Paolo Di~Lorenzo, et~al.
\newblock Position: Topological deep learning is the new frontier for relational learning.
\newblock In \emph{Forty-first International Conference on Machine Learning}, 2024.

\bibitem[Carlsson(2009)]{carlsson2009topology}
Gunnar Carlsson.
\newblock Topology and data.
\newblock \emph{Bulletin of the American Mathematical Society}, 46\penalty0 (2):\penalty0 255--308, 2009.

\bibitem[Edelsbrunner and Harer(2010)]{edelsbrunner2010computational}
Herbert Edelsbrunner and John Harer.
\newblock \emph{Computational topology: an introduction}.
\newblock American Mathematical Soc., 2010.

\bibitem[Nguyen et~al.(2020)Nguyen, Gao, Wang, and Wei]{nguyen2020mathdl}
Duc~Duy Nguyen, Kaifu Gao, Menglun Wang, and Guo-Wei Wei.
\newblock Mathdl: mathematical deep learning for d3r grand challenge 4.
\newblock \emph{Journal of computer-aided molecular design}, 34:\penalty0 131--147, 2020.

\bibitem[Ziou and Allili(2002)]{ziou2002generating}
Djemel Ziou and Madjid Allili.
\newblock Generating cubical complexes from image data and computation of the euler number.
\newblock \emph{Pattern Recognition}, 35\penalty0 (12):\penalty0 2833--2839, 2002.

\bibitem[Shen et~al.(2024)Shen, Feng, Li, Lei, Wu, and Wei]{shen2024knot}
Li~Shen, Hongsong Feng, Fengling Li, Fengchun Lei, Jie Wu, and Guo-Wei Wei.
\newblock Knot data analysis using multiscale gauss link integral.
\newblock \emph{Proceedings of the National Academy of Sciences}, 121\penalty0 (42):\penalty0 e2408431121, 2024.

\bibitem[Singh et~al.(2023)Singh, Farrelly, Hathaway, Leiner, Jagtap, Carlsson, and Erickson]{singh2023topological}
Yashbir Singh, Colleen~M Farrelly, Quincy~A Hathaway, Tim Leiner, Jaidip Jagtap, Gunnar~E Carlsson, and Bradley~J Erickson.
\newblock Topological data analysis in medical imaging: current state of the art.
\newblock \emph{Insights into Imaging}, 14\penalty0 (1):\penalty0 58, 2023.

\bibitem[Su et~al.(2024{\natexlab{a}})Su, Tong, and Wei]{su2024persistent}
Zhe Su, Yiying Tong, and Guo-Wei Wei.
\newblock Persistent de rham-hodge laplacians in eulerian representation for manifold topological learning.
\newblock \emph{AIMS Mathematics}, 9\penalty0 (10):\penalty0 27438--27470, 2024{\natexlab{a}}.

\bibitem[Su et~al.(2024{\natexlab{b}})Su, Tong, and Wei]{su2024hodge}
Zhe Su, Yiying Tong, and Guowei Wei.
\newblock Hodge decomposition of vector fields in cartesian grids.
\newblock In \emph{SIGGRAPH Asia 2024 Conference Papers}, pages 1--10, 2024{\natexlab{b}}.

\bibitem[Su et~al.(2024{\natexlab{c}})Su, Tong, and Wei]{su2024hodge2}
Zhe Su, Yiying Tong, and Guo-Wei Wei.
\newblock Hodge decomposition of single-cell rna velocity.
\newblock \emph{Journal of chemical information and modeling}, 64\penalty0 (8):\penalty0 3558--3568, 2024{\natexlab{c}}.

\bibitem[Yang et~al.(2023)Yang, Shi, Wei, Liu, Zhao, Ke, Pfister, and Ni]{yang2023medmnist}
Jiancheng Yang, Rui Shi, Donglai Wei, Zequan Liu, Lin Zhao, Bilian Ke, Hanspeter Pfister, and Bingbing Ni.
\newblock Medmnist v2-a large-scale lightweight benchmark for 2d and 3d biomedical image classification.
\newblock \emph{Scientific Data}, 10\penalty0 (1):\penalty0 41, 2023.

\bibitem[Yang et~al.(2021)Yang, Shi, and Ni]{yang2021medmnist}
Jiancheng Yang, Rui Shi, and Bingbing Ni.
\newblock Medmnist classification decathlon: A lightweight automl benchmark for medical image analysis.
\newblock In \emph{2021 IEEE 18th International Symposium on Biomedical Imaging (ISBI)}, pages 191--195. IEEE, 2021.

\bibitem[Al-Dhabyani et~al.(2020)Al-Dhabyani, Gomaa, Khaled, and Fahmy]{al2020dataset}
Walid Al-Dhabyani, Mohammed Gomaa, Hussien Khaled, and Aly Fahmy.
\newblock Dataset of breast ultrasound images.
\newblock \emph{Data in brief}, 28:\penalty0 104863, 2020.

\bibitem[Paw{\l}owska et~al.(2023)Paw{\l}owska, Karwat, and {\.Z}o{\l}ek]{pawlowska2023re}
Anna Paw{\l}owska, Piotr Karwat, and Norbert {\.Z}o{\l}ek.
\newblock re:“[dataset of breast ultrasound images by w. al-dhabyani, m. gomaa, h. khaled \& a. fahmy, data in brief, 2020, 28, 104863]”.
\newblock \emph{Data in Brief}, 48, 2023.

\bibitem[Liu et~al.(2022{\natexlab{a}})Liu, Li, Cao, Liu, and Cao]{liu2022feature}
Jinwei Liu, Yan Li, Guitao Cao, Yong Liu, and Wenming Cao.
\newblock Feature pyramid vision transformer for medmnist classification decathlon.
\newblock In \emph{2022 International joint conference on neural networks (IJCNN)}, pages 1--8. IEEE, 2022{\natexlab{a}}.

\bibitem[Manzari et~al.(2023)Manzari, Ahmadabadi, Kashiani, Shokouhi, and Ayatollahi]{manzari2023medvit}
Omid~Nejati Manzari, Hamid Ahmadabadi, Hossein Kashiani, Shahriar~B Shokouhi, and Ahmad Ayatollahi.
\newblock Medvit: a robust vision transformer for generalized medical image classification.
\newblock \emph{Computers in Biology and Medicine}, 157:\penalty0 106791, 2023.

\bibitem[Zheng and Jia(2023)]{zheng2023complex}
Zhuoran Zheng and Xiuyi Jia.
\newblock Complex mixer for medmnist classification decathlon.
\newblock \emph{arXiv preprint arXiv:2304.10054}, 2023.

\bibitem[Doerrich et~al.(2023)Doerrich, Di~Salvo, and Ledig]{doerrich2023unoranic}
Sebastian Doerrich, Francesco Di~Salvo, and Christian Ledig.
\newblock unoranic: Unsupervised orthogonalization of anatomy and image-characteristic features.
\newblock In \emph{International Workshop on Machine Learning in Medical Imaging}, pages 62--71. Springer, 2023.

\bibitem[Zhu et~al.(2024)Zhu, Cai, Wang, Chen, Fu, and Yao]{zhu2024bsda}
Yaoyao Zhu, Xiuding Cai, Xueyao Wang, Xiaoqing Chen, Zhongliang Fu, and Yu~Yao.
\newblock Bsda: Bayesian random semantic data augmentation for medical image classification.
\newblock \emph{Sensors}, 24\penalty0 (23):\penalty0 7511, 2024.

\bibitem[Zhemchuzhnikov and Grudinin(2024)]{zhemchuzhnikov2024ilpo}
Dmitrii Zhemchuzhnikov and Sergei Grudinin.
\newblock Ilpo-net: Network for the invariant recognition of arbitrary volumetric patterns in 3d.
\newblock In \emph{Joint European Conference on Machine Learning and Knowledge Discovery in Databases}, pages 352--368. Springer, 2024.

\bibitem[Kather et~al.(2018)Kather, Halama, and Marx]{kather_2018_1214456}
Jakob~Nikolas Kather, Niels Halama, and Alexander Marx.
\newblock 100,000 histological images of human colorectal cancer and healthy tissue, April 2018.
\newblock URL \url{https://doi.org/10.5281/zenodo.1214456}.

\bibitem[Kermany et~al.(2018{\natexlab{a}})Kermany, Zhang, and Goldbaum]{kermany2018large}
Daniel Kermany, Kang Zhang, and Michael Goldbaum.
\newblock Large dataset of labeled optical coherence tomography (oct) and chest x-ray images.
\newblock \emph{Mendeley Data}, 3\penalty0 (10.17632), 2018{\natexlab{a}}.

\bibitem[Tschandl et~al.(2018)Tschandl, Rosendahl, and Kittler]{tschandl2018ham10000}
Philipp Tschandl, Cliff Rosendahl, and Harald Kittler.
\newblock The ham10000 dataset, a large collection of multi-source dermatoscopic images of common pigmented skin lesions.
\newblock \emph{Scientific data}, 5\penalty0 (1):\penalty0 1--9, 2018.

\bibitem[Niu et~al.(2025)Niu, Lyu, Carothers, Kaviani, Tan, Yan, Kalra, Whitlow, and Wang]{niu2023specialty}
Chuang Niu, Qing Lyu, Christopher~D Carothers, Parisa Kaviani, Josh Tan, Pingkun Yan, Mannudeep~K Kalra, Christopher~T Whitlow, and Ge~Wang.
\newblock Specialty-oriented generalist medical ai for chest ct screening.
\newblock \emph{Nature Communications}, 2025.

\bibitem[Hodge(1989)]{hodge1989theory}
William Vallance~Douglas Hodge.
\newblock \emph{The theory and applications of harmonic integrals}.
\newblock CUP Archive, 1989.

\bibitem[Morrey(1956)]{morrey1956variational}
Charles~B Morrey.
\newblock A variational method in the theory of harmonic integrals, ii.
\newblock \emph{American Journal of Mathematics}, 78\penalty0 (1):\penalty0 137--170, 1956.

\bibitem[Shonkwiler(2009)]{shonkwiler2009poincare}
Clayton Shonkwiler.
\newblock \emph{Poincar{\'e} duality angles on Riemannian manifolds with boundary}.
\newblock PhD thesis, University of Pennsylvania, 2009.

\bibitem[Ribando-Gros et~al.(2024)Ribando-Gros, Wang, Chen, Tong, and Wei]{ribando2024combinatorial}
Emily Ribando-Gros, Rui Wang, Jiahui Chen, Yiying Tong, and Guo-Wei Wei.
\newblock Combinatorial and hodge laplacians: Similarities and differences.
\newblock \emph{SIAM Review}, 66\penalty0 (3):\penalty0 575--601, 2024.

\bibitem[Paszke et~al.(2019)Paszke, Gross, Massa, Lerer, Bradbury, Chanan, Killeen, Lin, Gimelshein, Antiga, et~al.]{paszke2019pytorch}
Adam Paszke, Sam Gross, Francisco Massa, Adam Lerer, James Bradbury, Gregory Chanan, Trevor Killeen, Zeming Lin, Natalia Gimelshein, Luca Antiga, et~al.
\newblock Pytorch: An imperative style, high-performance deep learning library.
\newblock \emph{Advances in neural information processing systems}, 32, 2019.

\bibitem[Loshchilov et~al.(2017)Loshchilov, Hutter, et~al.]{loshchilov2017fixing}
Ilya Loshchilov, Frank Hutter, et~al.
\newblock Fixing weight decay regularization in adam.
\newblock \emph{arXiv preprint arXiv:1711.05101}, 5, 2017.

\bibitem[Smith and Topin(2019)]{smith2019super}
Leslie~N Smith and Nicholay Topin.
\newblock Super-convergence: Very fast training of neural networks using large learning rates.
\newblock In \emph{Artificial intelligence and machine learning for multi-domain operations applications}, volume 11006, pages 369--386. SPIE, 2019.

\bibitem[Desbrun et~al.(2006)Desbrun, Kanso, and Tong]{desbrun2006discrete}
Mathieu Desbrun, Eva Kanso, and Yiying Tong.
\newblock Discrete differential forms for computational modeling.
\newblock In \emph{ACM SIGGRAPH 2006 Courses}, pages 39--54. 2006.

\bibitem[Kather et~al.(2019)Kather, Krisam, Charoentong, Luedde, Herpel, Weis, Gaiser, Marx, Valous, Ferber, et~al.]{kather2019predicting}
Jakob~Nikolas Kather, Johannes Krisam, Pornpimol Charoentong, Tom Luedde, Esther Herpel, Cleo-Aron Weis, Timo Gaiser, Alexander Marx, Nektarios~A Valous, Dyke Ferber, et~al.
\newblock Predicting survival from colorectal cancer histology slides using deep learning: A retrospective multicenter study.
\newblock \emph{PLoS medicine}, 16\penalty0 (1):\penalty0 e1002730, 2019.

\bibitem[Wang et~al.(2017)Wang, Peng, Lu, Lu, Bagheri, and Summers]{wang2017chestx}
Xiaosong Wang, Yifan Peng, Le~Lu, Zhiyong Lu, Mohammadhadi Bagheri, and Ronald~M Summers.
\newblock Chestx-ray8: Hospital-scale chest x-ray database and benchmarks on weakly-supervised classification and localization of common thorax diseases.
\newblock In \emph{Proceedings of the IEEE conference on computer vision and pattern recognition}, pages 2097--2106, 2017.

\bibitem[Codella et~al.(2019)Codella, Rotemberg, Tschandl, Celebi, Dusza, Gutman, Helba, Kalloo, Liopyris, Marchetti, et~al.]{codella2019skin}
Noel Codella, Veronica Rotemberg, Philipp Tschandl, M~Emre Celebi, Stephen Dusza, David Gutman, Brian Helba, Aadi Kalloo, Konstantinos Liopyris, Michael Marchetti, et~al.
\newblock Skin lesion analysis toward melanoma detection 2018: A challenge hosted by the international skin imaging collaboration (isic).
\newblock \emph{arXiv preprint arXiv:1902.03368}, 2019.

\bibitem[Kermany et~al.(2018{\natexlab{b}})Kermany, Goldbaum, Cai, Valentim, Liang, Baxter, McKeown, Yang, Wu, Yan, et~al.]{kermany2018identifying}
Daniel~S Kermany, Michael Goldbaum, Wenjia Cai, Carolina~CS Valentim, Huiying Liang, Sally~L Baxter, Alex McKeown, Ge~Yang, Xiaokang Wu, Fangbing Yan, et~al.
\newblock Identifying medical diagnoses and treatable diseases by image-based deep learning.
\newblock \emph{cell}, 172\penalty0 (5):\penalty0 1122--1131, 2018{\natexlab{b}}.

\bibitem[Liu et~al.(2022{\natexlab{b}})Liu, Wang, Wu, Dai, Fang, Yan, Son, Tang, Li, Gao, et~al.]{liu2022deepdrid}
Ruhan Liu, Xiangning Wang, Qiang Wu, Ling Dai, Xi~Fang, Tao Yan, Jaemin Son, Shiqi Tang, Jiang Li, Zijian Gao, et~al.
\newblock Deepdrid: Diabetic retinopathy—grading and image quality estimation challenge.
\newblock \emph{Patterns}, 3\penalty0 (6), 2022{\natexlab{b}}.

\bibitem[Organization(2006)]{world2006prevention}
World~Health Organization.
\newblock \emph{Prevention of blindness from diabetes mellitus: report of a WHO consultation in Geneva, Switzerland, 9-11 November 2005}.
\newblock World Health Organization, 2006.

\bibitem[Acevedo et~al.(2020)Acevedo, Merino, Alf{\'e}rez, Molina, Bold{\'u}, and Rodellar]{acevedo2020dataset}
Andrea Acevedo, Anna Merino, Santiago Alf{\'e}rez, {\'A}ngel Molina, Laura Bold{\'u}, and Jos{\'e} Rodellar.
\newblock A dataset of microscopic peripheral blood cell images for development of automatic recognition systems.
\newblock \emph{Data in brief}, 30:\penalty0 105474, 2020.

\bibitem[Woloshuk et~al.(2021)Woloshuk, Khochare, Almulhim, McNutt, Dean, Barwinska, Ferkowicz, Eadon, Kelly, Dunn, et~al.]{woloshuk2021situ}
Andre Woloshuk, Suraj Khochare, Aljohara~F Almulhim, Andrew~T McNutt, Dawson Dean, Daria Barwinska, Michael~J Ferkowicz, Michael~T Eadon, Katherine~J Kelly, Kenneth~W Dunn, et~al.
\newblock In situ classification of cell types in human kidney tissue using 3d nuclear staining.
\newblock \emph{Cytometry Part A}, 99\penalty0 (7):\penalty0 707--721, 2021.

\bibitem[Ljosa et~al.(2012)Ljosa, Sokolnicki, and Carpenter]{ljosa2012annotated}
Vebjorn Ljosa, Katherine~L Sokolnicki, and Anne~E Carpenter.
\newblock Annotated high-throughput microscopy image sets for validation.
\newblock \emph{Nature methods}, 9\penalty0 (7):\penalty0 637--637, 2012.

\bibitem[Bilic et~al.(2023)Bilic, Christ, Li, Vorontsov, Ben-Cohen, Kaissis, Szeskin, Jacobs, Mamani, Chartrand, et~al.]{bilic2023liver}
Patrick Bilic, Patrick Christ, Hongwei~Bran Li, Eugene Vorontsov, Avi Ben-Cohen, Georgios Kaissis, Adi Szeskin, Colin Jacobs, Gabriel Efrain~Humpire Mamani, Gabriel Chartrand, et~al.
\newblock The liver tumor segmentation benchmark (lits).
\newblock \emph{Medical Image Analysis}, 84:\penalty0 102680, 2023.

\bibitem[Armato~III et~al.(2011)Armato~III, McLennan, Bidaut, McNitt-Gray, Meyer, Reeves, Zhao, Aberle, Henschke, Hoffman, et~al.]{armato2011lung}
Samuel~G Armato~III, Geoffrey McLennan, Luc Bidaut, Michael~F McNitt-Gray, Charles~R Meyer, Anthony~P Reeves, Binsheng Zhao, Denise~R Aberle, Claudia~I Henschke, Eric~A Hoffman, et~al.
\newblock The lung image database consortium (lidc) and image database resource initiative (idri): a completed reference database of lung nodules on ct scans.
\newblock \emph{Medical physics}, 38\penalty0 (2):\penalty0 915--931, 2011.

\bibitem[Jin et~al.(2020)Jin, Yang, Kuang, Ni, Gao, Sun, Gao, Ma, Tan, Kang, et~al.]{jin2020deep}
Liang Jin, Jiancheng Yang, Kaiming Kuang, Bingbing Ni, Yiyi Gao, Yingli Sun, Pan Gao, Weiling Ma, Mingyu Tan, Hui Kang, et~al.
\newblock Deep-learning-assisted detection and segmentation of rib fractures from ct scans: Development and validation of fracnet.
\newblock \emph{EBioMedicine}, 62, 2020.

\bibitem[Yang et~al.(2020)Yang, Xia, Kin, and Igarashi]{yang2020intra}
Xi~Yang, Ding Xia, Taichi Kin, and Takeo Igarashi.
\newblock Intra: 3d intracranial aneurysm dataset for deep learning.
\newblock In \emph{Proceedings of the IEEE/CVF Conference on Computer Vision and Pattern Recognition}, pages 2656--2666, 2020.

\end{thebibliography}
\end{document}